\documentclass[12pt]{article}

\usepackage{graphicx}
\usepackage{amsmath,amssymb,amsfonts}
\usepackage{mathabx}
\usepackage{ascmac}
\usepackage{arydshln}
\usepackage{booktabs}
\usepackage{graphicx}
\usepackage{algorithm}
\usepackage{algpseudocode}
\usepackage{mathrsfs}
\usepackage{stmaryrd}
\usepackage{comment}
\usepackage{breakcites}
\usepackage{latexsym}
\usepackage{cases}
\usepackage{url}
\usepackage{pifont}
\usepackage{multirow}
\usepackage{setspace}
\usepackage{longtable}
\usepackage{booktabs}
\usepackage{threeparttable}
\usepackage{makecell}
\usepackage{booktabs}
\usepackage{graphicx}
\usepackage{threeparttable}
\usepackage{mathabx}
\usepackage{arydshln}
\usepackage{tikz}
\usetikzlibrary{patterns}
\usetikzlibrary{patterns.meta}
\usepackage{here}
\usepackage{natbib}
\usepackage[english]{babel}
\usepackage{threeparttable}
\usepackage[a4paper]{geometry}
\usepackage{amsthm}
\usepackage{prettyref}
\usepackage{natbib}
\usepackage{threeparttable}
\allowdisplaybreaks[1]

\usepackage[breaklinks=true]{hyperref}
\usepackage{amsthm}

\hypersetup{
  bookmarkstype=none,
  colorlinks=true,
  linkcolor=blue,
  citecolor=blue,
  filecolor=blue,
  urlcolor=blue
}

\usepackage{prettyref}

\tikzstyle{rednode} = [shape=rectangle, fill=red!50, line width=3]
\tikzstyle{bluenode} = [shape=rectangle, fill=blue!50, line width=3]
\tikzstyle{yellownode} = [shape=rectangle, fill=yellow!50, line width=3]
\tikzstyle{dotnode} = [dashed, pattern={Lines[angle=90,distance=3pt]}, pattern color=gray!150]
\tikzstyle{backslashnode} = [dashed, pattern={Lines[angle=45,distance=3pt]}, pattern color=gray!150]
\tikzstyle{slashnode} = [dashed, pattern={Lines[angle=-45,distance=3pt]}, pattern color=gray!150]

\usepackage{amsthm}
\newtheorem{theorem}{Theorem}[section]
\newtheorem{proposition}{Proposition}[section]

\newtheorem{assumption}{Assumption}[section]

\newtheorem{remark}[theorem]{Remark}
\newtheorem{definition}[theorem]{Definition}
\newtheorem{lemma}[theorem]{Lemma}
\newtheorem{example}[theorem]{Example}

\newcommand{\MS}{\mathcal{S}}
\newcommand{\MA}{\mathcal{A}}

\newcommand{\MZ}{\mathcal{Z}}

\newcommand{\MH}{\mathcal{H}}

\newcommand{\E}{\mathbb{E}}
\newcommand{\BP}{\mathbb{P}}

\newcommand{\I}{\mathbf{1}}
\newcommand{\Real}{\mathbb{R}}
\newcommand{\Natural}{\mathbb{N}}
\newcommand{\sign}{\mathrm{sign}}
\newcommand{\argmax}{\mathop{\rm arg~max}\limits}
\newcommand{\argmin}{\mathop{\rm arg~min}\limits}
\newcommand{\indep}{\perp \!\!\! \perp}

\usepackage{xr}
\makeatletter
\newcommand*{\addFileDependency}[1]{
  \typeout{(#1)}
  \@addtofilelist{#1}
  \IfFileExists{#1}{}{\typeout{No file #1.}}
}
\makeatother

\newcommand*{\myexternaldocument}[1]{
    \externaldocument{#1}
    \addFileDependency{#1.tex}
    \addFileDependency{#1.aux}
}

\myexternaldocument{supplement}

\makeatother

\begin{document}

\setstretch{1.2} 
\title{Policy Learning for Optimal Dynamic Treatment Regimes with Observational Data} 

\author{Shosei Sakaguchi\thanks{Faculty of Economics, The University of Tokyo, 7-3-1 Hongo, Bunkyo-ku, Tokyo 113-0033, Japan. Email: sakaguchi@e.u-tokyo.ac.jp.} }
\date{\today}

\maketitle


\begin{abstract}
\begin{spacing}{1.2}
   Public policies and medical interventions often involve dynamic treatment assignments, in which individuals receive a sequence of interventions over multiple stages. We study the statistical learning of optimal dynamic treatment regimes (DTRs) that determine the optimal treatment assignment for each individual at each stage based on their evolving history. We propose a novel, doubly robust, classification-based method for learning the optimal DTR from observational data under the sequential ignorability assumption. The method proceeds via backward induction: at each stage, it constructs and maximizes an augmented inverse probability weighting (AIPW) estimator of the policy value function to learn the optimal stage-specific policy. We show that the resulting DTR achieves an optimal convergence rate of $n^{-1/2}$ for welfare regret under mild convergence conditions on estimators of the nuisance components.
\end{spacing}
\end{abstract}

\thispagestyle{empty}
\clearpage
\addtocounter{page}{-1}
\setstretch{1.5}

\section{Introduction} \label{sec:introduction}

Public policies and medical interventions often involve dynamics in their treatment assignments. For example, some job training programs offer training sessions over multiple stages \citep[e.g.,][]{Lechner_2009,Rodriguez_et_al_2022}. In clinical medicine, physicians sequentially administer treatments in response to patients’ evolving conditions \citep[e.g.,][]{Wang_et_al_2012,Pelham_et_al_2016}. Multi-stage treatment assignments are also common in educational programs spanning multiple grades \citep[e.g.,][]{krueger_1999,Ding_Lehrer_2010} and in dynamic marketing strategies \citep[e.g.,][]{Liu_2023}.

This study focuses on the optimal allocation of sequential treatments \citep{Robins_1986}, in which individuals receive interventions over multiple stages. In this context, treatment effects at each stage are often heterogeneous depending on prior treatments and related states. Thus, adapting sequential treatment allocation to evolving information can substantially enhance the welfare gains of multi-stage interventions.

We study statistical learning for optimal sequential treatment assignment using data from quasi-experimental or observational studies. Throughout this paper, we assume sequential ignorability \citep{Robins_1997}, meaning that treatment assignment at each stage is independent of potential outcomes, conditional on the history of prior treatments and observed states. Under this assumption, we propose a novel method for learning the optimal dynamic treatment regime (DTR), a sequence of stage-specific policies that determines the optimal treatment for each individual at each stage based on their history up to that point.

In developing our approach, we build on recent advances in doubly robust policy learning \citep{Athey_Wager_2020, zhou2022offline} and extend them to dynamic settings. We propose a doubly robust, classification-based method for learning the optimal DTR via backward induction, which sequentially estimates the optimal policy from the final stage to the first. At each step, the method constructs an augmented inverse probability weighting (AIPW) estimator of the policy value function by combining estimators of the propensity score and the action-value function (Q-function) for future policies, while using cross-fitting. The Q-functions are estimated via fitted Q-evaluation (FQE) \citep{Munos_et_al_2008, Fonteneau_et_al_2013, Le_et_al_2019}, a method for offline policy evaluation in reinforcement learning. The optimal policy at each stage is then estimated by maximizing the estimated policy value function over a pre-specified class of stage-specific policies. This procedure yields the estimated DTR as a sequence of the estimated policies across all stages.

The proposed approach is computationally efficient due to its stepwise backward optimization, which is particularly advantageous given the complexity of optimizing DTRs over multiple stages. Furthermore, leveraging a classification-based framework \citep{Zhao_et_al_2012, Kitagawa_Tetenov_2018a}, our approach can enhance the interpretability of DTRs by allowing for the use of interpretable policy classes, such as decision trees, for each stage. It can also accommodate various dynamic treatment problems, including optimal stopping/starting problems, by appropriately constraining the class of feasible DTRs.
\footnote{An important example of an optimal stopping problem in economics is an unemployment insurance program in
which benefit levels are reduced after a certain duration \citep[e.g.,][]{Meyer_1995,Kolsrud_et_al_2018}. In this context, a DTR determines the timing of benefit reductions for each unemployed individual.}

We study the statistical properties of the proposed approach in terms of welfare regret, defined as the average outcome loss of the estimated DTR relative to the optimal one. This is a nontrivial task because the stage-specific policies within the DTR are estimated sequentially, rather than simultaneously, and state variables are influenced by past treatments. The main theoretical contribution of this paper is to establish the convergence rate for welfare regret, linking it to the convergence rates of the nuisance component estimators (propensity scores and Q-functions) in terms of the mean-squared error (MSE). Our key result identifies conditions on the nuisance component estimators and the class of DTRs under which the resulting DTR achieves the minimax optimal convergence rate of $n^{-1/2}$ for regret.
For instance, if all nuisance components are estimated with an MSE convergence rate of $n^{-1/2}$ -- a rate attainable by many machine learning methods under suitable structural assumptions -- and the complexity of the DTR class is appropriately constrained, the regret of the resulting DTR converges to zero at the optimal $n^{-1/2}$ rate. This result parallels those of \cite{Athey_Wager_2020} and \cite{zhou2022offline}, who study doubly robust policy learning in single-stage settings, and aligns with the principles of double machine learning \citep{chernozhukov_et_al_2018}.

We illustrate the proposed method through an empirical application to data from Project STAR \citep[e.g.,][]{krueger_1999}. Specifically, we learn the optimal DTR for allocating each student to either a regular-size class with a teacher aide or a small-size class without one during their early education (kindergarten and grade 1). The estimated DTR uses students’ intermediate academic performance to determine the optimal class type for the subsequent grade.  Our empirical results suggest that the estimated DTR improves cost-adjusted academic achievement relative to uniform class-type allocations.


\subsection*{Related Literature \label{sec:related literature}}

Although many studies have explored the statistical decision/learning of treatment choice, most have focused on the single-stage setting.\footnote{A partial list includes \cite{Manski_2004}, \cite{Hirano_Porter_2009}, \citeauthor{Stoye_2009} (\citeyear{Stoye_2009,Stoye_2012}), \cite{Qian_Murphy_2011}, \cite{Bhattacharya_Dupas_2012}, \cite{Tetenov_2012}, \cite{Zhao_et_al_2012}, \cite{Kitagawa_Tetenov_2018a}, \cite{Athey_Wager_2020}, \cite{Mbakop_Tabord-Meehan_2019}, \cite{kitagawa_et_al_2021}, \cite{zhou2022offline}, and \cite{viviano_2024}, among others.} 
Among these, this study is most closely related to \cite{Athey_Wager_2020} and \cite{zhou2022offline}, who develop doubly robust policy learning in single-stage settings and show that an upper bound of order $n^{-1/2}$ for regret can be achieved even in observational data settings. This paper seeks to extend their approach and results to the multi-stage dynamic treatment choice problem.

There is an expanding literature on the estimation of optimal DTRs, with various methods having been proposed.\footnote{\cite{Chakraborty_Murphy_2014}, \cite{Laber_et_al_2014}, \cite{Kosorok_et_al_2019}, and \cite{Li_et_al_2023} provide reviews of the literature.} Offline Q-learning \citep{Watkins_Dayan_1992} is arguably the most widely used method for estimating optimal DTRs \citep[e.g.,][]{Murphy_2005,Moodie_et_al_2012,Zhang_et_al_2018}. \cite{Murphy_2005} shows that the performance of the DTR obtained via Q-learning depends on the accuracy of the Q-function estimates; if the estimated Q-functions deviate from the true ones, the resulting DTR can be far from optimal.
Our approach also estimates Q-functions as part of its procedure. However, by leveraging propensity score models, it is more robust and accurate than Q-learning.

This study is also related to the classification-based inverse probability weighting approach for estimating optimal DTRs \citep[e.g.,][]{Zhao_et_al_2015, Sakaguchi_2021}. This approach uses inverse-propensity-weighted outcomes to estimate the value function of a DTR and maximizes it to estimate the optimal DTR over a pre-specified class of DTRs. However, the use of inverse-probability-weighted outcomes can sometimes lead to excessively high variance, which may result in suboptimal DTRs \citep{Doroudi_et_al_2018}. Our approach improves upon this method by incorporating models of Q-functions, thereby enhancing its overall performance.

Within this line of work, the closest predecessor to the present paper
is \cite{Sakaguchi_2021}, which develops IPW-based dynamic empirical
welfare maximization methods, including a backward-induction approach.
At a high level, the contribution of this paper is to develop a doubly
robust analogue of the backward dynamic empirical welfare maximization
approach. This extension is not straightforward.

In single-stage policy learning, the transition from an IPW objective
to an AIPW objective is relatively direct because the AIPW score
depends on treatment-specific outcome regressions and propensity scores
that can be estimated independently of the policy being optimized. In
dynamic policy learning, by contrast, the AIPW score at stage $t$
involves a Q-function associated with the future policy sequence, which
is recursively defined and policy-dependent. Adapting AIPW methods to
the backward approach therefore requires specifying how to estimate the
Q-functions associated with future policies and how to integrate their
estimation with policy optimization within the backward recursion. The
proposed method addresses this difficulty by combining doubly robust
policy learning with FQE for the future policies learned at later
stages.

The regret analysis also differs from the analyses in
\cite{Sakaguchi_2021} and \cite{Athey_Wager_2020}. Relative to
\cite{Sakaguchi_2021}, the analysis must additionally accommodate
estimation errors in the recursively estimated, policy-dependent
Q-functions that enter the AIPW scores, while relative to
\cite{Athey_Wager_2020}, it must handle the sequential learning
structure of DTRs and the resulting backward propagation of
policy-learning and nuisance-estimation errors. Despite these additional complications, we establish that the regret of
the proposed approach converges at the optimal $n^{-1/2}$ rate under
suitable convergence conditions on the nuisance estimators.

Doubly robust estimation for optimal DTRs has also been proposed by \cite{Zhang_et_2013}, \cite{Wallace_Moodie_2015}, and \cite{Ertefaie_et_al_2021}. \cite{Zhang_et_2013} suggest estimating the optimal DTR by maximizing an AIPW estimator of the welfare function over an entire class of DTRs. However, this approach faces computational challenges for two reasons: (i) the nuisance components to be estimated depend on each specific DTR, and (ii) the method maximizes the estimated welfare function simultaneously over the entire class of DTRs. Our approach addresses these issues by (i) ensuring that the nuisance components depend only on the estimated policies for future stages, and (ii) estimating the optimal DTR through stage-wise backward optimization. Section~\ref{sec:existing_approach} provides further details on this comparison.\footnote{\cite{zhang2018c} also propose a backward induction approach to estimate optimal DTRs using estimators of Q-functions and propensity scores, but their method is not robust to misspecification of the Q-functions. See \citeauthor{Sakaguchi_2025} (\citeyear{Sakaguchi_2025}, Footnote 9) for further discussion.}
\cite{Wallace_Moodie_2015} develop a doubly robust estimation method based on Q-learning and G-estimation \citep{Robins_2004}. \cite{Ertefaie_et_al_2021} propose a doubly robust approach for Q-learning and investigate the statistical properties of the estimated parameters in the Q-functions. By contrast, our focus is on the statistical properties of welfare regret for the estimated DTR.

In the context of policy learning for optimal stopping/starting problems, \cite{Nie_et_al_2021} develop a doubly robust learning approach with computational feasibility and show upper bounds on the associated regret. Our framework encompasses this problem as a specific case.

This study is also related to the offline reinforcement-learning literature on multistage decision problems. Most work in that
literature assumes a Markov decision process, an assumption that we do not impose. In the context of non-Markov decision processes, \cite{Jiang_Li_2016}, \cite{Thomas_et_al_2016}, and \cite{Kallus_Uehara_2020} propose doubly robust methods for evaluating DTRs, but they do not focus on optimizing DTRs.

Finally, in the econometric literature on dynamic treatment analysis, \cite{Heckman_Navarro_2007} and \cite{Heckman_et_al_2016} use exclusion restrictions to identify average dynamic treatment effects, though their focus is not on the identification of optimal DTRs. \cite{Han_2023} proposes a method to characterize the sharp partial ordering of counterfactual welfares of DTRs within an instrumental variable setting. \cite{Ida_et_al_2024} empirically demonstrate that the optimal DTR outperforms optimal static targeting policies in the context of energy-saving rebate programs.

\subsection*{Structure of the Paper }
The remainder of the paper is organized as follows. Section~\ref{sec:setup} outlines the dynamic treatment choice problem. Section~\ref{sec:doubly_robust_policy_learning} introduces the proposed doubly robust approach for learning the optimal DTR via backward induction. Section~\ref{sec:statistical_properties} establishes the statistical properties of the proposed approach. Section~\ref{sec:existing_approach} compares our approach with that of \cite{Zhang_et_2013}. Section~\ref{sec:simulation} presents a simulation study to evaluate its finite-sample performance. 
Section~\ref{sec:empirical_illustration} presents the empirical application. 
Section~\ref{sec:conclusion} concludes.
Appendix~\ref{app:main_proof} presents the proof of the main theorem along with some auxiliary lemmas. 
The Supplementary Appendix provides additional proofs, a numerical example, and additional simulation results.


\section{Setup\label{sec:setup}}

Section~\ref{sec:dynamic treatment framework} first introduces the dynamic treatment framework, following the dynamic potential outcome framework of \citet{Robins_1986,Robins_1997} and \cite{Murphy_2003}. Section~\ref{sec:dynamic treatment choice problem} then defines the dynamic treatment choice problem. 

\subsection{Dynamic Treatment Framework \label{sec:dynamic treatment framework}}

We consider a fixed number of stages, $T$ $(< \infty)$, for multiple treatment assignments.
For each $t=1,\ldots,T$, let $\MA_t \equiv \{0,\ldots,d_t-1\}$ denote
the set of possible treatment arms at stage $t$, where $d_t\geq 2$
denotes the number of treatment arms and may vary across stages. Let $S_{t}$ be a vector of state variables observed prior to treatment assignment at stage $t$, which may depend on past treatments. 
At each stage $t$, we observe the outcome $Y_{t}$ after the treatment intervention. The state vector $S_{t}$ (for $t \geq 2$) may include previous outcomes $(Y_{1},\ldots,Y_{t-1})$. 
Throughout this paper, for any time-dependent object $V_{t}$, we denote by $\text{\ensuremath{\text{\ensuremath{\underline{V}}}}}_{t}\equiv \left(V_{1},\ldots,V_{t}\right)$
its history up to stage $t$, and by $\text{\ensuremath{\underline{V}}}_{s:t} \equiv \left(V_{s},\ldots,V_{t}\right)$ the partial history from stage $s$ up to stage $t$ (for $s\leq t$).
For example, $\underline{A}_{t}$ denotes the treatment history up to stage $t$.

Let $Z\equiv(\underline{A}_{T},\underline{S}_T,\underline{Y}_{T})$ denote the vector of all observed variables. We define the history at stage $t$ as $H_t \equiv (\underline{A}_{t-1},\underline{S}_t)$, which is the information available to the decision-maker when selecting the treatment at stage $t$. Note that $H_1 = (S_1)$, where $S_1$ represents individual characteristics observed prior to the beginning of the sequential treatment intervention. We denote the supports of $H_{t}$ and $Z$ by ${\cal H}_{t}$ and $\MZ$, respectively.

To formalize our results, we adopt the dynamic potential outcomes framework \citep{Robins_1986, Hernan_et_al_2001, Murphy_2003}. Let $\underline{\MA}_t \equiv \MA_1 \times \cdots \times \MA_t$. For each $\underline{a}_t \in \underline{\MA}_t$, we define $Y_{t}\left(\text{\ensuremath{\underline{a}}}_{t}\right)$ as the potential outcome for stage $t$, representing the outcome that would be realized at stage $t$ if the treatment history up to that stage were $\underline{a}_{t}$. We assume that the outcome at each stage is not influenced by treatments in future stages. Since the state variables $S_t$ may also depend on past treatments, we define the potential state variables as $S_{t}(\underline{a}_{t-1})$ for each $t\geq 2$ and $\underline{a}_{t-1} \in \underline{\MA}_{t-1}$. For $t = 1$, we set $S_{1}(\underline{a}_{0}) = S_1$ for notational convenience. The observed outcomes and state variables are thus defined as $Y_{t} \equiv Y_{t}(\underline{A}_t)$ and $S_t \equiv S_t(\underline{A}_{t-1})$, respectively. 

Let $\underline{S}_t(\underline{a}_{t-1}) \equiv (S_1,S_2(\underline{a}_{1}),\ldots,S_t(\underline{a}_{t-1}))$ be the history of the potential state variables. We define the vector $H_{t}\left(\underline{a}_{t-1}\right) \equiv \left(\underline{a}_{t-1},\underline{S}_t(\underline{a}_{t-1})\right)$  as the potential history realized when the prior treatments are $\underline{a}_{t-1}$. For $t=1$, we set $H_{1}\left(\underline{a}_{0}\right) = H_1$. The observed history is then defined as $H_t \equiv H_t\left(\underline{A}_{t-1}\right)$.
We denote by $P$ the distribution of all underlying variables $\left(\underline{A}_T,\left\{\underline{S}_T(\underline{a}_{T-1})\right\}_{\underline{a}_{T-1} \in \underline{\MA}_{T-1}},\{\underline{Y}_{T}(\underline{a}_T)\}_{\underline{a}_T \in \underline{\MA}_T}\right)$, where  $\underline{Y}_T(\underline{a}_T) \equiv (Y_1(a_1),Y_2(\underline{a}_{2}),\ldots,Y_T(\underline{a}_{T}))$.

From an observational study, we observe $Z_{i}\equiv \left(\underline{A}_{i,T},\underline{S}_{i,T},\underline{Y}_{i,T}\right)$ for individuals $i=1,\ldots,n$, where $\underline{A}_{i,T}=(A_{i,1},\ldots,A_{i,T})$, $\underline{S}_{i,T}=(S_{i,1},\ldots,S_{i,T})$, and $\underline{Y}_{i,T}=(Y_{i,1},\ldots,Y_{i,T})$. The observed outcome $Y_{i,t}$ and state variables $S_{i,t}$ are defined as $Y_{i,t} \equiv Y_{i,t}(\underline{A}_{i,t})$ and $S_{i,t} \equiv S_{i,t}(\underline{A}_{i,t-1})$, respectively, with $Y_{i,t}\left(\underline{a}_{t}\right)$ and $S_{i,t}\left(\underline{a}_{t-1}\right)$ being the
potential outcome and state variables, respectively. Let $\underline{S}_{i,T}(\underline{a}_{T-1}) \equiv (S_{i,1},S_{i,2}(a_1),\ldots,S_{i,T}(\underline{a}_{T-1}))$ and $\underline{Y}_{i,T}(\underline{a}_{T}) \equiv (Y_{i,1}(a_1),\ldots,Y_{i,T}(\underline{a}_{T}))$. We assume that the vectors of random variables $\left(\underline{A}_{i,T},\left\{\underline{S}_{i,T}\left(\text{\ensuremath{\underline{a}}}_{T-1}\right)\right\}_{\text{\ensuremath{\underline{a}}}_{T-1}\in\underline{\MA}_{T-1}}, \left\{ \underline{Y}_{i,T}\left(\text{\ensuremath{\underline{a}}}_{T}\right)\right\} _{\text{\ensuremath{\underline{a}}}_{T}\in\underline{\MA}_{T}}\right)$, for $i=1,\ldots,n$, are independent and identically distributed (i.i.d.) under the distribution $P$. 
We denote by $H_{i,t}\equiv (\underline{A}_{i,t-1},\underline{S}_{i,t})$ the history of the $i$-th individual at stage $t$.

We define $e_{t}\left(h_{t},a_{t}\right)\equiv \BP\left(A_{t}=a_{t}\mid H_{t}=h_{t}\right)$ as the propensity score of treatment $a_t$ at stage $t$ with  history $h_t$. In the observational data setting we study, the propensity scores are unknown to the analyst. This contrasts with the experimental data setting, where the propensity
scores are known from the experimental design.\footnote{Even when data are obtained from a sequential multiple assignment randomized trial (SMART), propensity scores may be unknown due to non-compliance with assigned treatments or attrition at certain stages.}

Throughout the paper, we assume that the underlying distribution $P$ satisfies the following assumptions.

\medskip

\begin{assumption}[Sequential Ignorability]\label{asm:sequential independence} For any $t=1,\ldots,T$ and $\text{\ensuremath{\underline{a}}}_{T}\in\underline{\MA}_{T}$,
\begin{align*}
\{Y_{t}\left(\underline{a}_{t}\right),\ldots,Y_{T}\left(\underline{a}_{T}\right),S_{t+1}(\underline{a}_{t}),\ldots,S_{T}(\underline{a}_{T-1})\} \indep A_{t} \mid H_{t}.  
\end{align*}
\end{assumption}


\begin{assumption}[Bounded Outcomes]\label{asm:bounded outcome}
There exists $M <\infty$ such that the support of $Y_{t}(\underline{a}_{t})$ is
contained in $\left[-M/2,M/2\right]$ for all $t \in \{1,\ldots,T\}$ and $\underline{a}_{t} \in \underline{\MA}_{t}$.
\end{assumption}


\medskip

Assumption~\ref{asm:sequential independence} is also referred to as the dynamic unconfoundedness assumption or the sequential conditional independence assumption, and is commonly used in the literature on dynamic treatment effect analysis \citep{Robins_1997,Murphy_2003}. This assumption implies that the treatment assignment at each stage is independent of current and future potential outcomes, as well as future state variables, conditional on the history up to that point. In observational studies, this assumption holds when a sufficient set of confounders is controlled for at each stage. Assumption~\ref{asm:bounded outcome} is a standard assumption in the literature on estimating optimal DTRs.


\subsection{Dynamic Treatment Choice Problem\label{sec:dynamic treatment choice problem}}

The aim of this study is  to develop a method for learning optimal DTRs using data from an observational study. We define a policy for each stage $t$ as $\pi_{t}:\MH_{t}\to \MA_{t} $, a map from the history space for stage $t$ to the treatment space for stage $t$. A policy $\pi_t$ determines which treatment is assigned to each individual at stage $t$ based on their history $h_t$. We define a DTR as $\pi \equiv (\pi_{1}, \ldots, \pi_{T})$, a sequence of stage-specific policies. The DTR guides the treatment choice for each individual from the first to the final stage, based on their evolving history up to each stage.

Given a fixed DTR $\pi$, we define the welfare of $\pi$ as
\begin{align*}
W\left(\pi\right) &\equiv   
\E\left[\sum_{t=1}^{T}\sum_{\underline{a}_t \in \underline{\MA}_t}\left( Y_{t}(\underline{a}_t)\cdot \prod_{s=1}^{t}\mathbf{1}\{\pi_s(H_{s}(\underline{a}_{s-1}))=a_s\}\right)\right].
\end{align*}
This expression represents the expected total outcome realized when treatments are assigned sequentially according to the DTR $\pi$.\footnote{We can also define and consider welfare with weighted outcomes as follows: $W\left(\pi\right) =   
\E\left[\sum_{t=1}^{T}\sum_{\underline{a}_t \in \underline{\MA}_t}\left(\gamma_{t} Y_{t}(\underline{a}_t)\cdot \prod_{s=1}^{t}\mathbf{1}\{\pi_s(H_{s}(\underline{a}_{s-1}))=a_s\}\right)\right]$, where $\gamma_{t}$ represents the weight assigned to the outcome at stage $t$. For instance, $\gamma_t$ can be a discount factor $\gamma^t$ with $\gamma$ being a discount rate. If we focus solely on the final-stage outcome, we set $\gamma_{T}=1$ and $\gamma_{1}=\cdots=\gamma_{T-1}=0$.}

We consider choosing a DTR from a pre-specified class of DTRs denoted by $\Pi \equiv \Pi_{1}\times\cdots\times \Pi_{T}$, where $\Pi_{t}$ represents a class of policies for stage $t$ (i.e., a class of measurable functions $\pi_{t}:\MH_{t}\rightarrow \MA_{t}$). For example, \cite{Laber_Zhao_2015}, \cite{Tao_et_al_2018}, \cite{Sun_Wang_2021},  \cite{Blumlein_et_al_2022}, and \cite{zhou_et_al_2023} use a class of decision trees \citep{Breiman84book} for $\Pi_t$, and \cite{Zhang_et_al_2018} use a class of list-form policies for $\Pi_t$. These policy classes enhance the interpretability of the resulting DTRs.

Our framework also accommodates cases where only a part of the history, $\tilde{h}_t$, a subvector of $h_{t}$, is used for treatment choice by constraining $\Pi_t$ to be a class of functions of $\tilde{h}_t$. This is particularly relevant because the dimension of the full history $h_t$ increases over time, while only a sub-history may be informative for optimal treatment choice.
Additionally, our framework can encompass the optimal stopping/starting problem by constraining the class $\Pi$ of DTRs as follows.

\medskip

\begin{example}[Optimal Starting/Stopping Problem]\label{eg:optimal_start}
Suppose that the number of treatment arms $d_t$ is time-invariant ($d_t = K$ for some constant $K$) and that arm $0$ represents no treatment. Our framework can accommodate the optimal starting problem, in which the decision-maker determines when to start assigning one of the arms $a_t \in \{1,\ldots,K-1\}$ for each unit. This problem can be incorporated into our framework by constraining the class $\Pi$ of DTRs such that for any $t \in \{2,\ldots,T\}$ and $(\pi_{t}, h_{t}) \in \Pi_{t} \times \MH_{t}$, $\pi_{t}(h_t) = a_{t-1}$ if $a_{t-1} \neq 0$. The optimal stopping problem can also be specified in a similar manner.
\end{example}
\medskip

An important example of an optimal stopping problem in economics is an unemployment insurance program that reduces benefit levels after a certain duration \citep[e.g.,][]{Meyer_1995,Kolsrud_et_al_2018}. In this context, a DTR determines the timing of benefit reductions for each unemployed individual.

Given a pre-specified class $\Pi$ of DTRs, we impose an overlap condition on $\underline{A}_{T}$, related to the structure of $\Pi$, as follows.

\medskip

\begin{assumption}[Overlap Condition]\label{asm:overlap} 
For each $t \in \{1,\ldots,T\}$, there exists $\eta_{t} \in (0,1)$ for which $\eta_{t} \leq e_{t}(h_{t},a_{t})$ holds for any pair  $(h_{t},a_{t}) \in \MH_{t} \times \MA_{t}$ such that there exists $\pi_{t} \in \Pi_{t}$ that satisfies $\pi_{t}(h_t) = a_{t}$. 
\end{assumption}

\medskip

When $\Pi$ is structurally constrained, Assumption~\ref{asm:overlap} is weaker than the common overlap condition that requires $e_{t}(h_t, a_t) \in (0, 1)$ for all $(h_t, a_t) \in \MH_{t} \times \MA_{t}$ and $t$. For example, in the optimal stopping problem, Assumption~\ref{asm:overlap} does not require $e_{t}(h_t, 1) > 0$ for any $h_t$ such that $a_s$ in $h_t$ is 0 for some $s < t$.

The ultimate goal of our analysis is to choose an optimal DTR that maximizes the welfare $W\left(\cdot\right)$ over $\Pi$. 
We are especially interested in learning the optimal DTR from observational data that satisfy the sequential ignorability assumption (Assumption~\ref{asm:sequential independence}). The following section presents a novel step-wise approach to learning the optimal DTR.

\section{Learning the Optimal DTR}\label{sec:doubly_robust_policy_learning}

In this section, we propose a doubly robust approach to learning the optimal DTR through backward induction.
Section~\ref{sec:fitted_Q} introduces the population backward-induction procedure for solving the optimal DTR problem and discusses the conditions under which it identifies the optimal DTR.
Section~\ref{sec:learning_DTR} then presents our proposed approach to learning the optimal DTR from observational data.

\subsection{Backward Induction and Correct Specification}
\label{sec:fitted_Q}

For any DTR $\pi$ and class $\Pi$ of DTRs, we denote their partial sequences by $\pi_{s:t} \equiv (\pi_s,\ldots,\pi_t)$ and $\Pi_{s:t} \equiv \Pi_s \times \cdots \times \Pi_t$, respectively, for $s\leq t$.\footnote{For any object $v_{s:t}$ and $\underline{w}_{s:t}$ ($s\leq t$), $v_{t:t}$ and $\underline{w}_{t:t}$ correspond to $v_t$ and $w_t$, respectively.}
We define the policy value of $\pi_{t:T}$ for stage $t$ as 
\begin{align*}
    V_{t}\left(\pi_{t:T}\right) \equiv \E\left[\sum_{s=t}^{T}\sum_{\underline{a}_{s} \in \underline{\MA}_{s}}\left( Y_{s}(\underline{a}_{s})\cdot \mathbf{1}\{\underline{A}_{t-1} = \underline{a}_{t-1}\} \cdot \prod_{\ell=t}^{s}\mathbf{1}\{\pi_{\ell}\left(H_{\ell}(\underline{a}_{\ell-1})\right)=a_\ell\}\right)\right],
\end{align*}
where we assume $\mathbf{1}\{\underline{A}_{0} = \underline{a}_{0}\}=1$ for $t=1$. $V_{t}\left(\pi_{t:T}\right)$
represents the average total outcome from stage $t$ to stage $T$ that is realized when the treatment assignments before stage $t$ follow $\underline{A}_{t-1}$ (i.e., assignments in the observational data), and those from stage $t$ follow $\pi_{t:T}$.
With some abuse of terminology, we refer to $V_{t}\left(\pi_{t:T}\right)$ as the policy value function of $\pi_{t:T}$.
Note that $V_{1}(\pi_{1:T}) = W(\pi)$.

Given a fixed DTR $\pi$, we recursively define Q-functions (state-action-value functions) as follows:
\begin{align}
    Q_T(h_T,a_T) &\equiv \E\left[Y_{T}|H_T = h_T, A_T = a_T\right], \label{eq:action_value_func_1}\\
     Q_{T-1}^{\pi_{T}}(h_{T-1},a_{T-1}) &\equiv \E\left[Y_{T-1} + Q_T(H_T,\pi_T(H_{T}))|H_{T-1} = h_{T-1}, A_{T-1} = a_{T-1}\right],
     \end{align}
     and, for $t=T-2,\ldots,1$,
     \begin{align}
     Q_{t}^{\pi_{(t+1):T}}(h_t,a_t) &\equiv \E\left[Y_{t} + Q_{t+1}^{\pi_{(t+2):T}}(H_{t+1},\pi_{t+1}(H_{t+1}))|H_t = h_t, A_t = a_t\right]. \label{eq:state_value_func_2}
\end{align}
For a given future policy sequence $\pi_{(t+1):T}$, the Q-function $Q_t^{\pi_{(t+1):T}}(h_t,a_t)$ represents the average total outcome when $H_t=h_t$ and $A_t=a_t$ at stage $t$, and future treatments follow $\pi_{(t+1):T}$.
The recursive representation of the Q-functions also guides their sequential estimation via FQE
\citep{Munos_et_al_2008,Fonteneau_et_al_2013,Le_et_al_2019}, which we describe in Section~\ref{sec:learning_DTR}.

When $t=T$, we denote $Q_{T}^{\pi_{(T+1):T}}(\cdot,\cdot) = Q_{T}(\cdot,\cdot)$.  Note that $\E[Q_{t}^{\pi_{(t+1):T}}(H_{t},\pi_{t}(H_{t}))] = V_{t}(\pi_{t:T})$ and $\E[Q_{1}^{\pi_{2:T}}(H_{1},\pi_{1}(H_{1}))] = W(\pi)$ hold under Assumption~\ref{asm:sequential independence}.\footnote{See, for example, \citeauthor{Tsiatis_et_al_2019} (\citeyear{Tsiatis_et_al_2019}, Section 6.4) or Lemma~\ref{lemma:identification}.} 
In what follows, for any function $f(\cdot,\cdot):\MH_{t} \times \MA_{t} \rightarrow \Real$ and policy $\pi_{t}(\cdot):\MH_{t} \rightarrow \MA_{t}$, we abbreviate $f(h_{t},\pi_{t}(h_{t}))$ as $f(h_{t},\pi_{t})$ (e.g., $Q_{t}^{\pi_{(t+1):T}}(h_t,\pi_t(h_t))$ is denoted by $Q_{t}^{\pi_{(t+1):T}}(h_t,\pi_t)$).

Given the definitions of the Q-functions, we can optimize the DTR through backward induction. To present the idea, we here assume that the generative distribution function $P$ is known and that the pair  $(P,\Pi)$ satisfies Assumptions~\ref{asm:sequential independence}, \ref{asm:bounded outcome}, and \ref{asm:overlap}. The backward-induction approach in the population problem is a stepwise procedure. 
In the first step, the approach optimizes the final-stage policy over $\Pi_T$ by solving 
\begin{align*}
        \pi_{T}^{\ast,B} = \argmax_{\pi_{T} \in \Pi_{T}} \E\left[Q_{T}(H_{T},\pi_{T})\right].
\end{align*}
Then, recursively, from $t=T-1$ to $1$, the approach optimizes the $t$-th stage policy over $\Pi_{t}$ by solving
\begin{align*}
        \pi_{t}^{\ast,B} = \argmax_{\pi_{t} \in \Pi_{t}} \E\left[Q_{t}^{\pi_{(t+1):T}^{\ast,B}}(H_{t},\pi_{t})\right].
\end{align*}
Note that the objective function $\E\left[Q_{t}^{\pi_{(t+1):T}^{\ast,B}}(H_{t},\pi_{t})\right]$ corresponds to the policy value $V_{t}(\pi_{t},\pi_{(t+1):T}^{\ast,B})$ under the sequential ignorability assumption (Assumption~\ref{asm:sequential independence}).
The entire procedure yields the DTR $\pi^{\ast,B} \equiv (\pi_{1}^{\ast,B},\cdots,\pi_{T}^{\ast,B})$.

We denote the optimal DTR over $\Pi$ by $\pi^{\ast,opt} = \argmax_{\pi \in \Pi} W(\pi)$. Importantly, the DTR $\pi^{\ast,B}$ obtained through backward induction does not necessarily correspond to the optimal DTR $\pi^{\ast,opt}$ when $\Pi$ is structurally constrained, as noted by \cite{Li_et_al_2023} and \cite{Sakaguchi_2021}. 
To ensure that backward induction yields the optimal DTR, the policy class $\Pi_t$ for each stage $t$ ($\geq 2$) needs to be correctly specified. The following assumption gives a sufficient condition for the backward optimization to attain optimality.

\medskip

\begin{assumption}[Correct Specification] 
\label{asm:first-best} 
There exists $\pi_{2:T}^{\ast} = (\pi_{2}^{\ast},\ldots,\pi_{T}^{\ast})\in \Pi_{2:T}$ such that for any $t=2,\ldots,T$,
\begin{align*}
Q_{t}^{\pi_{(t+1):T}^{\ast}}\left(H_t,\pi_{t}^{\ast}\right) \geq \sup_{\pi_{t} \in \Pi_t}Q_{t}^{\pi_{(t+1):T}^{\ast}}\left(H_t,\pi_{t}\right) \mbox{\ a.s.}
\end{align*} 
\end{assumption}

\medskip

Assumption~\ref{asm:first-best} requires that for each stage $t \geq 2$, $\Pi_t$ includes a policy that can select the optimal arm among all feasible arms in $\{\pi_{t}(h_t):\pi_{t} \in \Pi_t\}$ $(\subseteq \MA_{t})$ for any history $h_t$. This assumption is satisfied when $\Pi_t$ (for $t=2,\ldots,T$) is flexible enough or correctly specified in relation to treatment effect heterogeneity.\footnote{\cite{Zhang_et_2013} discuss the specification of stage-specific policy classes based on treatment-effect heterogeneity, while \cite{Zhao_et_al_2015} use reproducing kernel Hilbert spaces to construct flexible policy classes.} Throughout the main analysis, we suppose that $\Pi$ satisfies
Assumption~\ref{asm:first-best}. 

Assumption~\ref{asm:first-best} is sufficient but not necessary for
backward induction to attain the welfare optimum over $\Pi$.
Supplemental Appendix~\ref{app:suboptimallity_backward_induction}
provides an example in which backward induction remains optimal even
when the assumption fails.

A stronger version of Assumption~\ref{asm:first-best} requires each
$\Pi_t$ ($t\geq2$) to contain a first-best policy; that is, there exists
$\pi_{2:T}^{\ast,FB}\in\Pi_{2:T}$ such that, for every $t=2,\ldots,T$,
\[
Q_t^{\pi_{(t+1):T}^{\ast,FB}}
\left(H_t,\pi_t^{\ast,FB}\right)
\geq
\max_{a_t\in\MA_t}
Q_t^{\pi_{(t+1):T}^{\ast,FB}}
\left(H_t,a_t\right)
\qquad\text{a.s.}
\]
Thus, the first-best policy selects an optimal action from the full
action set at every history. \cite{Li_et_al_2023} and
\cite{Sakaguchi_2021} impose related first-best feasibility conditions
to justify backward optimization. Assumption~\ref{asm:first-best} is
less restrictive because it requires optimality only relative to the
actions implementable by the constrained policy class.

For example, consider the optimal stopping problem in
Example~\ref{eg:optimal_start} and suppose that policies are represented
by decision trees subject to the restriction that treatment must remain
stopped once stopping occurs. This restriction may exclude the
unrestricted first-best DTR. Nevertheless,
Assumption~\ref{asm:first-best} can hold if the tree class is sufficiently
rich to represent the continuation-value-maximizing choice between
stopping and continuing before stopping occurs; afterward, only stopping
is implementable. Thus, a structurally constrained policy class can
satisfy Assumption~\ref{asm:first-best} even when it does not contain the
unrestricted first-best DTR. The practical consequences of possible
violations can be explored through sensitivity to richer policy classes;
Remark~\ref{rem:policy_class_approximation} discusses an
approximate-specification framework based on increasingly rich sieve
policy classes.

Assumption~\ref{asm:first-best} can also accommodate policies that use
only a sub-history of the observed history. Suppose that
$H_t=(\widetilde H_t,X_t)$, where policies in $\Pi_t$ depend only on
$\widetilde H_t$ and $X_t$ denotes additional observed information not
used for policy assignment. The presence of $X_t$ does not by itself
violate Assumption~\ref{asm:first-best}; the assumption can still hold
if, conditional on $\widetilde H_t$, the optimal feasible action does
not vary with $X_t$. For example, with binary treatments, it is
sufficient that the sign of
\[
Q_t^{\pi_{(t+1):T}^{\ast}}(\widetilde H_t,X_t,1)
-
Q_t^{\pi_{(t+1):T}^{\ast}}(\widetilde H_t,X_t,0)
\]
does not vary with $X_t$ conditional on $\widetilde H_t$. If $X_t$
instead changes the optimal feasible action, Assumption~\ref{asm:first-best}
may fail. Thus, $\widetilde H_t$ need not contain every variable
generating treatment-effect heterogeneity, but it must contain sufficient
information to determine the relevant optimal feasible action.

The following lemma formalizes the optimality of the backward-induction procedure under Assumption~\ref{asm:first-best}.

\medskip

\begin{lemma}\label{lem:optimality_backward_induction}
Under Assumptions~\ref{asm:sequential independence}, \ref{asm:overlap}, and \ref{asm:first-best}, $\pi^{\ast,B}$ is the optimal DTR over $\Pi$; i.e.,
\begin{align*}
    W(\pi^{\ast,B}) \geq W(\pi) \mbox{\ for all\ }\pi \in \Pi.
\end{align*}
\end{lemma}

\begin{proof}
See Appendix~\ref{app:preliminary_results} in the supplement.
\end{proof}


\subsection{Learning the Optimal DTR through Backward Induction}
\label{sec:learning_DTR}

This section presents a backward optimization procedure to learn the optimal DTR using an AIPW estimator of the policy value function. Following the doubly robust policy learning approaches of \cite{Athey_Wager_2020} and \cite{zhou2022offline}, we employ cross-fitting to separate nuisance-function estimation from
policy learning. We randomly divide the dataset $\{Z_i:i=1,\ldots,n\}$ into $K$ evenly-sized folds (e.g., $K=5$). Let $I_k \subseteq \{1,\ldots,n\}$ denote the set of indices of the data in the $k$-th fold, and $I_{-k}$ denote the set of indices of the data excluded from the $k$-th fold. In what follows, for any statistic $\hat{f}$, we denote by $\hat{f}^{-k}$ the corresponding statistic calculated using the data excluded from the $k$-th fold. Let $k(i)$ denote the fold number that contains the $i$-th observation.

The proposed approach is based on backward induction and thus consists of multiple steps. As a preliminary step, we estimate the propensity scores $\{e_t(\cdot,\cdot)\}_{t=1,\ldots,T}$ for all stages and the Q-function $Q_T(\cdot,\cdot)$ for the final stage using the data excluded in each cross-fitting fold. For each $k$, we denote by $\hat{e}_{t}^{-k}(\cdot,\cdot)$ and $\widehat{Q}_{T}^{-k}(\cdot,\cdot)$, respectively, the estimators of $e_t(\cdot,\cdot)$ and $Q_T(\cdot,\cdot)$ using data not contained in the $k$-th cross-fitting fold. Any regression method, including machine learning methods (e.g., random forests and neural networks), can be used to estimate $e_t(\cdot,\cdot)$ and $Q_{T}(\cdot,\cdot)$. 

Given $\{\hat{e}_{t}^{-k}(\cdot,\cdot)\}_{t=1,\ldots,T}$ and  $\widehat{Q}_{T}^{-k}(\cdot,\cdot)$ for each $k=1,\ldots,K$, we estimate the optimal DTR sequentially as follows. In the first step, regarding the final stage $T$, we construct a score function of the treatment $a_T$ for stage $T$ as 
\begin{align}
    \widehat{\Gamma}_{i,T}(a_T) \equiv \frac{Y_{i,T} - \widehat{Q}_{T}^{-k(i)}(H_{i,T},A_{i,T})}{\hat{e}_{T}^{-k(i)}(H_{i,T},A_{i,T})}\cdot \mathbf{1}\{A_{i,T}=a_T\} +
    \widehat{Q}_{T}^{-k(i)}(H_{i,T},a_{T}). \label{eq:score_func_T}
\end{align}
Given a policy $\pi_T$, the sample mean $(1/n)\sum_{i=1}^{n}\widehat{\Gamma}_{i,T}(\pi_T(H_{i,T}))$ is an AIPW estimator of the policy value $V_{T}(\pi_{T})$ for stage $T$.

We then find the best candidate policy for stage $T$ by solving
\begin{align*}
\hat{\pi}_{T} = \argmax_{\pi_{T}\in \Pi_{T}}\frac{1}{n}\sum_{i=1}^{n}\widehat{\Gamma}_{i,T}\left(\pi_T(H_{i,T})\right). 
\end{align*}

In the next step, we consider stage $T-1$. This step begins with FQE to estimate the Q-function associated with the final-stage policy $\hat{\pi}_T$ learned in the previous step. Specifically,
for each cross-fitting fold $k$, we apply the recursive definition of the
Q-function by replacing the continuation term
$Q_T(H_T,\hat{\pi}_T)$ with its fold-specific estimate
$\widehat Q_T^{-k}(H_T,\hat{\pi}_T)$. We then estimate
$Q_{T-1}^{\hat{\pi}_T}(\cdot,\cdot)$ by regressing the pseudo-outcome
$
Y_{i,T-1}
+
\widehat Q_T^{-k}
\bigl(H_{i,T},\hat{\pi}_T\bigr)
$
on $(H_{i,T-1},A_{i,T-1})$ using observations $i\in I_{-k}$.
We denote the resulting estimator by
$\widehat Q_{T-1}^{\hat{\pi}_T,-k}(\cdot,\cdot)$.
Any suitable regression method can be used for this regression.

We subsequently construct the score function of $a_{T-1}$ as follows:
\begin{align*}
    \widehat{\Gamma}_{i,T-1}^{\hat{\pi}_{T}}(a_{T-1})  &\equiv \frac{ Y_{i,T-1} + \widehat{\Gamma}_{i,T}(\hat{\pi}_{T}(H_{i,T})) - \widehat{Q}_{T-1}^{\hat{\pi}_{T},-k(i)}(H_{i,T-1},A_{i,T-1}) }{\hat{e}_{T-1}^{-k(i)}(H_{i,T-1},A_{i,T-1})} \cdot \mathbf{1}\{A_{i,T-1} = a_{T-1}\} \\
    &+ 
\widehat{Q}_{T-1}^{\hat{\pi}_{T},-k(i)}(H_{i,T-1},a_{T-1}).
\end{align*}
Given a policy $\pi_{T-1}$, the sample mean $(1/n)\sum_{i=1}^{n}\widehat{\Gamma}_{i,T-1}^{\hat{\pi}_{T}}(\pi_{T-1}(H_{i,T-1}))$ is an AIPW estimator of the policy value $V_{T-1}(\pi_{T-1},\hat{\pi}_{T})$.
We then find the best candidate policy for stage $T-1$ by solving
\begin{align*}
    \hat{\pi}_{T-1} = \argmax_{\pi_{T-1}\in \Pi_{T-1}} \frac{1}{n}\sum_{i=1}^{n}\widehat{\Gamma}_{i,T-1}^{\hat{\pi}_T}\left(\pi_{T-1}(H_{i,T-1})\right).
\end{align*}

Recursively, for $t=T-2,\ldots,1$, we learn the optimal policy as follows. For each cross-fitting fold $k$, we first estimate the Q-function $Q_{t}^{\hat{\pi}_{(t+1):T}}$ by regressing  $Y_{i,t} + \widehat{Q}_{t+1}^{\hat{\pi}_{(t+2):T},-k}(H_{i,t+1},\hat{\pi}_{t+1})$ on $(H_{i,t},A_{i,t})$ using observations $i\in I_{-k}$. This constitutes the FQE step. Any regression method can be applied at this step. 

We next construct the score function of $a_{t}$ as
\begin{align}
    \widehat{\Gamma}_{i,t}^{\hat{\pi}_{(t+1):T}}(a_t) &\equiv \frac{ Y_{i,t} + \widehat{\Gamma}_{i,t+1}^{\hat{\pi}_{(t+2):T}}(\hat{\pi}_{t+1}(H_{i,t+1})) - \widehat{Q}_{t}^{\hat{\pi}_{(t+1):T},-k(i)}(H_{i,t},A_{i,t}) }{\hat{e}_{t}^{-k(i)}(H_{i,t},A_{i,t})}\cdot \mathbf{1}\{A_{i,t} = a_t\} \nonumber \\
    &+ 
    \widehat{Q}_{t}^{\hat{\pi}_{(t+1):T},-k(i)}(H_{i,t},a_{t}). \label{eq:score_func_t}
\end{align}
We then find the best candidate policy for stage $t$ by solving
\begin{align*}
    \hat{\pi}_{t} = \argmax_{\pi_{t}\in \Pi_{t}} \frac{1}{n}\sum_{i=1}^{n} \widehat{\Gamma}_{i,t}^{\hat{\pi}_{(t+1):T}}\left(\pi_{t}(H_{i,t})\right), 
\end{align*}
where the objective function $(1/n)\sum_{i=1}^{n} \widehat{\Gamma}_{i,t}^{\hat{\pi}_{(t+1):T}}\left(\pi_{t}(H_{i,t})\right)$ is an AIPW estimator of the policy value $V_{t}(\pi_t,\hat{\pi}_{(t+1):T})$.

Throughout this procedure, we obtain the sequence $\hat{\pi} \equiv (\hat{\pi}_{1},\ldots,\hat{\pi}_{T})$, which serves as an estimator of the optimal DTR. Algorithm \ref{alg:proposed} summarizes the entire procedure. The following section establishes the statistical properties of $\hat{\pi}$ in terms of welfare regret. \\


\begin{algorithm}[h]
\caption{Doubly robust backward-induction learning}\label{alg:proposed}
\begin{algorithmic}[1]
\State \textbf{Input:} $K$ (the number of cross-fitting folds); $\{Z_{i}:i=1,\ldots,n\}$ (dataset) 
\State Construct $\hat{e}_{t}^{-k}(\cdot,\cdot)$ ($t=1,\ldots,T$) and $\widehat{Q}_{T}^{-k}(\cdot,\cdot)$ for each $k \in \{1,\ldots,K\}$
\State Compute $\widehat{\Gamma}_{i,T}(a_{T})$ (equation (\ref{eq:score_func_T})) for each $a_{T} \in \MA_{T}$ and $i \in \{1,\ldots,n\}$
\State Learn the optimal policy by setting $\hat{\pi}_{T} = \argmax_{\pi_{T}\in \Pi_{T}}(1/n)\sum_{i=1}^{n}\widehat{\Gamma}_{i,T}\left(\pi_T(H_{i,T})\right)$ 
\For{$t = T-1$ \textbf{to} $1$}
    \State Construct $\widehat{Q}_{t}^{\hat{\pi}_{(t+1):T},-k} (\cdot,\cdot)$ for each $k \in \{1,\ldots,K\}$ (FQE)
    \State Compute $\widehat{\Gamma}^{\hat{\pi}_{(t+1):T}}_{i,t}(a_{t})$ (equation (\ref{eq:score_func_t})) for each $a_{t} \in \MA_{t}$ and $i \in \{1,\ldots,n\}$
    \State Learn the optimal policy by setting $$\hat{\pi}_{t} = \argmax_{\pi_{t}\in \Pi_{t}} \frac{1}{n}\sum_{i=1}^{n} \widehat{\Gamma}_{i,t}^{\hat{\pi}_{(t+1):T}}\left(\pi_{t}(H_{i,t})\right)$$
\EndFor
\State \textbf{return} $\hat{\pi}=(\hat{\pi}_{1},\ldots,\hat{\pi}_{T})$
\end{algorithmic}
\end{algorithm}

\section{Statistical Properties}\label{sec:statistical_properties}

Given a DTR $\pi \in \Pi$, we define the regret of $\pi$ by $R(\pi) \equiv \max_{\tilde{\pi} \in \Pi}W(\tilde{\pi}) - W(\pi)$, the welfare loss of $\pi$ relative to the maximum welfare achievable in $\Pi$.
We study statistical properties of $\hat{\pi}$ with respect to its regret $R(\hat{\pi})$. This section shows the rate of convergence of $R(\hat{\pi})$ depending on the convergence rates of the estimators of the nuisance components, $\{\hat{e}_{t}^{-k}(\cdot,\cdot)\}_{t=1,\ldots,T}$ and  $\{\widehat{Q}_{t}^{\pi_{(t+1):T},-k}(\cdot,\cdot)\}_{t=1,\ldots,T}$, and the complexity of $\Pi$. 

Let $\mathcal S_n \equiv \{Z_i:i=1,\ldots,n\}$ be an estimation sample
randomly drawn from $P$. Let
$\widehat Q_t^{\pi_{(t+1):T},(n)}(\cdot,\cdot)$ and
$\hat e_t^{(n)}(\cdot,\cdot)$ denote estimators of
$Q_t^{\pi_{(t+1):T}}(\cdot,\cdot)$ and $e_t(\cdot,\cdot)$, respectively,
constructed from $\mathcal S_n$.
For $t=T$, we denote $\widehat{Q}_{T}^{\pi_{(T+1):T},(n)}(\cdot,\cdot) = \widehat{Q}_{T}^{(n)}(\cdot,\cdot)$ for notational convenience.
We suppose that $\{\widehat{Q}_{t}^{\pi_{(t+1):T},(n)}(\cdot,\cdot)\}_{t=1,\ldots,T}$ and $\{\hat{e}_{t}^{(n)}(\cdot,\cdot)\}_{t=1,\ldots,T}$ satisfy the following assumption.

\medskip{}

\begin{assumption}[Double Robustness]\label{asm:rate_of_convergence_backward}
(i) There exists $\tau >0$ such that the following holds: For all $t=1,\ldots,T$, $s=1,\ldots,t$, and $m \in \{0,1\}$,
\begin{align*}
    \sup_{\underline{a}_{s:t} \in \underline{\MA}_{s:t}} \sup_{\pi_{(t+1):T} \in \Pi_{(t+1):T}}& \E\left[ \left(\widehat{Q}_{t}^{\pi_{(t+1):T},(n)}(H_{t},a_{t}) - Q_{t}^{\pi_{(t+1):T}}(H_{t},a_{t})\right)^{2} \middle| \MS_{n}\right] \\
    &\times \E\left[\left(\frac{1}{\prod_{\ell=s}^{t-m}\hat{e}_{\ell}^{(n)}(H_{\ell},a_{\ell})} - \frac{1}{\prod_{\ell=s}^{t-m}e_{\ell}(H_{\ell},a_{\ell})} \right)^{2}\middle| \MS_{n} \right] = \frac{O_p(1)}{n^{\tau}}.
\end{align*}
(ii) There exists $n_0 \in \Natural$ such that for any $n \geq n_0$ and $t=1,\ldots,T$, 
\begin{align*}
     \sup_{a_{t} \in \MA_{t}, \pi_{(t+1):T} \in \Pi_{(t+1):T}}\left|\widehat{Q}_{t}^{\pi_{(t+1):T},(n)}(H_{t},a_{t})\right| < \infty \mbox{\ \ and\ \ }
     \min_{a_t \in \MA_t}\hat{e}_{t}^{(n)}(H_{t},a_{t}) > 0 
\end{align*}
hold a.s.

\end{assumption}

\medskip{}

As we will see later, the regret converges to zero at the $n^{-1/2}$ rate when Assumption~\ref{asm:rate_of_convergence_backward}~(i) holds with $\tau=1$. This is not very strong or restrictive. For example, Assumption~\ref{asm:rate_of_convergence_backward}~(i) is satisfied with $\tau = 1$ when
\begin{align}
&\sup_{a_{t} \in \MA_{t}}  \sup_{\pi_{(t+1):T} \in \Pi_{(t+1):T}}\E\left[\left(\widehat{Q}_{t}^{\pi_{(t+1):T},(n)}(H_{t},a_{t}) - Q_{t}^{\pi_{(t+1):T}}(H_{t},a_{t})\right)^{2}\middle| \MS_n\right] = \frac{O_p(1)}{\sqrt{n}} \label{eq:Qest_convergence_rate}\\
    \mbox{and}& \quad \sup_{\underline{a}_{s:t} \in \underline{\MA}_{s:t}} \E\left[\left(\frac{1}{\prod_{\ell=s}^{t}\hat{e}_{\ell}^{(n)}(H_{\ell},a_{\ell})} - \frac{1}{\prod_{\ell=s}^{t}e_{\ell}(H_{\ell},a_{\ell})} \right)^{2}\middle| \MS_n\right] = \frac{O_p(1)}{\sqrt{n}} \nonumber
\end{align}
hold for all $t=1,\ldots,T$ and $s=1,\ldots,t$. 

The uniform MSE convergence rate of $\widehat{Q}_{t}^{\pi_{(t+1):T},(n)}(H_{t},a_{t})$ for the FQE is not a standard result, but some existing results are applicable with some modifications \citep[e.g.,][]{Zhang_et_al_2018,Kallus_Uehara_2020}.
For example, \cite{Kallus_Uehara_2020} argue that the FQE estimators $\{\widehat{Q}_{t}^{\pi_{(t+1):T},(n)}:t=1,\ldots,T\}$ for a fixed DTR $\pi$ can be viewed as M-estimators that minimize $$n^{-1}\sum_{t=1}^{T}\sum_{i=1}^{n}\left(Y_{i,t} + Q_{t+1}^{\pi_{(t+2):T}}(H_{i,t+1},\pi_{t+1}) - Q_{t}^{\pi_{(t+1):T}}(H_{i,t},\pi_{t}) \right)^2$$ over (semi/non)parametric classes of Q-functions, where $Q_{T+1}^{\pi_{(T+2):T}}(H_{i,T+1},\pi_{T+1})=0$ by convention. This formulation makes it possible to apply existing results on (semi/nonparametric) M-estimation to establish convergence rates for FQE estimators.

In Supplemental Appendix~\ref{app:convergence_FQE}, we provide a fully worked example of FQE based on B-spline series least-squares regression. In particular, we derive primitive sufficient conditions under which the uniform MSE condition in \eqref{eq:Qest_convergence_rate} holds.

Note also that Assumption~\ref{asm:rate_of_convergence_backward}~(i) encompasses the property of double robustness; that is, Assumption~\ref{asm:rate_of_convergence_backward}~(i) holds if either $\widehat{Q}_{t}^{\pi_{(t+1):T},(n)}(\cdot,\cdot)$ is uniformly consistent or $\hat{e}_{t}^{(n)}(\cdot,\cdot)$ is consistent, provided that Assumption~\ref{asm:rate_of_convergence_backward}~(ii) holds.

We next consider the complexity of the class $\Pi$ of DTRs and the classes $\Pi_t$ of stage-specific policies. Following \cite{zhou2022offline}, we use the $\epsilon$-Hamming covering number to measure the complexity of the class of policy sequences $\pi_{s:t} \in \Pi_{s}\times \cdots \times \Pi_{t}$ for each $s$ and $t$ such that $s \leq t$. 

\medskip{}

\begin{definition}
(i) For any stages $s$ and $t$ such that $s \leq t$, given a set of history points $\{h_{t}^{(1)},\ldots,h_{t}^{(n)}\} \subseteq \MH_{t}$, we define the Hamming distance between two sequences of policies  $\pi_{s:t}, \pi_{s:t}^{\prime} \in \Pi_{s:t}$ as $d_{h}(\pi_{s:t},\pi_{s:t}^{\prime}) \equiv n^{-1}\sum_{i=1}^{n}\mathbf{1}\{\pi_s(h_{s}^{(i)})\neq \pi_{s}^{\prime}(h_{s}^{(i)}) \vee \cdots \vee \pi_t(h_{t}^{(i)})\neq \pi_{t}^{\prime}(h_{t}^{(i)})\}$,
where we denote $h_{t}^{(i)} = (\underline{a}_{t-1}^{(i)},\underline{s}_{t}^{(i)})$ and $h_{\ell}^{(i)}$ ($\ell =s,\ldots,t-1$) is the subvector of $h_{t}^{(i)}$ such that $h_{\ell}^{(i)} = (\underline{a}_{\ell-1}^{(i)},\underline{s}_{\ell}^{(i)}) \in \MH_{\ell}$. 
Let $N_{d_h}\left(\epsilon,\Pi_{s:t},\left\{h_{t}^{(1)},\ldots,h_{t}^{(n)}\right\}\right)$ be the smallest number of sequences of policies $\pi_{s:t}^{(1)},\pi_{s:t}^{(2)},\ldots$ in $\Pi_{s:t}$ such that for any $\pi_{s:t} \in \Pi_{s:t}$, there exists $\pi_{s:t}^{(i)}$ satisfying $d_h(\pi_{s:t},\pi_{s:t}^{(i)}) \leq \epsilon$.
We define the $\epsilon$-Hamming covering number of $\Pi_{s:t}$ as
\begin{align*}
    N_{d_h}(\epsilon,\Pi_{s:t})  \equiv  \sup\left\{N_{d_h}\left(\epsilon,\Pi_{s:t},\left\{h_{t}^{(1)},\ldots,h_{t}^{(n)}\right\}\right)\middle| n \geq 1, h_{t}^{(1)},\ldots,h_{t}^{(n)} \in \MH_t\right\}.
\end{align*}
(ii) We define the entropy integral of $\Pi_{s:t}$ as $\kappa(\Pi_{s:t})= \int_{0}^{1}\sqrt{\log N_{d_h}\left(\epsilon^2,\Pi_{s:t}\right)}d\epsilon$.
\end{definition}

\medskip{}

Note that when $s=t$, $N_{d_h}(\epsilon,\Pi_{s:t}) = N_{d_h}(\epsilon,\Pi_t)$ and $\kappa(\Pi_{s:t}) = \kappa(\Pi_t)$. 
We assume that the policy class $\Pi_t$ for each $t$ is not excessively complex in terms of the covering number.
\medskip{}

\begin{assumption}[Complexity of $\Pi_t$]\label{asm:bounded entropy}
For all $t=1,\ldots,T$, $N_{d_h}(\epsilon,\Pi_{t}) \leq C\exp(D(1/\epsilon)^{\omega})$ holds for any $\epsilon>0$ and some constants $C,D>0$ and $0<\omega<0.5$.
\end{assumption}

\medskip{}

This assumption implies that the covering number of $\Pi_t$ does not grow too quickly, but allows $\log N_{d_h}(\epsilon,\Pi_{t})$ to grow at the rate $\epsilon^{-\omega}$.
This assumption is satisfied, for example, by a class of finite-depth trees (see \citeauthor{zhou2022offline} (\citeyear{zhou2022offline}, Lemma 4)). In the case of a binary action set (i.e., $|\MA_t|=2$), a VC-class of $\pi_{t}$ also satisfies Assumption~\ref{asm:bounded entropy}. 
\citeauthor{zhou2022offline} (\citeyear{zhou2022offline}, Remark 4) show that the entropy integral $\kappa(\Pi_t)$ is finite under Assumption~\ref{asm:bounded entropy}. 

Assumptions~\ref{asm:first-best} and \ref{asm:bounded entropy} impose distinct but interacting restrictions on the policy class. Assumption~\ref{asm:first-best} requires the stage-specific policy classes to be sufficiently rich, or correctly specified, for backward induction to attain the welfare optimum over $\Pi$, whereas Assumption~\ref{asm:bounded entropy} restricts their statistical complexity. Thus, the two conditions jointly create a trade-off between specification and statistical complexity and require policy-relevant treatment-effect heterogeneity to admit a sufficiently low-complexity
representation. Remark~\ref{rem:policy_class_approximation} below discusses this trade-off and an approximate-specification framework based on increasingly rich policy classes.


Regarding the class $\Pi$ of entire DTRs, its entropy integral $\kappa(\Pi)$ is finite as well under Assumption~\ref{asm:bounded entropy}. 

\medskip

\begin{lemma}\label{lem:entropy_integral_bound}
Suppose that Assumption~\ref{asm:bounded entropy} holds.
For any $1\leq s\leq t\leq T$,
\[
\kappa(\Pi_{s:t})
\leq
\sqrt{(t-s+1)\log C}
+
\frac{\sqrt{D}}{1-\omega}
(t-s+1)^{(1+\omega)/2}.
\]
In particular,
\[
\kappa(\Pi)
\leq
\sqrt{T\log C}
+
\frac{\sqrt{D}}{1-\omega}
T^{(1+\omega)/2}
<\infty.
\]
\end{lemma}

\begin{proof}
See Appendix~\ref{app:preliminary_results} in the supplement.
\end{proof}

\medskip

This lemma makes explicit how the upper bound on the entropy integral of the full DTR class depends on the number of stages.

For \(1\leq s\leq t\leq T\), define $\eta_{s:t}:=\prod_{\ell=s}^t\eta_\ell$, and let \(\boldsymbol{\eta}:=(\eta_1,\ldots,\eta_T)\).
The following theorem presents our main regret result for the DTR $\hat{\pi}$ obtained by the proposed backward-induction approach, making explicit its dependence on the treatment horizon, policy-class complexity, and policy-relevant overlap.

\medskip

\begin{theorem}\label{thm:main_theorem_backward}
Suppose that Assumptions
\ref{asm:sequential independence}--\ref{asm:overlap},
\ref{asm:first-best},
\ref{asm:rate_of_convergence_backward}, and
\ref{asm:bounded entropy} hold.
Then the regret of the estimated DTR $\hat{\pi}$ satisfies
\[
R(\hat\pi)
=
O_p\left(
M\sum_{t=1}^T
B_t(\boldsymbol{\eta})\,
\kappa(\Pi_{t:T})
\sqrt{
\frac{\mathcal V_{t:T}(\boldsymbol{\eta})}{n}
}
\right)
+
O_p\left(
n^{-\min\{1/2,\tau/2\}}
\sum_{t=1}^T
B_t(\boldsymbol{\eta})(T-t+1)
\right),
\]
where
\[
B_1(\boldsymbol{\eta})=1,
\qquad
B_t(\boldsymbol{\eta})
=
\frac{2^{t-2}}{\eta_{1:(t-1)}},
\quad t=2,\ldots,T,
\]
and
\[
\mathcal V_{t:T}(\boldsymbol{\eta})
=
1+\frac{2}{\eta_{t:T}}
+3\sum_{s=t}^{T-1}\frac{1}{\eta_{t:s}},
\quad t=1,\ldots,T.
\]
In particular, when \(T\) is fixed, the bound reduces to
\[
R(\hat\pi)
=
O_p\left(\kappa(\Pi)n^{-1/2}\right)
+
O_p\left(n^{-\min\{1/2,\tau/2\}}\right).
\]
\end{theorem}

\begin{proof}
See Appendix~\ref{app:main_proof}.
\end{proof}

\medskip

When Assumption~\ref{asm:rate_of_convergence_backward}(i) holds with $\tau=1$, the proposed method attains the $n^{-1/2}$ rate for regret, which is minimax-optimal in $n$ under fixed policy-class complexity.\footnote{
In the case of binary treatment at each stage, \cite{Sakaguchi_2021} shows that the minimax optimal rate of convergence
for the regret is $\sqrt{V_{1:T}/n}$, where $V_{1:T}$ is the VC dimension of the class of DTRs.
}
This parallels the $n^{-1/2}$ regret guarantees obtained by \cite{Athey_Wager_2020} and \cite{zhou2022offline} for doubly robust policy learning in single-stage settings.\footnote{
In the single-stage binary-treatment setting,
\cite{Athey_Wager_2020} exploit the specific entropy growth of a VC class
and obtain a slightly sharper bound than one based on a general finite
entropy integral.
}

The bound also makes explicit how the statistical performance of the proposed method depends on the treatment horizon and policy-relevant overlap. The factor $B_t(\boldsymbol{\eta})$ propagates stage-$t$ policy-learning errors back to first-stage welfare, while $\mathcal V_{t:T}(\boldsymbol{\eta})$ captures the variance inflation associated with cumulative inverse propensity weights from stage $t$ onward. Thus, although the convergence rate in $n$ remains $n^{-1/2}$
for fixed $T$ and overlap constants bounded away from zero, the leading
constants in the bound can become large as the treatment horizon
increases or some $\eta_t$ approaches zero.

Deriving the asymptotic upper bound on $R(\hat\pi)$ is nontrivial
because the stage-specific policies are estimated sequentially rather
than simultaneously. If the DTR were estimated simultaneously across all stages, one could adapt the theoretical analysis of \citet{zhou2022offline}. However, the sequential nature of the estimation precludes a direct application of their approach. Despite this challenge, Appendix~\ref{app:main_proof} develops novel analytical techniques for evaluating the regret of the sequentially estimated DTR.


\medskip

\begin{remark}[Policy-class complexity and approximate specification]
\label{rem:policy_class_approximation}

The specification requirement in Assumption~\ref{asm:first-best} can be
interpreted geometrically. Suppose, for example, that treatment is
binary and define, for a relevant future policy sequence
$\pi_{(t+1):T}$,
\[
    \Delta_t^{\pi_{(t+1):T}}(h_t)
    =
    Q_t^{\pi_{(t+1):T}}(h_t,1)
    -
    Q_t^{\pi_{(t+1):T}}(h_t,0).
\]
If $\Pi_t$ consists of linear-threshold policies, the relevant
optimal-action region must be representable by a linear decision
boundary; for finite-depth trees, it must be representable by a tree of
the prescribed depth. Thus, together with
Assumption~\ref{asm:bounded entropy}, Assumption~\ref{asm:first-best}
imposes a trade-off between specification and statistical complexity.

One way to relax reliance on a single fixed policy class is to consider
nested sieve classes
\[
    \Pi_{t,1}\subseteq\Pi_{t,2}\subseteq\cdots\subseteq\overline{\Pi}_t,
    \qquad
    \Pi^{(K)}
    =
    \Pi_{1,K}\times\cdots\times\Pi_{T,K},
\]
where $\overline{\Pi}
=\overline{\Pi}_1\times\cdots\times\overline{\Pi}_T$
is a richer class of admissible DTRs. This is analogous to sieve
empirical welfare maximization; for example,
\cite{Mbakop_Tabord-Meehan_2019} approximate a rich class of monotone
treatment rules by finite-complexity subclasses with increasingly
flexible boundaries.\footnote{
With binary treatments, one may similarly
use polynomial-threshold sieves:
$
\Pi_{t,K}
=
\left\{
h_t\mapsto
\mathbf 1\{p_K(h_t)^\prime\beta\geq0\}:
\beta\in\mathbb R^{d_K}
\right\},
$
where $p_K(h_t)$ is a nested sequence of finite-dimensional polynomial
bases.
For each fixed $K$, these are
finite-dimensional VC classes satisfying
Assumption~\ref{asm:bounded entropy}, while increasing $K$
allows richer treatment boundaries at the cost of greater complexity.
}

Let
\[
    W_{\overline{\Pi}}^*
    :=
    \sup_{\pi\in\overline{\Pi}}W(\pi),
    \qquad
    W_K^*
    :=
    \sup_{\pi\in\Pi^{(K)}}W(\pi),
\]
and let $\pi^{B,K}$ denote the population backward-induction solution
over $\Pi^{(K)}$. Then
\[
    W_{\overline{\Pi}}^*-W(\pi^{B,K})
    =
    \underbrace{W_{\overline{\Pi}}^*-W_K^*}
    _{\text{sieve approximation error}}
    +
    \underbrace{W_K^*-W(\pi^{B,K})}
    _{\text{backward-induction suboptimality}}.
\]
The first term is the usual sieve approximation error. The second term
measures the possible welfare loss from backward rather than joint
optimization within $\Pi^{(K)}$. If
Assumption~\ref{asm:first-best} holds with $\Pi$ replaced by
$\Pi^{(K)}$, this term is zero.

An approximate-specification framework may therefore require
\[
    W_{\overline{\Pi}}^*-W_K^*\to0
    \qquad\text{and}\qquad
    W_K^*-W(\pi^{B,K})\to0.
\]
The second condition can be viewed as an approximate counterpart of
Assumption~\ref{asm:first-best}: rather than requiring backward
induction to attain the within-class welfare optimum exactly for each
$K$, it only requires the corresponding welfare gap to vanish as
$K\to\infty$. Increasing $K$, however, generally increases the
statistical complexity of $\Pi^{(K)}$, yielding the usual trade-off
between approximation error and statistical complexity. A formal
analysis with growing policy sieves would require extending
Theorem~\ref{thm:main_theorem_backward}, which is stated for a fixed
policy class.
\end{remark}

\medskip

\begin{remark}[Practical guidance for longer horizons]
\label{rem:practical_guidance_long_horizon}
In applications with long time horizons, overlap can become increasingly
fragile. As $t$ increases, the history $H_t=(\underline A_{t-1},\underline S_t)$ becomes
higher-dimensional, potentially leading to sparse history--treatment
combinations. Moreover, because the AIPW scores involve cumulative inverse-propensity
weights across stages, poor overlap at even a single stage can lead to
extreme weights and unstable score values.

A common finite-sample practice is to truncate estimated
propensity scores from below when constructing the AIPW scores, for
example by replacing $\widehat e_t(h_t,a_t)$ with
$\max\{\widehat e_t(h_t,a_t),\underline e\}$ for a small threshold
$\underline e>0$. Such truncation should be viewed as a stabilization or
sensitivity device rather than as a substitute for overlap.

Substantive knowledge about the dynamic structure can also improve stability. If sufficiently remote past treatments and state variables affect subsequent treatment assignments and relevant future outcomes only
through a lower-dimensional summary of the history, the propensity scores and policy-specific Q-functions can be modeled as functions of that reduced history. Doing so can alleviate sparsity, mitigate overlap problems, and simplify nuisance-function estimation.

Alternative weighting and debiasing methods may provide further stabilization. For example, \citet{viviano_Bradix_2026} propose dynamic covariate balancing, while \citet{Chernozhukov_et_al_2022} develop recursive debiased machine-learning methods for dynamic treatment effects. Adapting these approaches to dynamic policy learning would likely require constructing policy-dependent balancing weights or recursive orthogonal scores and establishing uniform control over the policy class. A formal analysis of such extensions is beyond the scope of this paper.
\end{remark}


\section{Existing Approach}\label{sec:existing_approach}

An alternative doubly robust approach to estimating the optimal DTR is to maximize an AIPW estimator of the welfare function simultaneously over the entire DTR class.  Specifically, we have 
\begin{align}
    \hat{\pi}^{AIPW} = \argmax_{\pi \in \Pi} & \ \widehat{W}^{AIPW}(\pi) \mbox{\ \ with} \label{eq:simultaneous_maximization}\\
    \widehat{W}^{AIPW}(\pi) = \frac{1}{n}\sum_{i=1}^{n}\sum_{t=1}^{T}&\left(\hat{\psi}_{it}^{-k(i)}\left(\underline{\pi}_{t}\right)\cdot Y_{it} \right. \nonumber \\
      - &\left.\left(\hat{\psi}_{it}^{-k(i)}\left(\underline{\pi}_{t}\right)- \hat{\psi}_{i,t-1}^{-k(i)}\left(\underline{\pi}_{t-1}\right)\right)\cdot \widehat{Q}_{t}^{\pi_{(t+1):T},-k(i)}\left(H_{it},\pi_t\right)\right) \nonumber
\end{align}
and $\hat{\psi}_{it}^{-k(i)}(\underline{\pi}_{t}) \equiv \left(\prod_{s=1}^{t}\mathbf{1}\left\{ A_{is}=\pi_{s}(H_{is})\right\}\right)/(\prod_{s=1}^{t}\hat{e}_{s}^{-k(i)}\left(H_{is},A_{is}\right))$, where we set
$\widehat\psi_{i,0}^{-k(i)}=1$ by convention. The quantity $\widehat{W}^{AIPW}(\pi)$ is an AIPW estimator of $W(\pi)$. This approach was originally proposed by \citet{Zhang_et_2013} without cross-fitting and for binary treatments at each stage, although its statistical properties were not established.\footnote{\cite{Jiang_Li_2016}, \cite{Thomas_et_al_2016}, and \cite{Kallus_Uehara_2020} propose AIPW estimators of welfare functions for evaluating fixed DTRs, but their focus is not on optimizing DTRs.}
An advantage of this method is that it does not require the correct specification of $\Pi$ (Assumption~\ref{asm:first-best}) for consistent estimation of the optimal DTR.

However, this approach faces two computational challenges. First, since the nuisance components $\{Q_{t}^{\pi_{(t+1):T}}\}_{t=1,\ldots,T}$ depend on each specific DTR $\pi$, implementing the method requires estimating $\{Q_{t}^{\pi_{(t+1):T}}\}_{t=1,\ldots,T}$ for every candidate DTR $\pi$.\footnote{\label{fn:heuristic_simultaneous}
A heuristic alternative to solve this difficulty, proposed by \cite{Zhang_et_2013}, is to first estimate the optimal policy by Q-learning, denoted by $\hat{\pi}_{t}^{Q}$, and then use the optimal Q-function estimate $\widehat{Q}_{t}^{\hat{\pi}_{(t+1):T}^{Q},-k(i)}\left(H_{it},a_{t}\right)$ instead of $\widehat{Q}_{t}^{\pi_{(t+1):T},-k(i)}\left(H_{it},a_t\right)$ when constructing and maximizing $\widehat{W}^{AIPW}(\pi)$. This approach avoids the need to estimate $Q_{t}^{\pi_{(t+1):T}}$ for each $\pi_{(t+1):T}$.
} \citet{Nie_et_al_2021} highlight this computational burden.
Second, maximizing $\widehat{W}^{AIPW}(\pi)$ simultaneously across all stages is computationally demanding, especially when $T$ is not very small, as the problem is non-convex. These computational issues make the approach intractable unless the class of DTRs $\Pi$ is sufficiently small.

In contrast, our proposed approach offers notable computational advantages: (i) the nuisance components $\{Q_{t}^{\hat{\pi}_{(t+1):T}}\}_{t=1,\ldots,T}$ depend only on the continuation policies
$\hat{\pi}_{(t+1):T}$ estimated at earlier steps of the backward
recursion, and (ii) the backward optimization is computationally more efficient than joint optimization across all stages. While our approach requires estimating the Q-function $Q_{t}^{\hat{\pi}_{(t+1):T}}$ at each step of the sequential procedure, the associated computational cost is relatively small.

Although the simultaneous maximization approach (\ref{eq:simultaneous_maximization}) is not our primary proposal, we show its statistical properties as follows.\footnote{\citeauthor{Sakaguchi_2021} (\citeyear{Sakaguchi_2021}, Theorem E.1) shows statistical properties of this approach in the case of binary treatment.}

\medskip

\begin{theorem}\label{thm:theorem_simlutaneous}
Suppose that Assumptions
\ref{asm:sequential independence}--\ref{asm:overlap},
\ref{asm:rate_of_convergence_backward}, and
\ref{asm:bounded entropy} hold.
Then the regret of the simultaneous AIPW estimator satisfies
\[
R(\widehat{\pi}^{AIPW})
=
O_p\left(
M\kappa(\Pi)
\sqrt{
\frac{
\mathcal V_{1:T}(\boldsymbol{\eta})
}{n}
}
\right)
+
O_p\left(
n^{-\min\{1/2,\tau/2\}}
\right),
\]
where $\mathcal V_{1:T}(\boldsymbol{\eta})$ is defined in Theorem~\ref{thm:main_theorem_backward}.
\end{theorem}

\begin{proof}
See Appendix~\ref{app:statistical_property_AIPW} in the supplement.
\end{proof}

\medskip

This theorem shows that the simultaneous AIPW estimator $\widehat{\pi}^{AIPW}$ attains the same convergence rate in $n$ as the
proposed backward-induction estimator under the same requirements on the MSE convergence rates of the nuisance-component estimators and the complexity of the DTR class. Because the simultaneous approach directly
optimizes over the entire DTR class, it does not require Assumption~\ref{asm:first-best}. Moreover, unlike the regret bound for the backward-induction estimator, its bound does not contain the propagation factors $B_t(\boldsymbol{\eta})$, which arise from the propagation of stage-specific policy-learning errors. The factor $\mathcal V_{1:T}(\boldsymbol{\eta})$ remains, however, so cumulative weak overlap can still inflate the leading constant.

\begin{remark}[Computational burden]\label{rem:computational_burden}
We provide a simple comparison of the policy-optimization burden of the
backward and simultaneous approaches. To isolate this difference, we
abstract from nuisance-function estimation and compare the proposed
approach with the heuristic simultaneous approach of
\citet{Zhang_et_2013} described in
Footnote~\ref{fn:heuristic_simultaneous}, which avoids estimating
policy-specific $Q$-functions.

Suppose that each stage-specific policy class $\Pi_t$ is finite, with
cardinality $L_t$, and that optimization is performed by exhaustive
search. The proposed backward approach solves $T$ stage-specific
optimization problems and therefore evaluates
$
    \sum_{t=1}^T L_t
$
candidate policies. By contrast, the heuristic simultaneous approach
jointly optimizes over the entire DTR class and therefore evaluates
$
    \prod_{t=1}^T L_t
$
candidate DTRs. Thus, when $L_t=L$ for all $t$, the respective numbers
of candidate evaluations are $TL$ and $L^T$. Consequently, unless both
the time horizon and the stage-specific policy classes are small,
simultaneous optimization can become computationally prohibitive,
whereas backward optimization may remain tractable. An analogous
computational contrast applies to structured policy classes such as finite-depth policy trees. Supplemental Appendix~\ref{app:linear_computational_comparison} compares learning performance and computation times of the two approaches in a simulation study.
\end{remark}


\section{Simulation Study}\label{sec:simulation}

We conduct a simulation study to examine the finite-sample performance of the approach presented in Section~\ref{sec:doubly_robust_policy_learning}. We consider three data-generating processes (DGPs), labeled DGPs 1--3.
DGPs 1 and 2 are stylized designs that allow us to study the performance of the proposed method under controlled forms of dynamic treatment effect heterogeneity, whereas DGP3 is an empirical Monte Carlo design calibrated to the Project STAR data.

All three DGPs consist of two stages of binary treatment assignment, $(A_1,A_2)\in\{0,1\}^2$, second-stage potential outcomes
$\{Y_2(a_1,a_2):(a_1,a_2)\in\{0,1\}^2\}$, and state variables $S_1$ and $S_2$.

\subsection{Stylized Simulation Designs: DGPs 1 and 2}

We first consider two stylized DGPs that feature different forms of dynamic treatment effect heterogeneity. In both DGPs, the first-stage state vector consists of 20 variables, $S_1=(S_1^{(1)},\ldots,S_1^{(20)})$, and a scalar state variable $S_2$ is observed at the second stage. The DGPs are specified as follows:
\begin{align*}
&(S_{1}^{(1)},\ldots,S_{1}^{(20)})^{\prime} \thicksim N(\mathbf{0},I_{20});\\
&S_{2}\left(a_{1}\right)=  \sign\left(S_{1}^{(1)}\right)\cdot a_{1}+ S_{1}^{(2)} + \left(S_{1}^{(3)}\right)^2 + S_{1}^{(4)} + \varepsilon_{1} \mbox{\ with\ }\varepsilon_{1} \sim N(0,1);\\
&Y_{2}\left(a_{1},a_{2}\right) =  \phi(a_1,S_2(a_1))\cdot a_{2} 
    + 0.5 \cdot S_{2}(a_1) + S_{1}^{(4)} - \left(S_{1}^{(5)}\right)^2 + S_{1}^{(6)} + \varepsilon_{2} \mbox{\ with\ } \varepsilon_{2} \sim N(0,1);\\
&A_{1} \thicksim Ber\left(1/(1+e^{0.5 S_{1}^{(2)}-0.5S_{1}^{(3)}-S_{1}^{(5)}})\right),\ A_{2} \thicksim Ber\left(1/(1+e^{0.5S_{1}^{(5)}+0.5 S_{2}-0.2A_{1}})\right).
\end{align*}
In DGPs 1 and 2, we specify $\phi(a_1,S_2(a_1))$ as $\phi(a_1,S_2(a_1)) = \sign(S_2(a_1)\cdot (a_1 - 1/2))$ and $\phi(a_1,S_2(a_1)) = S_2(a_1) + (a_1 - 1/2)$, respectively. In each DGP, the first-stage treatment $a_1$ influences the outcome $Y_2$ through both direct and indirect channels: (i) a direct effect on $Y_2$ via treatment effect heterogeneity $Y_2(a_1,1)-Y_2(a_1,0)$; (ii) an indirect effect on $Y_2$ via the second-stage state $S_2(a_1)$. 

We compare the proposed approach (labeled ``DR'') with three existing methods:  Q-learning without and with policy search (labeled ``Q-learn'' and ``Q-search,'' respectively) and the IPW classification-based approach with backward optimization (labeled ``IPW'').\footnote{Following \cite{Murphy_2005}, we separately consider Q-learning with and without policy search. In Q-learning with policy search, the optimal policy for each stage $t$ is chosen from a pre-specified policy class $\Pi_t$, specifically estimated as $\hat{\pi}_{t} = \argmax_{\pi_{t}\in \Pi_{t}}\sum_{i=1}^{n}\widehat{Q}_{t}^{\hat{\pi}_{(t+1):T}}(H_{it},\pi_{t})$. 
In contrast, Q-learning without policy search optimizes the policy for each stage $t$ over all measurable policies, such as $\hat{\pi}_{t}(h_t) = \argmax_{a_{t}\in \MA_{t}}\sum_{i=1}^{n}\widehat{Q}_{t}^{\hat{\pi}_{(t+1):T}}(h_{t},a_{t})$ for any $h_t$. This approach consistently estimates the first-best DTR unless the Q-functions are misspecified.}
For each method, we use generalized random forests \citep{Athey_et_al_2019} to estimate nuisance components. We set $K=5$ for cross-fitting in the proposed approach.

For DR, Q-search, and IPW, we use a class of DTRs $\Pi=\Pi_{1} \times \Pi_{2}$ with $\Pi_1$ being the class of depth-1 decision trees based on $H_1$, and $\Pi_2$ being the class of depth-2 decision trees based on $H_2$. In DGP1, $\Pi$ is correctly specified in the sense of Assumption~\ref{asm:first-best}, whereas in DGP2, $\Pi$ is misspecified, potentially leading to a loss of optimality in backward optimization. Note that Q-learn consistently estimates the optimal DTR in both DGPs. We solve the stage-wise optimization problems involving decision trees using the exact learning algorithm proposed by \cite{zhou2022offline}.

Tables~\ref{table:mc_dgp1} and \ref{table:mc_dgp2} report mean welfare (with Monte Carlo standard deviations), regret, and average computation times from 500 replications for sample sizes $n=250,500,1000,2000,$ and $4000$.\footnote{All simulations were conducted on a machine with a 13th Gen Intel(R) Core(TM) i9-13900 (2.00 GHz) processor and 64 GB of RAM. For DGPs 1--3, Monte Carlo replications were parallelized across 10 workers. The reported computation times are average elapsed times for individual replications.} Welfare and regret are evaluated using an independent test sample of 50,000 observations from the corresponding DGP. DR achieves the highest mean welfare and the lowest regret among the methods considered at every sample size in both DGPs.
For example, in DGP1 at $n=500$, its mean welfare is about 40\% higher than that of the next-best method. In DGP2, DR also outperforms Q-learn in finite samples even though Q-learn consistently estimates the first-best DTR.

\subsection{DGP 3: Empirical Monte Carlo Simulation}

DGP3 is tailored to the empirical application in Section~\ref{sec:empirical_illustration}. Specifically, we apply the
Wasserstein generative adversarial network (WGAN)-based empirical Monte Carlo approach of \cite{Athey_et_al_2024} to the Project STAR data and adapt their static-treatment procedure to a two-stage dynamic treatment setting. The purpose of this design is to evaluate the finite-sample performance of the proposed method in a data environment that more closely resembles the empirical application.

The treatment timing and variable construction in DGP3 follow those in the empirical application. The first-stage treatment
$A_1$ denotes the kindergarten class type, and the second-stage treatment $A_2$ denotes the grade-1 class type. We set $A_t=0$ for a regular-size class with a full-time teacher aide and $A_t=1$ for a small-size class without a teacher aide. The final outcome $Y_2$ is the cost-adjusted grade-1 outcome used in the empirical application. The state variables are based on the corresponding student characteristics and intermediate achievement measures used in Section~\ref{sec:empirical_illustration}.

Using the fitted WGANs, we construct DGP3 to approximate the data environment of the Project STAR application. For each synthetic individual, we generate the treatment assignments $A_1$ and $A_2$, the baseline state variables $S_1$, the potential intermediate states $\{S_2(a_1):a_1\in\{0,1\}\}$, and the potential final outcomes $\{Y_2(a_1,a_2):(a_1,a_2)\in\{0,1\}^2\}$. The observed trajectory used for policy learning is obtained by setting $S_2=S_2(A_1)$ and $Y_2=Y_2(A_1,A_2)$, while the full set of potential variables is used for policy evaluation. Supplemental Appendix~\ref{app:wgan_dgp} provides full details of how the WGAN-based empirical Monte Carlo procedure is adapted to the dynamic treatment setting.

Table~\ref{table:mc_dgp3} reports mean welfare (with Monte Carlo standard
deviations), regret, and average computation times across 500 Monte
Carlo replications. Welfare and regret are evaluated using the same independent-sample
procedure as in DGPs 1 and 2. Among the methods restricted to the policy class
$\Pi$, DR achieves the highest mean welfare and the lowest regret at
every sample size considered. Q-learn achieves higher welfare at larger
sample sizes, but unlike the other methods, it is not restricted to
$\Pi$.

To assess the extent to which the regret of DR reflects the population
suboptimality of backward optimization rather than finite-sample
learning and estimation error, we also numerically approximate the
population welfare gap between joint and backward optimization over
$\Pi$. This gap is about 0.38 in DGP3, which is substantially smaller
than the regret of DR at all sample sizes considered. Thus, in this
application-calibrated design, the welfare loss attributable to backward
optimization accounts for only a relatively small portion of the
finite-sample regret of DR.

\medskip

\begin{remark}[Additional Simulation Studies]
Supplemental Appendix~\ref{app:additional_simulation_results} presents
three additional simulation studies examining the robustness and
practical performance of the proposed method.

First, we examine the finite-sample consequences of misspecification of
either the Q-functions or the propensity scores. The results show that
the performance of the proposed doubly robust method remains relatively
stable when either nuisance component is misspecified.

Second, we compare the proposed backward-induction approach with the
simultaneous optimization approach under linear policy classes. The two
approaches achieve similar welfare and regret, while simultaneous
optimization is substantially more computationally demanding.

Third, we study the effects of longer treatment horizons and weaker
policy-relevant overlap, including sensitivity to propensity-score
clipping. Normalized regret generally increases as the treatment horizon
becomes longer and overlap becomes weaker.
\end{remark}


\section{Empirical Application}\label{sec:empirical_illustration}

We apply the proposed approach to data from Project STAR \citep[e.g.,][]{krueger_1999, Ding_Lehrer_2010}.
We study the optimal allocation of students to regular-size classes (22 to 25 students per teacher) with a full-time teacher aide and small-size classes (13 to 17 students per teacher) without one in their early education (kindergarten and grade 1).\footnote{\cite{krueger_1999} reports that the presence of a teacher aide did not have a significant impact on student test scores. However, the effects of teacher aides on academic attainment have not been examined in a framework that accounts for multistage treatment and treatment-effect heterogeneity.} We use a dataset of 1,877 students who were assigned to either regular-size classes with a full-time teacher aide or small-size classes without one in kindergarten. Among these students, 702 were randomly assigned to regular-size classes with a teacher aide, while the remaining students were assigned to small-size classes without a teacher aide in kindergarten (labeled ``grade K''). Upon their progression to grade 1, students were expected to remain in the same class type. However, about 10\% of students switched class types (see, e.g., \cite{Ding_Lehrer_2010} for a detailed discussion). We leverage this variation to estimate the optimal DTR and consider this empirical task in the observational data setting.

We investigate the optimal allocation of students to two class types in grades K and 1, based on their socioeconomic backgrounds, educational environments, and intermediate academic achievement. Each student’s final academic achievement is measured by their mathematics test score at the end of grade 1. 
We define the first and second stages ($t=1$ and $t=2$) as grades K and 1, respectively. 
We define the action set $\MA_{t} = \{\text{aide},\text{small}\}$,
where the treatment variable $A_{t}$ is labeled ``aide'' if a student is in a regular-size class with a teacher aide at stage $t$, and ``small'' if the student is in a small-size class. 

Because both small classes and regular-size classes with a full-time teacher aide require
additional resources, we incorporate these costs into the welfare objective. Since the
outcome is measured in test-score units, we convert the monetary cost of each class type
into grade-1 mathematics test-score units, following the discussion in
\citeauthor{krueger_1999} (\citeyear{krueger_1999}, Section IV). Supplemental Appendix~\ref{appendix:cost_benefit} provides a detailed discussion of this conversion, which accounts for the monetary costs of both a full-time teacher aide and small class size, as well as the present value, at age 6, of lifetime earnings gains associated with improvements in mathematics test scores. The cost of a full-time teacher aide per student is estimated to be equivalent to 0.806 points in kindergarten and 0.782 points in grade 1 on the grade-1 mathematics test, whereas the corresponding costs of small class size are 2.120 points in kindergarten and 2.059 points in grade 1.

We define the potential outcome as the grade-1 mathematics test score net of the incremental costs of the assigned class types:
\[
Y_2(a_1,a_2)
=
\widetilde{Y}_2(a_1,a_2)
-c_1\mathbf{1}\{a_1=\text{small}\}
-c_2\mathbf{1}\{a_2=\text{small}\},
\]
where $\widetilde{Y}_2(a_1,a_2)$ denotes the mathematics test score at the end of grade 1 under the class-type sequence $(a_1,a_2)$. The parameters $c_1=2.120-0.806 \simeq 1.31$ and $c_2=2.059-0.782 \simeq 1.28$ represent the incremental costs,
measured in grade-1 mathematics test-score units, of assigning a student to a small class rather than to a regular-size class with a full-time teacher aide in kindergarten and grade 1, respectively. We do not include the first-stage test score in the welfare objective function $W(\pi)$.

We use seven variables in $H_1 (=S_{1})$: student gender, student ethnicity (White/Asian or other), eligibility for free or reduced-price school lunch, school location type (rural or non-rural), teacher’s degree (bachelor’s or higher), years of teaching experience, and teacher ethnicity (White or other). The second-stage state vector $S_2$ includes the mathematics and reading test scores at the end of grade K. Recall that $H_2 = (A_1, S_1, S_2)$. We assume that the academic and socioeconomic information contained in $H_2$ satisfies the conditional ignorability assumption (Assumption~\ref{asm:sequential independence}) for grade 1, as these factors are strongly associated with the self-selection into class types.  Since there is no self-selection in grade K, the ignorability condition for the first-stage holds.

The class of DTRs $ \Pi = \Pi_1 \times \Pi_2$ is defined as follows. For the policy class $\Pi_1$ associated with class allocation in grade K, we employ a class of depth-1 trees that may take splitting variables from eligibility for free or reduced-price school lunch and school location type. For the policy class $\Pi_2$ associated with class allocation in grade 1, we use a class of depth-2 trees that may take splitting variables from math test scores at the end of grade K and the grade-K class type. 
This specification reflects the maintained empirical restriction that the  policy-relevant treatment-effect heterogeneity for grade-1 class-type allocation  is summarized by prior mathematics achievement and the previous class type.  Under this restriction, a depth-2 tree is sufficiently rich to represent the  threshold-based continuation rules that we view as substantively plausible in this  application, including rules with different mathematics-score cutoffs by grade-K class type. Thus, the baseline policy class provides a plausible specification for  Assumption~\ref{asm:first-best} in this empirical setting.  We also examine the sensitivity of the empirical results to a  richer depth-3 specification for $\Pi_2$. 

We exclude student gender and student ethnicity because using these variables for treatment choice would be discriminatory. We also exclude teacher characteristics from the policy variables, because the empirical exercise concerns class-type allocation rather than teacher assignment. We also do not impose an explicit capacity constraint on the number of students assigned to small-size classes. Instead, the empirical exercise should be interpreted as a cost-benefit class-type allocation problem.

In applying the proposed approach, we employ 5-fold cross-fitting and use generalized random forests \citep{Athey_et_al_2019} to estimate the nuisance components. The decision trees are optimized using the exact learning algorithm of \cite{zhou2022offline}.

Figure~\ref{fig:estimated_decision_trees} shows the DTR estimated using the proposed approach under the baseline depth-2 specification for $\Pi_2$.  The estimated policy for class allocation in grade K uses eligibility for free or 
reduced-price lunch to determine class type: students eligible for free or reduced-price lunch are assigned to small-size classes, while other students are  assigned to regular-size classes with a teacher aide. 
The estimated policy for grade 1 class allocation uses the grade-K mathematics  test score and the grade-K class type.  For example, students who were assigned to a regular-size class with a teacher aide in grade K and whose grade-K mathematics test score is no greater than 494 are assigned to a small-size class in grade 1.
Supplemental Appendix~\ref{app:depth3_tree} presents the corresponding
estimated DTR under the depth-3 specification.

Table \ref{table:empirical_results} shows estimates of the welfare gains for the optimal DTR, $\pi^{\ast,opt}$, relative to uniformly assigning all students either to regular-size classes with teacher aides or to small classes in both grades (i.e., $W(\pi^{\ast,opt}) - \E[Y_2(\text{aide},\text{aide})]$ and $W(\pi^{\ast,opt}) - \E[Y_2(\text{small},\text{small})]$).\footnote{We use a 5-fold sample-splitting procedure. For each fold, the DTR is
learned using the other four folds and evaluated on the held-out fold
using a doubly robust policy-value estimator. The nuisance components
used for policy evaluation are themselves estimated by cross-fitting
within the held-out fold. The reported welfare estimates are obtained
by aggregating the corresponding out-of-sample evaluations across the
five folds.} The table also reports 95\% confidence intervals and the shares of
students assigned to each allocation arm $(a_1,a_2)\in\{\mathrm{aide},\mathrm{small}\}^2$. 

For the baseline depth-2 specification for $\Pi_2$, the estimated welfare gain is 11.74 test-score points (0.26 SD) relative to the all-aide benchmark and 3.19 test-score points (0.07 SD) relative to the all-small benchmark. The former gain is statistically distinguishable from zero, whereas the latter is positive but imprecisely estimated.
Thus, the point estimates suggest that the estimated DTR improves cost-adjusted academic achievement relative to both uniform benchmark policies, although the improvement relative to uniform assignment to small classes is not statistically conclusive.

Under the estimated DTR, more than half of the students are assigned to a regular-size class with a teacher aide in at least one of grades K and 1. Given the higher cost of small-size classes relative to regular-size classes with a teacher aide on a per-student basis \citep{Word_et_al_1990}\footnote{According to \cite{Word_et_al_1990}, adding a full-time aide in Grades K-3 across Tennessee cost approximately 75 million dollars annually, while reducing class sizes by one-third cost around 196 to 205 million dollars per year.}, this finding suggests that allocation based on the DTR can reduce the costs associated with class-size reduction.

The sensitivity check using depth-3 grade-1 trees yields very similar
welfare estimates. The estimated welfare gain is 11.33 test-score
points (0.25 SD) relative to the all-aide benchmark and 2.78 test-score
points (0.06 SD) relative to the all-small benchmark. Although the
allocation shares change somewhat, allowing an additional split in the
grade-1 policy tree does not materially change the estimated welfare.
This similarity provides some support for using the depth-2 tree as a
parsimonious and interpretable baseline specification, although the
sensitivity analysis does not constitute a formal verification of
Assumption~\ref{asm:first-best}.

\section{Conclusion}\label{sec:conclusion}
We studied the statistical learning of the optimal DTR using observational data and developed a novel doubly robust approach for learning it under the assumption of sequential ignorability. Based on backward induction, the approach learns the optimal DTR sequentially, ensuring computational tractability. Our main result shows that the resulting DTR achieves the optimal convergence rate of $n^{-1/2}$ for welfare regret under mild conditions on the MSE convergence rates of the propensity-score and Q-function estimators. The simulation study shows that the proposed approach generally outperforms the competing methods in finite-sample settings. Applying the proposed approach to Project STAR data, we estimate the optimal DTR for the sequential allocation of students to regular-size classes with a teacher aide and small-size classes without one in early education.


\newpage
\appendix
\section*{Tables}

\begin{table}[!htbp]
\centering
\caption{Monte Carlo simulation results for DGP1}
\label{table:mc_dgp1}
\begin{threeparttable}
\small
\setlength{\tabcolsep}{4.5pt}
\begin{tabular}{llccccc}
\toprule
Method & Metric
& \multicolumn{5}{c}{Sample size $(n)$} \\
\cmidrule(lr){3-7}
& & 250 & 500 & 1000 & 2000 & 4000 \\
\midrule
\multirow{3}{*}{Q-learn}
& Welfare (SD) & 0.20 (0.11) & 0.31 (0.11) & 0.44 (0.10) & 0.58 (0.07) & 0.69 (0.03) \\
& Regret       & -- & -- & -- & -- & -- \\
& Time (sec.)  & 2.67 & 3.98 & 7.47 & 15.06 & 28.91 \\
\addlinespace[0.3em]

\multirow{3}{*}{Q-search}
& Welfare (SD) & 0.16 (0.18) & 0.28 (0.22) & 0.41 (0.21) & 0.54 (0.16) & 0.64 (0.10) \\
& Regret       & 0.58 & 0.46 & 0.34 & 0.21 & 0.10 \\
& Time (sec.)  & 2.41 & 3.63 & 8.02 & 25.04 & 81.48 \\
\addlinespace[0.3em]

\multirow{3}{*}{IPW}
& Welfare (SD) & 0.23 (0.13) & 0.30 (0.15) & 0.37 (0.16) & 0.46 (0.14) & 0.55 (0.12) \\
& Regret       & 0.51 & 0.45 & 0.38 & 0.28 & 0.20 \\
& Time (sec.)  & 1.87 & 3.05 & 7.34 & 24.27 & 79.37 \\
\addlinespace[0.3em]

\multirow{3}{*}{DR}
& Welfare (SD) & 0.28 (0.17) & 0.43 (0.17) & 0.61 (0.11) & 0.70 (0.04) & 0.72 (0.02) \\
& Regret       & 0.46 & 0.31 & 0.14 & 0.05 & 0.02 \\
& Time (sec.)  & 17.36 & 22.72 & 45.66 & 103.99 & 247.79 \\
\bottomrule
\end{tabular}
\begin{tablenotes}\footnotesize
\item Notes: Results are based on 500 Monte Carlo replications. Welfare (SD)
reports mean welfare, with Monte Carlo standard deviations in
parentheses. Regret is defined as
$R_\Pi(\hat\pi)=W(\pi_\Pi^\ast)-W(\hat\pi)$, where $\pi_\Pi^\ast$ is the
oracle welfare-maximizing DTR within the policy class $\Pi$. We
numerically approximate $\pi_\Pi^\ast$ by maximizing simulated welfare
over $\Pi$ using an independent sample of 50,000 observations from DGP1.
The welfare of both the resulting oracle policy and each estimated
policy is evaluated on a separate independent test sample of 50,000
observations from DGP1, which is also used to compute regret. Because
Q-learning without policy search is not restricted to $\Pi$, regret
relative to $\pi_\Pi^\ast$ is not reported for Q-learn. Time is the
average elapsed computation time per replication in seconds.
\end{tablenotes}
\end{threeparttable}
\end{table}

\begin{table}[!htbp]
\centering
\caption{Monte Carlo simulation results for DGP2}
\label{table:mc_dgp2}
\begin{threeparttable}
\small
\setlength{\tabcolsep}{4.5pt}
\begin{tabular}{llccccc}
\toprule
Method & Metric
& \multicolumn{5}{c}{Sample size $(n)$} \\
\cmidrule(lr){3-7}
& & 250 & 500 & 1000 & 2000 & 4000 \\
\midrule
\multirow{3}{*}{Q-learn}
& Welfare (SD) & 1.30 (0.34) & 1.70 (0.12) & 1.78 (0.08) & 1.83 (0.05) & 1.89 (0.03) \\
& Regret       & -- & -- & -- & -- & -- \\
& Time (sec.)  & 2.50 & 3.98 & 7.47 & 14.56 & 27.28 \\
\addlinespace[0.3em]

\multirow{3}{*}{Q-search}
& Welfare (SD) & 1.26 (0.47) & 1.60 (0.20) & 1.54 (0.10) & 1.52 (0.06) & 1.53 (0.09) \\
& Regret       & 0.80 & 0.46 & 0.53 & 0.54 & 0.53 \\
& Time (sec.)  & 2.47 & 4.33 & 8.01 & 25.48 & 82.97 \\
\addlinespace[0.3em]

\multirow{3}{*}{IPW}
& Welfare (SD) & 1.33 (0.20) & 1.41 (0.18) & 1.53 (0.19) & 1.67 (0.21) & 1.81 (0.19) \\
& Regret       & 0.74 & 0.65 & 0.54 & 0.40 & 0.25 \\
& Time (sec.)  & 1.91 & 3.35 & 7.27 & 24.60 & 81.23 \\
\addlinespace[0.3em]

\multirow{3}{*}{DR}
& Welfare (SD) & 1.51 (0.26) & 1.77 (0.24) & 1.94 (0.14) & 2.00 (0.07) & 2.04 (0.03) \\
& Regret       & 0.55 & 0.29 & 0.12 & 0.06 & 0.03 \\
& Time (sec.)  & 17.17 & 24.43 & 45.58 & 102.96 & 242.29 \\
\bottomrule
\end{tabular}
\begin{tablenotes}\footnotesize
\item Notes: Results are based on 500 Monte Carlo replications. Welfare (SD)
reports mean welfare, with Monte Carlo standard deviations in
parentheses. Regret is defined as
$R_\Pi(\hat\pi)=W(\pi_\Pi^\ast)-W(\hat\pi)$, where $\pi_\Pi^\ast$ is the
oracle welfare-maximizing DTR within the policy class $\Pi$. We
numerically approximate $\pi_\Pi^\ast$ by maximizing simulated welfare
over $\Pi$ using an independent sample of 50,000 observations from DGP2.
The welfare of both the resulting oracle policy and each estimated
policy is evaluated on a separate independent test sample of 50,000
observations from DGP2, which is also used to compute regret. Because
Q-learning without policy search is not restricted to $\Pi$, regret
relative to $\pi_\Pi^\ast$ is not reported for Q-learn. Time is the
average elapsed computation time per replication in seconds.
\end{tablenotes}
\end{threeparttable}
\end{table}

\begin{table}[!htbp]
\centering
\caption{Monte Carlo simulation results for DGP3}
\label{table:mc_dgp3}
\begin{threeparttable}
\small
\setlength{\tabcolsep}{4.5pt}
\begin{tabular}{llccccc}
\toprule
Method & Metric
& \multicolumn{5}{c}{Sample size $(n)$} \\
\cmidrule(lr){3-7}
& & 250 & 500 & 1000 & 2000 & 4000 \\
\midrule
\multirow{3}{*}{Q-learn}
& Welfare (SD) & 4.79 (1.94) & 6.55 (1.53) & 8.08 (1.07) & 9.14 (0.81) & 9.87 (0.60) \\
& Regret       & -- & -- & -- & -- & -- \\
& Time (sec.)  & 1.59 & 2.66 & 4.47 & 8.37 & 16.40 \\
\addlinespace[0.3em]

\multirow{3}{*}{Q-search}
& Welfare (SD) & 3.60 (3.12) & 5.08 (3.32) & 6.23 (2.96) & 7.49 (1.95) & 7.91 (1.08) \\
& Regret       & 5.75 & 4.26 & 3.11 & 1.85 & 1.44 \\
& Time (sec.)  & 1.15 & 1.75 & 3.47 & 6.81 & 12.94 \\
\addlinespace[0.3em]

\multirow{3}{*}{IPW}
& Welfare (SD) & 2.93 (3.69) & 3.51 (3.77) & 3.69 (3.55) & 4.72 (2.98) & 4.81 (2.32) \\
& Regret       & 6.41 & 5.83 & 5.65 & 4.62 & 4.53 \\
& Time (sec.)  & 0.81 & 1.20 & 2.01 & 3.83 & 8.01 \\
\addlinespace[0.3em]

\multirow{3}{*}{DR}
& Welfare (SD) & 5.37 (2.31) & 6.42 (1.85) & 7.13 (1.27) & 7.61 (0.94) & 8.03 (0.82) \\
& Regret       & 3.97 & 2.92 & 2.22 & 1.73 & 1.31 \\
& Time (sec.)  & 9.85 & 14.49 & 25.47 & 48.40 & 99.52 \\
\bottomrule
\end{tabular}
\begin{tablenotes}\footnotesize
\item Notes: Results are based on 500 Monte Carlo replications. Welfare (SD)
reports mean welfare, with Monte Carlo standard deviations in
parentheses. Regret is defined as
$R_\Pi(\hat\pi)=W(\pi_\Pi^\ast)-W(\hat\pi)$, where $\pi_\Pi^\ast$ is the
oracle welfare-maximizing DTR within the policy class $\Pi$. We
numerically approximate $\pi_\Pi^\ast$ by maximizing simulated welfare
over $\Pi$ using an independent sample of 50,000 observations from DGP3.
The welfare of both the resulting oracle policy and each estimated
policy is evaluated on a separate independent test sample of 50,000
observations from DGP3, which is also used to compute regret. Because
Q-learning without policy search is not restricted to $\Pi$, regret
relative to $\pi_\Pi^\ast$ is not reported for Q-learn. Time is the
average elapsed computation time per replication in seconds.
\end{tablenotes}
\end{threeparttable}
\end{table}

\begin{table}[!htbp]
\centering
\caption{Empirical results for optimal class-type allocation}
\label{table:empirical_results}
\smallskip
\begin{threeparttable}
\small
\setlength{\tabcolsep}{5.5pt}
\renewcommand{\arraystretch}{1.15}

\begin{tabular}{@{}ccccccc@{}}
\toprule
& \multicolumn{2}{c}{Welfare gain} 
& \multicolumn{4}{c}{Allocation share by sequence} \\
\cmidrule(lr){2-3} \cmidrule(l){4-7}
 \makecell[c]{Policy class\\ ($\Pi_2$)}
& \makecell[c]{vs.\\ all-aide}
& \makecell[c]{vs.\\ all-small}
& \multicolumn{1}{c}{(aide, aide)}
& \multicolumn{1}{c}{(small, aide)}
& \multicolumn{1}{c}{(aide, small)}
& \multicolumn{1}{c}{(small, small)} \\
\midrule

Depth-2
& \makecell[c]{11.74\\{\footnotesize (0.26 SD)}\\{\footnotesize [5.03, 18.45]}}
& \makecell[c]{3.19\\{\footnotesize (0.07 SD)}\\{\footnotesize [-3.40, 9.77]}}
& 21.6\% & 14.9\% & 34.4\% & 29.1\% \\

\addlinespace[0.35em]

Depth-3
& \makecell[c]{11.33\\{\footnotesize (0.25 SD)}\\{\footnotesize [4.87, 17.82]}}
& \makecell[c]{2.78\\{\footnotesize (0.06 SD)}\\{\footnotesize [-3.42, 9.00]}}
& 20.1\% & 21.7\% & 24.9\% & 33.3\% \\

\bottomrule
\end{tabular}
\smallskip
\begin{tablenotes}[flushleft]

\footnotesize
\item \textit{Notes:}
The grade-K policy class $\Pi_1$ is held fixed as the class of depth-1 trees, 
whereas the depth of the grade-1 policy class $\Pi_2$ is varied. 
The columns labeled ``vs. all-aide'' and ``vs. all-small'' report estimates of
$W(\pi^{\ast,\mathrm{opt}}) - \mathrm{E}[Y_2(\mathrm{aide},\mathrm{aide})]$
and
$W(\pi^{\ast,\mathrm{opt}}) - \mathrm{E}[Y_2(\mathrm{small},\mathrm{small})]$,
respectively. Welfare gains are measured in grade-1 mathematics test-score points.
Standardized gains are reported in parentheses, and 95\% confidence intervals are
reported in square brackets. Allocation shares are reported in percent.
\end{tablenotes}

\end{threeparttable}
\end{table}

\bigskip
\bigskip
\bigskip
\newpage
\newpage
\appendix
\section*{Figure}

\medskip

\begin{figure}[h]
\caption{Estimated DTR for class-type allocation in grades K and 1 under the baseline policy-class specification} \label{fig:estimated_decision_trees}
\vspace{.5cm}
\begin{minipage}[t]{0.45\hsize}
\begin{center}

{\small  (a) Policy for grade K} \\
\vspace{0.5cm}

\begin{tikzpicture}
[   scale = 1.1,
every node/.style={font=\small, align=center},
    level 1/.style = {sibling distance = 3cm},
    level 2/.style = {sibling distance = 3cm},
    edge from parent/.style={draw, -latex},
]
 
\node {Free/reduced-price lunch eligibility}
    child {node {Small} 
    edge from parent node [left] {True}} 
    child {node {Aide}
    edge from parent node [right] {False}
    };
 
\end{tikzpicture} 

\end{center}
\end{minipage}
\begin{minipage}[t]{0.45\hsize}
\begin{center}
{\small (b) Policy for grade 1} \\
\vspace{.5cm}

\begin{tikzpicture}
[   scale = 1.1,
every node/.style={font=\small, align=center},
    level 1/.style = {sibling distance = 4.5cm},
    level 2/.style = {sibling distance = 2cm},
    edge from parent/.style={draw, -latex},
]
 
\node {Math test score $\leq$ 494}
    child {node  {Grade-K class type = Aide} 
    child {node   {Small } edge from parent node [left] {True}}
    child {node {Aide } edge from parent node [right] { False}}
    edge from parent node [left] {True}} 
    child {node  {Math test score $\leq$ 528}
    child {node  {Small } edge from parent node [left] { True}}
    child {node {Aide } edge from parent node [right] { False}}
    edge from parent node [right] {False}
    };
 
\end{tikzpicture}
\end{center}
\end{minipage}

\vspace{.3cm}
\begin{tablenotes}
\footnotesize
\item[] Notes: This figure displays the estimated DTR from Section~\ref{sec:empirical_illustration} 
under the baseline specification. The grade-K policy class $\Pi_1$ is restricted to 
depth-1 trees, and the grade-1 policy class $\Pi_2$ is restricted to depth-2 trees. 
Panels (a) and (b) show the estimated class-type allocation rules for grade K and 
grade 1, respectively. ``Aide'' denotes assignment to a regular-size class with a 
full-time teacher aide, and ``Small'' denotes assignment to a small-size class without 
a teacher aide.
\end{tablenotes}
\end{figure}
\medskip

\appendix
\part*{Appendix}

\setcounter{equation}{0}
\renewcommand{\theequation}{A.\arabic{equation}}


\section{Proof of Theorem~\ref{thm:main_theorem_backward}}\label{app:main_proof}
\renewcommand{\theequation}{A.\arabic{equation}}
This appendix presents the proof of Theorem~\ref{thm:main_theorem_backward}, along with some auxiliary lemmas. Our goal is to derive an asymptotic upper bound on $R(\hat{\pi})$. This is, however, a non-trivial task because the components of the DTR, $\hat{\pi} = (\hat{\pi}_1, \dots, \hat{\pi}_T)$, are estimated sequentially rather than simultaneously. We therefore cannot directly apply the theoretical analysis of \cite{Athey_Wager_2020} and \cite{zhou2022offline}, who study doubly robust policy learning in single-stage settings. To address this, we present a novel analysis to derive an asymptotic upper bound on $R(\hat{\pi})$.

We begin by noting that Lemma~\ref{lem:optimality_backward_induction} allows us to analyze the welfare regret, $R(\hat{\pi}) = W(\pi^{\ast,opt}) - W(\hat{\pi})$, by evaluating the welfare difference $W(\pi^{\ast,B}) - W(\hat{\pi})$ between the DTR $\pi^{\ast,B}$ derived from the population backward optimization problem and the estimated DTR $\hat{\pi}$. This reformulation facilitates the analysis of $R(\hat{\pi})$.

Given the estimated DTR $\hat{\pi}$, we define $R_{t}^{\hat{\pi}_{t:T}}(\pi_t) \equiv V_{t}(\pi_t,\hat{\pi}_{(t+1):T}) - V_{t}(\hat{\pi}_{t:T})$ for any $\pi_t \in \Pi_{t}$ and $t=1,\ldots,T$. $R_{t}^{\hat{\pi}_{t:T}}(\pi_t)$ measures the deviation of policy $\pi_{t}$ from the sequence of estimated policies $\hat{\pi}_{t:T}$ at stage $t$ with respect to the value function. For $t=T$, we denote $R_{T}^{\hat{\pi}_{T:T}}(\pi_T)= V_T(\pi_T) - V_T(\hat{\pi}_T)$.

The following lemma provides a useful result for analyzing the regret $R(\hat{\pi})$, relating the regret of the entire DTR to the stage-specific regrets.

\medskip

\begin{lemma}\label{lem:helpful_lemma}
Under Assumptions~\ref{asm:sequential independence}, \ref{asm:bounded outcome}, \ref{asm:overlap}, and \ref{asm:first-best}, the regret of $\hat{\pi}$ is bounded from above as
\begin{align}
    R(\hat{\pi}) \leq \sum_{t=1}^{T} B_t(\boldsymbol{\eta}) R_{t}^{\hat{\pi}_{t:T}}(\pi_{t}^{\ast,B}). \label{eq:decomposition_bound}
\end{align}
\end{lemma}

\begin{proof}
See Appendix~\ref{app:proofs_of_main_lemmas} in the supplement.
\end{proof}

\medskip
The result (\ref{eq:decomposition_bound}) enables us to evaluate $R(\hat{\pi})$ by evaluating each stage-specific regret $R_{t}^{\hat{\pi}_{t:T}}(\pi_{t}^{\ast,B})$ ($t=1,\ldots,T$), which is simpler to analyze as shown below.

Given a fixed DTR $\pi=(\pi_1,\ldots,\pi_T)$, we define
\begin{align*}
\widetilde{V}_{i,T}(\pi_{T}) &\equiv \frac{ Y_{i,T} - Q_{T}(H_{i,T},A_{i,T}) }{e_{T}(H_{i,T},A_{i,T})}\cdot \mathbf{1}\{A_{i,T}=\pi_{T}(H_{i,T})\} + Q_{T}(H_{i,T},\pi_{T}),\\
\widehat{V}_{i,T}(\pi_T) &\equiv \frac{ Y_{i,T}- \widehat{Q}_{T}^{-k(i)}(H_{i,T},A_{i,T}) }{\hat{e}_{T}^{-k(i)}(H_{i,T},A_{i,T})}\cdot\mathbf{1}\{A_{i,T} = \pi_{T}(H_{i,T})\} 
+  \widehat{Q}_{T}^{-k(i)}\left(H_{i,T},\pi_{T}\right),
\end{align*}
and recursively for $t=T-1,\ldots,1$:
\begin{align}
\widetilde{V}_{i,t}(\pi_{t:T}) &\equiv \frac{ Y_{i,t} + \widetilde{V}_{i,t+1}(\pi_{(t+1):T}) - Q_{t}^{\pi_{(t+1):T}}(H_{i,t},A_{i,t}) }{e_{t}(H_{i,t},A_{i,t})}\cdot \mathbf{1}\{A_{i,t}=\pi_{t}(H_{i,t})\} \nonumber \\
&+ Q_{t}^{\pi_{(t+1):T}}(H_{i,t},\pi_{t}), \nonumber\\
\widehat{V}_{i,t}(\pi_{t:T}) &\equiv \frac{Y_{i,t} + \widehat{V}_{i,t+1}(\pi_{(t+1):T}) - \widehat{Q}_{t}^{\pi_{(t+1):T},-k(i)}(H_{i,t},A_{i,t}) }{\hat{e}_{t}^{-k(i)}(H_{i,t},A_{i,t})}\cdot \mathbf{1}\{A_{i,t}=\pi_{t}(H_{i,t})\} \nonumber\\
&+ \widehat{Q}_{t}^{\pi_{(t+1):T},-k(i)}(H_{i,t},\pi_{t}). \label{eq:definition_hat_V}
\end{align}
Note that the sample mean $(1/n)\sum_{i=1}^{n}\widetilde{V}_{i,t}\left(\pi_{t:T}\right)$ is an oracle estimator of the policy value function $V_{t}(\pi_{t:T})$ with oracle access to $\{Q_{s}^{\pi_{(s+1):T}}(\cdot,\cdot)\}_{s=t,\ldots,T}$ and $\{e_{s}(\cdot,\cdot)\}_{s=t,\ldots,T}$. Lemma~\ref{lemma:identification} in the supplement shows that $(1/n)\sum_{i=1}^{n}\widetilde{V}_{i,t}\left(\pi_{t:T}\right)$ is an unbiased estimator of the policy value $V_{t}(\pi_{t:T})$ under sequential ignorability (Assumption~\ref{asm:sequential independence}).

Following the analysis of \cite{zhou2022offline}, for each stage $t=1,\ldots,T$, we define the policy value difference function $\Delta_{t}(\cdot;\cdot):\Pi_{t:T} \times \Pi_{t:T} \rightarrow \Real$, the oracle influence difference function $\widetilde{\Delta}_{t}(\cdot;\cdot):\Pi_{t:T} \times \Pi_{t:T} \rightarrow \Real$, and the estimated policy value difference function $\widehat{\Delta}_{t}(\cdot;\cdot):\Pi_{t:T} \times \Pi_{t:T} \rightarrow \Real$, as follows. For $\pi_{t:T}^{a}=(\pi_{t}^{a},\ldots,\pi_{T}^{a}) \in \Pi_{t:T}$ and $\pi_{t:T}^{b}=(\pi_{t}^{b},\ldots,\pi_{T}^{b}) \in \Pi_{t:T}$,
\begin{align}
\Delta_{t}(\pi_{t:T}^{a};\pi_{t:T}^{b})&\equiv V_{t}(\pi_{t:T}^{a}) - V_{t}(\pi_{t:T}^{b}), \label{eq:Delta} \\
 \widetilde{\Delta}_{t}(\pi_{t:T}^{a};\pi_{t:T}^{b}) &\equiv \frac{1}{n}\sum_{i=1}^{n} \widetilde{V}_{i,t}\left(\pi_{t:T}^{a}\right) - \frac{1}{n}\sum_{i=1}^{n} \widetilde{V}_{i,t}\left(\pi_{t:T}^{b}\right), \label{eq:Delta_tilde}\\
 \widehat{\Delta}_{t}(\pi_{t:T}^{a};\pi_{t:T}^{b}) &\equiv  \frac{1}{n}\sum_{i=1}^{n} \widehat{V}_{i,t}\left(\pi_{t:T}^{a}\right) -  \frac{1}{n}\sum_{i=1}^{n} \widehat{V}_{i,t}\left(\pi_{t:T}^{b}\right).\nonumber
\end{align}
From the definitions, the stage-specific regret $R_{t}^{\hat{\pi}_{t:T}}(\pi_{t}^{\ast,B})$ is expressed as
\begin{align*}
    R_{t}^{\hat{\pi}_{t:T}}(\pi_{t}^{\ast,B}) = \Delta_{t}\left(\pi_{t}^{\ast,B},\hat{\pi}_{(t+1):T};\hat{\pi}_{t:T}\right).
\end{align*}

In what follows, we evaluate $R_{t}^{\hat{\pi}_{t:T}}(\pi_{t}^{\ast,B})$ for each $t$. 
A standard argument from statistical learning theory  \citep[e.g.,][]{Lugosi_2002} gives
\begin{align}
    R_{t}^{\hat{\pi}_{t:T}}(\pi_{t}^{\ast,B})  
    &=\Delta_{t}\left(\pi_{t}^{\ast,B},\hat{\pi}_{(t+1):T};\hat{\pi}_{t:T}\right) 
    \nonumber\\
    &\leq \Delta_{t}\left(\pi_{t}^{\ast,B},\hat{\pi}_{(t+1):T};\hat{\pi}_{t:T}\right) - \widehat{\Delta}_{t}\left(\pi_{t}^{\ast,B},\hat{\pi}_{(t+1):T};\hat{\pi}_{t:T}\right) \nonumber \\
    &\leq 
    \sup_{\pi_{(t+1):T} \in \Pi_{(t+1):T}} \sup_{\pi_{t}^{a},\pi_{t}^{b} \in \Pi_{t}} |\Delta_{t}(\pi_{t}^{a},\pi_{(t+1):T};\pi_{t}^{b},\pi_{(t+1):T}) - \widehat{\Delta}_{t}(\pi_{t}^{a},\pi_{(t+1):T};\pi_{t}^{b},\pi_{(t+1):T})| \nonumber \\
    & \leq  \sup_{\pi_{(t+1):T} \in \Pi_{(t+1):T}} \sup_{\pi_{t}^{a},\pi_{t}^{b} \in \Pi_{t}}|\Delta_{t}(\pi_{t}^{a},\pi_{(t+1):T};\pi_{t}^{b},\pi_{(t+1):T}) - \widetilde{\Delta}_{t}(\pi_{t}^{a},\pi_{(t+1):T};\pi_{t}^{b},\pi_{(t+1):T})| \nonumber \\
    &    + \sup_{\pi_{(t+1):T} \in \Pi_{(t+1):T}} \sup_{\pi_{t}^{a},\pi_{t}^{b} \in \Pi_{t}} |\widehat{\Delta}_{t}(\pi_{t}^{a},\pi_{(t+1):T};\pi_{t}^{b},\pi_{(t+1):T}) - \widetilde{\Delta}_{t}(\pi_{t}^{a},\pi_{(t+1):T};\pi_{t}^{b},\pi_{(t+1):T})|, \label{eq:standard_inequality}
\end{align}
where the first inequality follows because $\hat{\pi}_{t}$ maximizes  $(1/n)\sum_{i=1}^{n} \widehat{\Gamma}_{i,t}^{\hat{\pi}_{(t+1):T}}\left(\pi_{t}(H_{i,t})\right)$ over $\Pi_t$; hence, $\widehat{\Delta}_{t}\left(\pi_{t}^{\ast,B},\hat{\pi}_{(t+1):T};\hat{\pi}_{t:T}\right) \leq 0$. 

We can now evaluate $R_{t}^{\hat{\pi}_{t:T}}(\pi_{t}^{\ast,B})$ by evaluating
\begin{align}
  \sup_{\pi_{(t+1):T} \in \Pi_{(t+1):T}} \sup_{\pi_{t}^{a},\pi_{t}^{b} \in \Pi_{t}} |\Delta_{t}(\pi_{t}^{a},\pi_{(t+1):T};\pi_{t}^{b},\pi_{(t+1):T}) - \widetilde{\Delta}_{t}(\pi_{t}^{a},\pi_{(t+1):T};\pi_{t}^{b},\pi_{(t+1):T})|  
  \label{eq:former}
\end{align}
and 
\begin{align}
    \sup_{\pi_{(t+1):T} \in \Pi_{(t+1):T}} \sup_{\pi_{t}^{a},\pi_{t}^{b} \in \Pi_{t}} |\widehat{\Delta}_{t}(\pi_{t}^{a},\pi_{(t+1):T};\pi_{t}^{b},\pi_{(t+1):T}) - \widetilde{\Delta}_{t}(\pi_{t}^{a},\pi_{(t+1):T};\pi_{t}^{b},\pi_{(t+1):T})|.
    \label{eq:latter}
\end{align}
For the former term, we apply the uniform concentration result of \citeauthor{zhou2022offline} (\citeyear{zhou2022offline}, Lemma 2) for the oracle influence difference function to obtain the following lemma.

\medskip
\begin{lemma}\label{lem:bound_influence_difference_function}
Suppose that Assumptions~\ref{asm:sequential independence}, \ref{asm:bounded outcome}, \ref{asm:overlap}, and \ref{asm:bounded entropy} hold. Then, for any stage $t \in \{1,2,\ldots,T\}$ and $\delta \in (0,1)$, with probability at least $1-2\delta$, the following holds:
\begin{align}
    &\sup_{\pi_{(t+1):T} \in \Pi_{(t+1):T}} \sup_{\pi_{t}^{a},\pi_{t}^{b} \in \Pi_{t}} |\Delta_{t}(\pi_{t}^{a},\pi_{(t+1):T};\pi_{t}^{b},\pi_{(t+1):T}) - \widetilde{\Delta}_{t}(\pi_{t}^{a},\pi_{(t+1):T};\pi_{t}^{b},\pi_{(t+1):T})| \nonumber\\
    &\leq \left(54.4 \sqrt{2}\kappa(\Pi_{t:T}) + 435.2 + \sqrt{2 \log \frac{1}{\delta}}\right)\sqrt{\frac{M^2\mathcal V_{t:T}(\boldsymbol{\eta})}{n}}  + o\left(\frac{1}{\sqrt{n}}\right). \label{eq:bound_influence_difference_function}
\end{align}
\end{lemma}

\begin{proof}
See Appendix~\ref{app:preliminary_results} in the supplement.
\end{proof}

\medskip

To evaluate the latter (equation (\ref{eq:latter})), we consider extending the analytical approach of \cite{Athey_Wager_2020} and \cite{zhou2022offline}, which leverages orthogonal moments and cross-fitting, to the sequential multi-stage setting. A key challenge arises from the recursive structure of  $\widehat{V}_{i,t}(\pi_{t:T})$, which depends on $\widehat{V}_{i,t+1}(\pi_{(t+1):T})$ as defined in equation (\ref{eq:definition_hat_V}). This recursive dependency prevents a direct application of existing analytical approaches. To address this, we develop an extension that accommodates the recursive nature of the value function.

    
Specifically, we further decompose $\widehat{V}_{i,t}(\pi_{t:T})$ by introducing the following functionals:
\begin{align*}
\check{V}_{i,T}(\pi_T) &\equiv \frac{ Y_{i,T}- Q_{T}(H_{i,T},A_{i,T}) }{\hat{e}_{T}^{-k(i)}(H_{i,T},A_{i,T})}\cdot\mathbf{1}\{A_{i,T} = \pi_{T}(H_{i,T})\} 
+  Q_{T}\left(H_{i,T},\pi_{T}\right);  \\
\check{V}_{i,t}(\pi_{t:T}) &\equiv \frac{Y_{i,t} + \widetilde{V}_{i,t+1}(\pi_{(t+1):T}) - \widehat{Q}_{t}^{\pi_{(t+1):T},-k(i)}(H_{i,t},A_{i,t}) }{\hat{e}_{t}^{-k(i)}(H_{i,t},A_{i,t})}\cdot \mathbf{1}\{A_{i,t}=\pi_{t}(H_{i,t})\} \nonumber\\
&+ \widehat{Q}_{t}^{\pi_{(t+1):T},-k(i)}(H_{i,t},\pi_{t}),
\end{align*}
for $t=1,\ldots,T-1$. The functional $\check{V}_{i,t}(\pi_{t:T})$ is similar to $\widehat{V}_{i,t}(\pi_{t:T})$, but it replaces the estimated future value function $\widehat{V}_{i,t+1}(\pi_{(t+1):T})$ with the oracle counterpart $\widetilde{V}_{i,t+1}(\pi_{(t+1):T})$. As a result, $\check{V}_{i,t}(\pi_{t:T})$ does not have a recursive structure.

We can then decompose $\widehat{V}_{i,t}(\pi_{t:T})$ as follows:
\begin{align*}
    \widehat{V}_{i,t}(\pi_{t:T}) &= \check{V}_{i,t}(\pi_{t:T}) + \frac{1}{\hat{e}_{t}^{-k(i)}(H_{i,t},\pi_{t})}\left(\widehat{V}_{i,t+1}(\pi_{(t+1):T}) - \widetilde{V}_{i,t+1}(\pi_{(t+1):T})\right) \\
    &= \check{V}_{i,t}(\pi_{t:T}) + \sum_{s=t}^{T-1}\frac{1}{\prod_{\ell=t}^{s} \hat{e}_{\ell}^{-k(i)}(H_{i,\ell},\pi_{\ell})}\left(\check{V}_{i,s+1}(\pi_{(s+1):T}) - \widetilde{V}_{i,s+1}(\pi_{(s+1):T})\right).
\end{align*}
We hence have
\begin{align}
   \widehat{V}_{i,t}(\pi_{t:T}) - \widetilde{V}_{i,t}(\pi_{t:T}) = \sum_{s=t}^{T}\left(\check{V}_{i,t,s}^{\dagger}(\pi_{t:T}) - \widetilde{V}_{i,t,s}^{\dagger}(\pi_{t:T})\right) \label{eq:V_dif_decomposition}
\end{align}
with  
\begin{align*}
    \check{V}_{i,t,s}^{\dagger}(\pi_{t:T}) \equiv \frac{1}{\prod_{\ell=t}^{s-1} \hat{e}_{\ell}^{-k(i)}(H_{i,\ell},\pi_{\ell})}\check{V}_{i,s}(\pi_{s:T}) \mbox{\ \ and\ \ }
    \widetilde{V}_{i,t,s}^{\dagger}(\pi_{t:T}) &\equiv \frac{1}{\prod_{\ell=t}^{s-1} \hat{e}_{\ell}^{-k(i)}(H_{i,\ell},\pi_{\ell})} \widetilde{V}_{i,s}(\pi_{s:T}),
\end{align*}
where we define $\prod_{\ell=t}^{s-1} \hat{e}_{\ell}^{-k(i)}(H_{i,\ell},\pi_{\ell}) = 1$ for $s \leq t$.

Using the result (\ref{eq:V_dif_decomposition}), we can bound (\ref{eq:latter}) from above as
\begin{align}
    &\sup_{\pi_{(t+1):T} \in \Pi_{(t+1):T}} \sup_{\pi_{t}^{a},\pi_{t}^{b} \in \Pi_{t}} \left|\widehat{\Delta}_{t}(\pi_{t}^{a},\pi_{(t+1):T};\pi_{t}^{b},\pi_{(t+1):T}) - \widetilde{\Delta}_{t}(\pi_{t}^{a},\pi_{(t+1):T};\pi_{t}^{b},\pi_{(t+1):T})\right| \nonumber \\
    &\leq \sum_{s=t}^{T}\sup_{\pi_{(t+1):T} \in \Pi_{(t+1):T}} \sup_{\pi_{t}^{a},\pi_{t}^{b} \in \Pi_{t}} \left|\check{\Delta}_{t,s}^{\dagger}(\pi_{t}^{a},\pi_{(t+1):T};\pi_{t}^{b},\pi_{(t+1):T}) - \widetilde{\Delta}_{t,s}^{\dagger}(\pi_{t}^{a},\pi_{(t+1):T};\pi_{t}^{b},\pi_{(t+1):T})\right|, \label{eq:latter_2}
\end{align}
where 
\begin{align*}
  \check{\Delta}_{t,s}^{\dagger}(\pi_{t:T}^{a};\pi_{t:T}^{b}) &\equiv \frac{1}{n}\sum_{i=1}^{n} \check{V}_{i,t,s}^{\dagger}\left(\pi_{t:T}^{a}\right) - \frac{1}{n}\sum_{i=1}^{n} \check{V}_{i,t,s}^{\dagger}\left(\pi_{t:T}^{b}\right) \mbox{\ \ and\ }\\
  \widetilde{\Delta}_{t,s}^{\dagger}(\pi_{t:T}^{a};\pi_{t:T}^{b}) &\equiv \frac{1}{n}\sum_{i=1}^{n} \widetilde{V}_{i,t,s}^{\dagger}\left(\pi_{t:T}^{a}\right) - \frac{1}{n}\sum_{i=1}^{n} \widetilde{V}_{i,t,s}^{\dagger}\left(\pi_{t:T}^{b}\right).
\end{align*}

The bound in (\ref{eq:latter_2}) is easier to analyze than (\ref{eq:latter}) because $\check{V}_{i,t}(\pi_{t:T})$ does not have a recursive structure, though some complexity arises from the dependence of $\widetilde{V}_{i,t,s}^{\dagger}\left(\pi_{t:T}^{a}\right)$ on the estimated propensity scores $\left\{\hat{e}_{\ell}^{-k(i)}(\cdot,\cdot)\right\}_{\ell=t,\ldots,s-1}$. The extended analysis in Supplemental Appendix~\ref{app:proof_of_main_lemma_2} leads to the following lemma.

\medskip

\begin{lemma}\label{lem:asymptotic_estimated_policy_difference_function}
Suppose that Assumptions~\ref{asm:sequential independence}, \ref{asm:bounded outcome}, \ref{asm:overlap}, and \ref{asm:rate_of_convergence_backward} hold. Then, for any integers $s$ and $t$ such that $1\leq t \leq s \leq T$, the following holds:
\begin{align*}
    &\sup_{\pi_{(t+1):T} \in \Pi_{(t+1):T}}\sup_{\pi_{t}^{a},\pi_{t}^{b} \in \Pi_{t}} \left|\check{\Delta}_{t,s}^{\dagger}(\pi_{t}^{a},\pi_{(t+1):T};\pi_{t}^{b},\pi_{(t+1):T}) - \widetilde{\Delta}_{t,s}^{\dagger}(\pi_{t}^{a},\pi_{(t+1):T};\pi_{t}^{b},\pi_{(t+1):T})\right|\\
    &= O_{p}(n^{-\min\{1/2,\tau/2\}}).
\end{align*}
\end{lemma}

\begin{proof}
See Appendix~\ref{app:proof_of_main_lemma_2} in the supplement.
\end{proof}

\medskip

We now complete the proof of Theorem~\ref{thm:main_theorem_backward}
by combining the above results.

\begin{proof}[Proof of Theorem~\ref{thm:main_theorem_backward}]
Combining inequalities \eqref{eq:standard_inequality} and
\eqref{eq:latter_2} with Lemmas
\ref{lem:bound_influence_difference_function} and
\ref{lem:asymptotic_estimated_policy_difference_function}, we obtain, for
each $t=1,\ldots,T$,
\[
R_{t}^{\hat{\pi}_{t:T}}(\pi_{t}^{\ast,B})
=
O_p\left(
M\kappa(\Pi_{t:T})
\sqrt{
\frac{\mathcal V_{t:T}(\boldsymbol{\eta})}{n}
}
\right)
+
O_p\left(
(T-t+1)n^{-\min\{1/2,\tau/2\}}
\right).
\]
Substituting this bound into \eqref{eq:decomposition_bound} yields
\[
R(\hat{\pi})
=
O_p\left(
M\sum_{t=1}^{T}
B_t(\boldsymbol{\eta})
\kappa(\Pi_{t:T})
\sqrt{
\frac{\mathcal V_{t:T}(\boldsymbol{\eta})}{n}
}
\right)
+
O_p\left(
n^{-\min\{1/2,\tau/2\}}
\sum_{t=1}^{T}
B_t(\boldsymbol{\eta})(T-t+1)
\right),
\]
which establishes the first result.
For fixed $T$ and overlap constants bounded away from zero,
$B_t(\boldsymbol{\eta})$ and
$\mathcal V_{t:T}(\boldsymbol{\eta})$ are bounded, and the second claim
follows from $\kappa(\Pi_{t:T})\leq\kappa(\Pi)$.
\end{proof}

\smallskip
\subsection*{Acknowledgements}
I thank the editor, Bo Honor{\'e}, three anonymous referees, {\'A}ureo de Paula, Toru Kitagawa, Takahide Yanagi, and participants at various seminars and conferences for their helpful comments and suggestions. I gratefully acknowledge financial support from JSPS KAKENHI Grant (number 22K20155) and ERC Grant (number 715940).


\medskip

\setstretch{1.5}

\bibliographystyle{ecta}
\bibliography{ref_DTR,ref_surrogate_loss}

\part*{Supplementary Appendix}




\section{Preliminary Results and Proofs of Lemmas \ref{lem:optimality_backward_induction}, \ref{lem:entropy_integral_bound}, \ref{lem:helpful_lemma}, \ref{lem:bound_influence_difference_function}, and \ref{lem:asymptotic_estimated_policy_difference_function}}
\label{app:proofs_of_lemmas}

\setcounter{equation}{0}
\renewcommand{\theequation}{B.\arabic{equation}}

\subsection{Preliminary Results and Proofs of Lemmas \ref{lem:optimality_backward_induction}, \ref{lem:entropy_integral_bound}, and \ref{lem:bound_influence_difference_function}}\label{app:preliminary_results}

This section presents several preliminary results and proofs of Lemmas \ref{lem:optimality_backward_induction}, \ref{lem:entropy_integral_bound}, and \ref{lem:bound_influence_difference_function}.
We begin by providing the proof of Lemma~\ref{lem:entropy_integral_bound}.

Lemma~\ref{lem:covering_number_of_product} below relates the
$\epsilon$-Hamming covering number of a class of policy sequences to
those of the corresponding stage-specific policy classes and will be
used to prove Lemma~\ref{lem:entropy_integral_bound}.

\medskip

\begin{lemma}\label{lem:covering_number_of_product}
Given a class of DTRs $\Pi=\Pi_1 \times \cdots \times \Pi_T$, for any integers $s$ and $t$ such that $1\leq s \leq t \leq T$, the following inequality holds: $$N_{d_h}((t-s + 1)\epsilon,\Pi_{s:t}) \leq \prod_{\ell=s}^{t}N_{d_h}(\epsilon,\Pi_{\ell}).$$
\end{lemma}

\begin{proof}
Fix a set of history points $\{h_{t}^{(1)},\ldots,h_{t}^{(n)}\} \subseteq \MH_t$. For each $\ell=s,\ldots,t$, let $h_{\ell}^{(i)}$ denote the partial history contained in $h_t^{(i)}$ up to stage $\ell$. Let $K_\ell  \equiv  N_{d_h}(\epsilon,\Pi_\ell,\{h_{\ell}^{(1)},\ldots,h_{\ell}^{(n)}\})$. For each $\ell\in \{s,\ldots,t\}$, we denote by $\widetilde{\Pi}_{\ell} \equiv \left\{\pi_{\ell}^{(1)},\ldots,\pi_{\ell}^{(K_{\ell})}\right\}$ the set of policies such that for any $\pi_{\ell} \in \Pi_{\ell}$, there exists $\pi_{\ell}^{(i)} \in \widetilde{\Pi}_{\ell}$ satisfying $d_{h}(\pi_{\ell},\pi_{\ell}^{(i)}) \leq \epsilon$. The existence of such a set follows from the definition of $N_{d_h}(\epsilon,\Pi_\ell,\{h_{\ell}^{(1)},\ldots,h_{\ell}^{(n)}\})$.

Fix $\pi_{s:t} \in \Pi_{s:t}$, and define $\widetilde{\Pi}_{s:t} \equiv \widetilde{\Pi}_{s}\times \cdots \times \widetilde{\Pi}_{t}$. Let $\tilde{\pi}_{s:t}=(\tilde{\pi}_s,\ldots,\tilde{\pi}_t) \in\widetilde{\Pi}_{s:t}$ be such that for any $\ell \in \{s,\ldots,t\}$, $d_{h}(\pi_{\ell},\tilde{\pi}_{\ell}) \leq \epsilon$. Then
\begin{align*}
    d_h(\pi_{s:t},\tilde{\pi}_{s:t}) & = \frac{1}{n}\sum_{i=1}^{n}\mathbf{1}\{\pi_{s}(h_{s}^{(i)})\neq \tilde{\pi}_{s}(h_{s}^{(i)}) \vee \cdots \vee \pi_{t}(h_{t}^{(i)})\neq \tilde{\pi}_{t}(h_{t}^{(i)})\} \\
    & \leq \sum_{\ell=s}^{t} \left(\frac{1}{n}\sum_{i=1}^{n}\mathbf{1}\{\pi_{\ell}(h_{\ell}^{(i)})\neq \tilde{\pi}_{\ell}(h_{\ell}^{(i)})\}\right)\\
    &= \sum_{\ell=s}^{t} d_h(\pi_{\ell},\tilde{\pi}_{\ell}) \leq (t-s+1)\epsilon.
\end{align*}
Therefore, for any $\pi_{s:t} \in \Pi_{s:t}$, there exists $\tilde{\pi}_{s:t}\in \widetilde{\Pi}_{s:t}$ such that $d_h(\pi_{s:t},\tilde{\pi}_{s:t}) \leq (t-s+1)\epsilon$. Since $\left|\widetilde{\Pi}_{s:t}\right| = \prod_{\ell=s}^{t}\left|\widetilde{\Pi}_{\ell}\right|=\prod_{\ell=s}^{t}N_{d_h}(\epsilon,\Pi_\ell,\{h_{\ell}^{(1)},\ldots,h_{\ell}^{(n)}\})$, where $|\cdot|$ denotes the cardinality, we have
\begin{align*}
    N_{d_h}\left((t-s+1)\epsilon,\Pi_{s:t},\{h_{t}^{(1)},\ldots,h_{t}^{(n)}\}\right) \leq \prod_{\ell=s}^{t}N_{d_h}\left(\epsilon,\Pi_\ell,\{h_{\ell}^{(1)},\ldots,h_{\ell}^{(n)}\}\right).
\end{align*}
As this holds for any $n$ and any set of history points $\{h_{t}^{(1)},\ldots,h_{t}^{(n)}\}$, the result in the statement holds.
\end{proof}

\medskip

Using Lemma~\ref{lem:covering_number_of_product}, we present the proof of Lemma~\ref{lem:entropy_integral_bound} below.

\medskip

\begin{proof}[Proof of Lemma~\ref{lem:entropy_integral_bound}]
Fix $s$ and $t$ such that $s \leq t$. 
Let $m:=t-s+1$. Applying Lemma~\ref{lem:covering_number_of_product} to $\Pi_{s:t}$, we have
$
N_{d_h}
\left(
m\delta,\Pi_{s:t}
\right)
\leq
\prod_{\ell=s}^t
N_{d_h}(\delta,\Pi_\ell)
$
for every $\delta>0$. Setting $\delta=\epsilon^2/m$ gives
\[
N_{d_h}
\left(
\epsilon^2,\Pi_{s:t}
\right)
\leq
\prod_{\ell=s}^t
N_{d_h}
\left(
\frac{\epsilon^2}{m},\Pi_\ell
\right).
\]
By Assumption~\ref{asm:bounded entropy},
\begin{align*}
\log N_{d_h}
\left(
\epsilon^2,\Pi_{s:t}
\right)
&\leq
\sum_{\ell=s}^t
\left\{
\log C
+
D
\left(
\frac{m}{\epsilon^2}
\right)^\omega
\right\} \\
&=
m\log C
+
D m^{1+\omega}\epsilon^{-2\omega}.
\end{align*}
Therefore,
\begin{align*}
\kappa(\Pi_{s:t})
&=
\int_0^1
\sqrt{
\log N_{d_h}(\epsilon^2,\Pi_{s:t})
}
\,d\epsilon\\
&\leq
\int_0^1
\sqrt{
m\log C
+
D m^{1+\omega}\epsilon^{-2\omega}
}
\,d\epsilon\\
&\leq
\sqrt{m\log C}
+
\sqrt{D}\,
m^{(1+\omega)/2}
\int_0^1\epsilon^{-\omega}\,d\epsilon\\
&=
\sqrt{m\log C}
+
\frac{\sqrt{D}}{1-\omega}
m^{(1+\omega)/2},
\end{align*}
where the last integral is finite since $\omega<1$.
Taking $s=1$ and $t=T$ yields the second result.
\end{proof}

\medskip

Next, we give several preliminary results. For any $h_t=(\underline a_{t-1},\underline s_t)\in\mathcal H_t$, define the conditional policy value of $\pi_{t:T}$ at stage $t$ by
\begin{align*}
    V_{t}\left(\pi_{t:T};h_{t}\right) \equiv \E\left[\sum_{s=t}^{T}\sum_{\underline{a}_{t:s}^{\prime} \in \underline{\MA}_{t:s}}\left( Y_{s}(\underline{a}_{t-1},\underline{a}_{t:s}^{\prime})\cdot \prod_{\ell=t}^{s}\mathbf{1}\{\pi_{\ell}\left(H_{\ell}(\underline{a}_{t-1},\underline{a}_{t:(\ell-1)}^{\prime})\right)=a_{\ell}^{\prime}\}\right)\middle| H_{t}=h_{t}\right],
\end{align*}
where $H_{\ell}(\underline{a}_{t-1},\underline{a}_{t:(\ell-1)}^{\prime})=H_{t}(\underline{a}_{t-1})$ for $\ell=t$.
Note that $\E\left[V_t(\pi_{t:T};H_{t})\right] = V_{t}(\pi_{t:T})$. For ease of notation, for any stages $s$ and $t$ such that $s \geq t$ and any policy sequence $\pi_{t:s}\in \Pi_{t:s}$, we define
\begin{align*}
    \widetilde{Y}_{t:s}\left(\underline{a}_{t-1},\pi_{t:s}\right) \equiv \sum_{\underline{a}_{t:s}^{\prime} \in \underline{\MA}_{t:s}}\left( Y_{s}(\underline{a}_{t-1},\underline{a}_{t:s}^{\prime})\cdot \prod_{\ell=t}^{s}\mathbf{1}\{\pi_{\ell}\left(H_{\ell}(\underline{a}_{t-1},\underline{a}_{t:(\ell-1)}^{\prime})\right)=a_{\ell}^{\prime}\}\right).
\end{align*}
Let $\widetilde{Y}_{t}\left(\underline{a}_{t-1},\pi_{t}\right)$ denote $\widetilde{Y}_{t:s}\left(\underline{a}_{t-1},\pi_{t:s}\right)$ when $t=s$. Note that, for any $\pi_{t:T} \in \Pi_{t:T}$ and $h_{t} = (\underline{a}_{t-1},\underline{s}_{t}) \in \MH_{t}$, $V_{t}(\pi_{t:T};h_{t})$ can be written as  $$\E\left[\sum_{s=t}^{T}\widetilde{Y}_{t:s}\left(\underline{a}_{t-1},\pi_{t:s}\right) \middle| H_{t}=h_{t}\right] = V_{t}(\pi_{t:T};h_{t}).$$

The following lemma will be used in the proofs of Lemmas \ref{lem:optimality_backward_induction}, \ref{lem:bound_influence_difference_function}, and \ref{lem:asymptotic_estimated_policy_difference_function}.

\medskip{}

\begin{lemma}\label{lemma:identification}
Suppose that Assumption \ref{asm:sequential independence} holds. Then, for any stage $t$ and DTR $\pi \in \Pi$, the following hold:
\begin{itemize}
\item[(i)] $Q_{t}^{\pi_{(t+1):T}}(h_{t},\pi_{t}) = V_{t}(\pi_{t:T};h_{t})$ for any $h_{t} \in \MH_{t}$;
    \item[(ii)] $\E\left[\widetilde{V}_{i,t}(\pi_{t:T})\right] = V_{t}(\pi_{t:T})$ for any $i=1,\ldots,n$. 
\end{itemize}
\end{lemma}
\medskip

\begin{proof}
    We first prove (i), following essentially the proof of
\citeauthor{Tsiatis_et_al_2019}
(\citeyear{Tsiatis_et_al_2019}, equation (6.53)). For $t=T$, the following holds for any $h_{T} = (\underline{a}_{T-1},\underline{s}_{T}) \in \MH_{T}$:   
    \begin{align}
        Q_{T}(h_{T},\pi_{T}) 
        &= \E\left[Y_{T}|A_{T}=\pi_{T}(h_{T}),H_{T}=h_T\right] \nonumber \\
        &= \E\left[Y_{T}(\underline{A}_{T})|A_{T}=\pi_{T}(h_{T}),H_{T}=h_{T}\right] \nonumber \\
        &= \E\left[Y_{T}(\underline{a}_{T-1},\pi_{T}(h_{T}))|A_{T}=\pi_{T}(h_{T}),H_{T}=h_{T}\right] \nonumber \\
        &= \sum_{a_{T} \in \MA_{T}}\E\left[Y_{T}(\underline{a}_{T-1},a_{T}) \cdot \mathbf{1}\{a_{T} = \pi_{T}(h_{T})\}\middle|A_{T}=a_{T},H_{T}=h_{T}\right] \nonumber \\
         &= \sum_{a_{T} \in \MA_{T}} \E\left[Y_{T}(\underline{a}_{T-1},a_{T}) \cdot \mathbf{1}\{a_{T} = \pi_{T}(h_{T})\}\middle|H_{T}=h_{T}\right] \nonumber \\
        &= \E\left[\sum_{a_{T} \in \MA_{T}}Y_{T}(\underline{a}_{T-1},a_{T}) \cdot \mathbf{1}\{a_{T}
        = \pi_{T}(H_{T}(\underline{a}_{T-1}))\}\middle| H_{T}=h_{T}\right]  \nonumber \\
        &= V_{T}(\pi_{T};h_{T}), \label{eq:Q_T_V_T}
    \end{align}
    where the fifth equality follows from Assumption \ref{asm:sequential independence}. 
    
For any integer $t$ such that $t <T$, the following holds for any $h_{t} = (\underline{a}_{t-1},\underline{s}_{t}) \in \MH_{t}$:
\begin{align}
   & \E\left[Y_{t} + V_{t+1}(\pi_{(t+1):T};H_{t+1}) | A_{t}=\pi_{t}(h_{t}),H_{t}=h_{t}\right] \nonumber\\
      &=\E\left[Y_{t}\left(\underline{a}_{t}\right) + \sum_{s=t+1}^{T}\widetilde{Y}_{(t+1):s}(\underline{a}_{t-1},\pi_{t}(h_{t}),\pi_{(t+1):s})\middle| A_{t}=\pi_{t}(h_{t}),H_{t}=h_{t}\right] \nonumber \\
            &=\sum_{a_{t} \in \MA_{t}}\left(\E\left[Y_{t}\left(\underline{a}_{t-1},a_{t}\right) + \sum_{s=t+1}^{T}\widetilde{Y}_{(t+1):s}(\underline{a}_{t-1},a_{t},\pi_{(t+1):s})\middle| A_{t}=a_{t},H_{t}=h_{t}\right] \cdot \mathbf{1}\{\pi_{t}(h_{t}) = a_{t}\}\right) \nonumber \\
             &=\sum_{a_{t} \in \MA_{t}}\left(\E\left[Y_{t}\left(\underline{a}_{t-1},a_{t}\right) + \sum_{s=t+1}^{T}\widetilde{Y}_{(t+1):s}(\underline{a}_{t-1},a_{t},\pi_{(t+1):s})\middle| H_{t} = h_{t}\right] \cdot \mathbf{1}\{\pi_{t}(h_{t}) = a_{t}\}\right) \nonumber \\
            &=\E\left[Y_{t}\left(\underline{a}_{t-1},\pi_{t}(h_{t})\right) + \sum_{s=t+1}^{T}\widetilde{Y}_{(t+1):s}(\underline{a}_{t-1},\pi_{t}(h_{t}),\pi_{(t+1):s})\middle| H_{t}=h_{t}\right] \nonumber \\
        &=\E\left[Y_{t}\left(\underline{a}_{t-1},\pi_{t}(h_{t})\right) + \sum_{s=t+1}^{T}\widetilde{Y}_{(t+1):s}(\underline{A}_{t-1},\pi_{t}(h_{t}),\pi_{(t+1):s})\middle| H_{t}=h_{t}\right] \nonumber \\
   & =\E\left[\sum_{s=t}^{T}\widetilde{Y}_{t:s}(\underline{a}_{t-1},\pi_{t:s})\middle|H_{t} = h_{t}\right] \nonumber \\
   &=V_{t}(\pi_{t:T};h_{t}), \label{eq:V_t+1_V_t}
\end{align}
where the first equality follows from the law of total expectations, and the third equality follows from Assumption \ref{asm:sequential independence}. 

When $t=T-1$,
\begin{align*}
    Q_{T-1}^{\pi_{T}}(h_{T-1},\pi_{T-1})
    &= \E\left[Y_{T-1} + Q_{T}(H_{T},\pi_{T})|A_{T-1}=\pi_{T-1}(h_{T-1}),H_{T-1}=h_{T-1}\right] \\
    &= \E\left[Y_{T-1} + V_{T}(\pi_{T};H_{T})|A_{T-1}=\pi_{T-1}(h_{T-1}),H_{T-1}=h_{T-1}\right]\\
    &= V_{T-1}(\pi_{(T-1):T};h_{T-1}),
\end{align*}
where the second and third equalities follow from equations (\ref{eq:Q_T_V_T}) and (\ref{eq:V_t+1_V_t}), respectively. Recursively applying the same argument from $t=T-2$ to $1$, we have $Q_{t}^{\pi_{(t+1):T}}(h_{t},\pi_{t}) = V_{t}(\pi_{t:T};h_{t})$ for any $t$ and $h_{t} \in \MH_{t}$,  which leads to the result (i). 

We proceed to prove (ii). Given a fixed DTR $\pi=(\pi_1,\ldots,\pi_T)$, to simplify notation, we define the independent copies of $\{\widetilde{V}_{i,t}(\pi_{t:T}):t=1,\ldots,T\}$ as follows:
\begin{align*}
\widetilde{V}_{T}(\pi_{T}) &\equiv \frac{ Y_{T} - Q_{T}(H_{T},A_{T}) }{e_{T}(H_{T},A_{T})}\cdot \mathbf{1}\{A_{T}=\pi_{T}(H_{T})\} + Q_{T}(H_{T},\pi_{T}(H_{T})),
\end{align*}
and, recursively for $t=T-1,\ldots,1$,
\begin{align*}
\widetilde{V}_{t}(\pi_{t:T}) &\equiv \frac{ Y_{t} + \widetilde{V}_{t+1}(\pi_{(t+1):T}) - Q_{t}^{\pi_{(t+1):T}}(H_{t},A_{t}) }{e_{t}(H_{t},A_{t})}\cdot \mathbf{1}\{A_{t}=\pi_{t}(H_{t})\} \\
&+ Q_{t}^{\pi_{(t+1):T}}(H_{t},\pi_{t}(H_{t})).
\end{align*}
Note that $\E\left[\widetilde{V}_{T}(\pi_{T})\right]=\E\left[\widetilde{V}_{i,T}(\pi_{T})\right]$ and $\E\left[\widetilde{V}_{t}(\pi_{t:T})\right]=\E\left[\widetilde{V}_{i,t}(\pi_{t:T})\right]$ for any $i$ and $t$.

We first consider the case that $t=T$. For the first term in $\widetilde{V}_{T}(\pi_{T})$, for any $h_{T} \in \MH_{T}$,
\begin{align*}
    &\E\left[Y_{T} \cdot \frac{\mathbf{1}\{A_{T}=\pi_{T}(H_{T})\}}{e_{T}(H_{T},A_{T})} \middle| H_{T}=h_{T}\right]\\
        &= \E\left[\E\left[Y_{T}\middle|A_{T}=\pi_{T}(H_{T}), H_{T} \right] \cdot \frac{\mathbf{1}\{A_{T}=\pi_{T}(H_{T})\}}{e_{T}(H_{T},A_{T})} \middle| H_{T}=h_{T}\right]\\
    &= \E\left[Q_{T}(H_{T},\pi_{T}) \cdot \frac{\mathbf{1}\{A_{T}=\pi_{T}(H_{T})\}}{e_{T}(H_{T},A_{T})} \middle| H_{T}=h_{T}\right],
\end{align*}
where the last equality follows from the definition of $Q_{T}(H_{T},\pi_{T})$.

Therefore, we have
\begin{align}
    &\E\left[\widetilde{V}_{T}(\pi_{T}) \middle| H_{T} = h_{T}\right] \nonumber\\
    &= \E\left[\left(Q_{T}(H_{T},A_{T}) - Q_{T}(H_{T},A_{T})\right) \cdot \frac{\mathbf{1}\{A_{T}=\pi_{T}(H_{T})\}}{e_{T}(H_{T},A_{T})} \middle| H_{T}=h_{T}\right] + Q_{T}(h_{T},\pi_{T}) \nonumber \\
    &= Q_{T}(h_{T},\pi_{T})  = V_{T}(\pi_{T};h_{T}), \label{eq:V_tile_T_to_V_T}
\end{align}
where the last equality follows from the result (i).
We consequently have $\E\left[\widetilde{V}_{T}(\pi_{T})\right] = \E[V_{T}(\pi_{T};H_{T})] = V_{T}(\pi_{T})$.

When $t=T-1$, for any $h_{T-1} \in \MH_{T-1}$,
\begin{align*}
    &\E\left[\left(Y_{T-1} + \widetilde{V}_{T}(\pi_{T})\right) \cdot \frac{\mathbf{1}\{A_{T-1}=\pi_{T-1}(H_{T-1})\}}{e_{T-1}(H_{T-1},A_{T-1})} \middle| H_{T-1}=h_{T-1}\right]\\
    &=\E\left[\left(Y_{T-1} + \E\left[\widetilde{V}_{T}(\pi_{T})\middle| H_{T}\right]\right) \cdot \frac{\mathbf{1}\{A_{T-1}=\pi_{T-1}(H_{T-1})\}}{e_{T-1}(H_{T-1},A_{T-1})} \middle| H_{T-1}=h_{T-1}\right]\\
    &= \E\left[\left(Y_{T-1} + V_{T}(\pi_{T};H_{T})\right) \cdot \frac{\mathbf{1}\{A_{T-1}=\pi_{T-1}(H_{T-1})\}}{e_{T-1}(H_{T-1},A_{T-1})} \middle| H_{T-1}=h_{T-1}\right]\\
    &= \E\left[\left(Y_{T-1} + Q_{T}(H_{T},\pi_{T})\right) \cdot \frac{\mathbf{1}\{A_{T-1}=\pi_{T-1}(H_{T-1})\}}{e_{T-1}(H_{T-1},A_{T-1})} \middle| H_{T-1}=h_{T-1}\right]\\
     &= \E\left[\E\left[Y_{T-1} + Q_{T}(H_{T},\pi_{T})\middle| A_{T-1}=\pi_{T-1}(H_{T-1}),H_{T-1}\right]\right.\\
     &\left. \times \frac{\mathbf{1}\{A_{T-1}=\pi_{T-1}(H_{T-1})\}}{e_{T-1}(H_{T-1},A_{T-1})} \middle| H_{T-1}=h_{T-1}\right]\\
     &= \E\left[Q_{T-1}^{\pi_{T}}(H_{T-1},A_{T-1}) \cdot \frac{\mathbf{1}\{A_{T-1}=\pi_{T-1}(H_{T-1})\}}{e_{T-1}(H_{T-1},A_{T-1})} \middle| H_{T-1}=h_{T-1}\right],
\end{align*}
where the second equality follows from the result (\ref{eq:V_tile_T_to_V_T}); the third equality follows from the result (i); the last equality follows from the definition of the Q-function $Q_{T-1}^{\pi_T}(\cdot,\cdot)$.

Therefore, for any $h_{T-1} \in \MH_{T-1}$,
\begin{align*}
    &\E\left[\widetilde{V}_{T-1}(\pi_{(T-1):T})\middle| H_{T-1} = h_{T-1}\right]\\
    & = \E\left[\left(Y_{T-1} + \widetilde{V}_{T}(\pi_{T}) - Q_{T-1}^{\pi_{T}}(H_{T-1},A_{T-1})\right) \cdot \frac{\mathbf{1}\{A_{T-1}=\pi_{T-1}(H_{T-1})\}}{e_{T-1}(H_{T-1},A_{T-1})} \middle| H_{T-1}=h_{T-1}\right] \\
    &+ Q_{T-1}^{\pi_{T}}(h_{T-1},\pi_{T-1})\\
    & = \E\left[\left(Q_{T-1}^{\pi_{T}}(H_{T-1},A_{T-1}) - Q_{T-1}^{\pi_{T}}(H_{T-1},A_{T-1})\right) \cdot \frac{\mathbf{1}\{A_{T-1}=\pi_{T-1}(H_{T-1})\}}{e_{T-1}(H_{T-1},A_{T-1})} \middle| H_{T-1}=h_{T-1}\right] \\
    &+ Q_{T-1}^{\pi_{T}}(h_{T-1},\pi_{T-1})\\
& =  Q_{T-1}^{\pi_{T}}(h_{T-1},\pi_{T-1}) = V_{T-1}(\pi_{(T-1):T};h_{T-1}),
\end{align*}
where the last equality follows from the result (i). Hence, we have $\E\left[\widetilde{V}_{T-1}(\pi_{(T-1):T})\right] = \E\left[V_{T-1}(\pi_{(T-1):T};H_{T-1})\right] = V_{T-1}(\pi_{(T-1):T})$.

Recursively applying the same argument from $t=T-2$ to $1$, we have $\E\left[\widetilde{V}_{t}(\pi_{t:T})\right] = V_{t}(\pi_{t:T})$ for any $t$, which proves part (ii).
\end{proof}

\medskip

We next provide the proof of Lemma~\ref{lem:optimality_backward_induction}, which extends the proof of \citeauthor{Tsiatis_et_al_2019} (\citeyear{Tsiatis_et_al_2019}, equation (7.21)) to the case under Assumption \ref{asm:first-best}.

\medskip

\begin{proof}[Proof of Lemma~\ref{lem:optimality_backward_induction}]
Let $\pi \in \Pi$ be fixed. Under Assumptions~\ref{asm:overlap} and \ref{asm:first-best}, $\pi^{\ast,B}$ satisfies the condition in Assumption \ref{asm:first-best}.
For any $t=1,\ldots,T$,  comparing the welfares of $(\pi_{1:(t-1)},\pi_{t:T}^{\ast,B})$ and $(\pi_{1:t},\pi_{(t+1):T}^{\ast,B})$ yields:
\begin{align*}
    &W(\pi_{1:(t-1)},\pi_{t:T}^{\ast,B}) - W(\pi_{1:t},\pi_{(t+1):T}^{\ast,B}) \\
    &= V_{1}(\pi_{1:(t-1)},\pi_{t:T}^{\ast,B}) - V_{1}(\pi_{1:t},\pi_{(t+1):T}^{\ast,B}) \\
    &= \E\left[Q_{1}^{\pi_{2:(t-1)},\pi_{t:T}^{\ast,B}}(H_{1},\pi_{1}) - Q_{1}^{\pi_{2:t},\pi_{(t+1):T}^{\ast,B}}(H_{1},\pi_{1})\right] \\
    &=\E\left[\E\left[Q_{t}^{\pi_{(t+1):T}^{\ast,B}}(H_{t},\pi_{t}^{\ast,B}) - Q_{t}^{\pi_{(t+1):T}^{\ast,B}}(H_{t},\pi_{t})\middle| A_{1}=\pi_{1}(H_1),\ldots, A_{t-1}=\pi_{t-1}(H_{t-1}), H_{t-1} \right] \right] \\
    &\geq 0,
\end{align*}
where we denote $W(\pi_{1:(0)},\pi_{1:T}^{\ast,B})=W(\pi^{\ast,B})$ and $W(\pi_{1:T},\pi_{(T+1):T}^{\ast,B})=W(\pi)$ for $t=1$ and $t=T$, respectively; the second equality follows from Lemma~\ref{lemma:identification} (i); the third equality follows from the definitions of the Q-functions; the inequality follows from Assumption \ref{asm:first-best}.   We therefore obtain $W(\pi_{1:(t-1)},\pi_{t:T}^{\ast,B}) \geq W(\pi_{1:t},\pi_{(t+1):T}^{\ast,B})$ for any $t=1,\ldots,T$.

Applying this result, we have 
\begin{align*}
    W(\pi^{\ast,B}) - W(\pi) = \sum_{t=1}^{T}\left(W(\pi_{1:(t-1)},\pi_{t:T}^{\ast,B}) - W(\pi_{1:t},\pi_{(t+1):T}^{\ast,B})\right)
     \geq 0.
\end{align*} Since this result holds for any $\pi \in \Pi$, we obtain the result in Lemma~\ref{lem:optimality_backward_induction}. 
\end{proof}
\medskip

The following lemma, which follows from Lemma 2 of
\cite{zhou2022offline} and its proof, plays an important role in the
proof of Theorem~\ref{thm:main_theorem_backward}.
\medskip{}

\begin{lemma} \label{lem:concentration inequality_influence difference function}
Fix integers $s$ and $t$ such that $1\leq s \leq t \leq T$. For any $\underline{a}_{s:t}\in \underline{\MA}_{s:t}$, let $\{\Gamma_{i}^{\dag}(\underline{a}_{s:t})\}_{i=1}^{n}$ be i.i.d. random variables with bounded support. 
For any $\pi_{s:t} \in \Pi_{s:t}$, we define $\widetilde{Q}(\pi_{s:t}) \equiv (1/n) \sum_{i=1}^{n}  \Gamma_{i}^{\dag}(\pi_{s:t})$, where $\Gamma_{i}^{\dag}(\pi_{s:t}) \equiv \Gamma_{i}^{\dag}((\pi_{s}(H_{i,s}),\ldots,\pi_{t}(H_{i,t})))$ and $Q(\pi_{s:t}) \equiv  \E[\widetilde{Q}(\pi_{s:t})]$. For any $\pi_{s:t}^{a},\pi_{s:t}^{b} \in \Pi_{s:t}$, let $\widetilde{\Delta}(\pi_{s:t}^{a},\pi_{s:t}^{b}) \equiv \widetilde{Q}(\pi_{s:t}^{a}) - \widetilde{Q}(\pi_{s:t}^{b})$ and $\Delta(\pi_{s:t}^{a},\pi_{s:t}^{b}) \equiv Q(\pi_{s:t}^{a}) - Q(\pi_{s:t}^{b})$. Then, when $\kappa(\Pi_{s:t}) < \infty$, the following holds: For any $\delta \in (0,1)$, with probability at least $1-2\delta$,
\begin{align*}
    \sup_{\pi_{s:t}^{a},\pi_{s:t}^{b} \in \Pi_{s:t}} \left|\widetilde{\Delta}(\pi_{s:t}^{a},\pi_{s:t}^{b})-\Delta(\pi_{s:t}^{a},\pi_{s:t}^{b})\right| &\leq \left(54.4 \sqrt{2}\kappa(\Pi_{s:t}) + 435.2 + \sqrt{2 \log \frac{1}{\delta}}\right)\sqrt{\frac{V_{s:t}^{\ast}}{n}} \\
    &+ o\left(\frac{1}{\sqrt{n}}\right),
\end{align*}
where 
    $V_{s:t}^{\ast}  \equiv  \sup_{\pi_{s:t}^{a},\pi_{s:t}^{b} \in \Pi_{s:t}}\E
    \left[\left(\Gamma_{i}^{\dag}(\pi_{s:t}^{a}) - \Gamma_{i}^{\dag}(\pi_{s:t}^{b}) \right)^2\right] < \infty$.
\end{lemma}

\medskip

Using Lemma~\ref{lem:concentration inequality_influence difference function}, we present the proof of Lemma~\ref{lem:bound_influence_difference_function} as follows.

\medskip

\begin{proof}[Proof of  Lemma~\ref{lem:bound_influence_difference_function}]
Given a fixed DTR $\pi=(\pi_1,\ldots,\pi_T)$, we define the recursive expression for $t=T,\ldots,1$ as
\begin{align*}
\widetilde{\Gamma}_{i,t}^{\pi_{(t+1):T}}(\underline{a}_{t:T}) &\equiv \frac{Y_{i,t} + \widetilde{\Gamma}_{i,t+1}^{\pi_{(t+2):T}}(\underline{a}_{(t+1):T}) - Q_{t}^{\pi_{(t+1):T}}(H_{i,t},A_{i,t}) }{e_{t}(H_{i,t},A_{i,t})}\cdot \mathbf{1}\{A_{i,t}=a_{t}\} \\
&+ Q_{t}^{\pi_{(t+1):T}}(H_{i,t},a_{t}),
\end{align*}
where we set $\widetilde{\Gamma}_{i,T+1}=0$ by convention.
For any ${\pi}_{(t+1):T} \in \Pi_{(t+1):T}$ and ${\pi}_{t:T}^{\prime} = (\pi_{t}^{\prime},\ldots,\pi_{T}^{\prime}) \in \Pi_{t:T}$, let $\widetilde{\Gamma}_{i,t}^{{\pi}_{(t+1):T}}(\pi_{t:T}^{\prime}) \equiv \widetilde{\Gamma}_{i,t}^{{\pi}_{(t+1):T}}(\pi_{t}^{\prime}(H_{i,t}),\ldots,\pi_{T}^{\prime}(H_{i,T}))$.
Note that $\widetilde{\Gamma}_{i,t}^{{\pi}_{(t+1):T}}(\pi_{t:T}) = \widetilde{V}_{i,t}(\pi_{t:T})$. 

Fix $\pi_{(t+1):T} \in \Pi_{(t+1):T}$. For any $\pi_{t:T}^{a}, \pi_{t:T}^{b} \in \Pi_{t:T}$, we define 
\begin{align*}
    \widetilde{\Delta}^{\pi_{(t+1):T}} (\pi_{t:T}^{a},\pi_{t:T}^{b}) \equiv \frac{1}{n}\sum_{i=1}^{n}\left(\widetilde{\Gamma}_{i,t}^{\pi_{(t+1):T}}(\pi_{t:T}^{a}) - \widetilde{\Gamma}_{i,t}^{\pi_{(t+1):T}}(\pi_{t:T}^{b})\right)
\end{align*}
and
\begin{align*}
    \Delta^{\pi_{(t+1):T}} (\pi_{t:T}^{a},\pi_{t:T}^{b}) \equiv \E\left[\frac{1}{n}\sum_{i=1}^{n}\left(\widetilde{\Gamma}_{i,t}^{\pi_{(t+1):T}}(\pi_{t:T}^{a}) - \widetilde{\Gamma}_{i,t}^{\pi_{(t+1):T}}(\pi_{t:T}^{b})\right)\right].
\end{align*}
Note that 
\begin{align*}
\widetilde{\Delta}^{\pi_{(t+1):T}}(\pi_{t}^{a},\pi_{(t+1):T};\pi_{t}^{b},\pi_{(t+1):T})
    = \widetilde{\Delta}_{t}(\pi_{t}^{a},\pi_{(t+1):T};\pi_{t}^{b},\pi_{(t+1):T}),
\end{align*}
where $\widetilde{\Delta}_{t}(\cdot;\cdot)$ is defined in (\ref{eq:Delta_tilde}). Noting that 
\begin{align*}
    \Delta^{\pi_{(t+1):T}}(\pi_{t}^{a},\pi_{(t+1):T};\pi_{t}^{b},\pi_{(t+1):T}) = \E\left[\widetilde{V}_{i,t}(\pi_{t}^{a},\pi_{(t+1):T})\right] - \E\left[\widetilde{V}_{i,t}(\pi_{t}^{b},\pi_{(t+1):T})\right],
\end{align*}
Lemma~\ref{lemma:identification} leads to
\begin{align*}
    \Delta^{\pi_{(t+1):T}}(\pi_{t}^{a},\pi_{(t+1):T};\pi_{t}^{b},\pi_{(t+1):T}) = \Delta_{t}(\pi_{t}^{a},\pi_{(t+1):T};\pi_{t}^{b},\pi_{(t+1):T}),
\end{align*}
where $\Delta_{t}(\cdot;\cdot)$ is defined in (\ref{eq:Delta}). 

Therefore, it follows that 
\begin{align}
   & \sup_{\pi_{(t+1):T} \in \Pi_{(t+1):T}}\sup_{\pi_{t}^{a},\pi_{t}^{b} \in \Pi_{t}} |\Delta_{t}(\pi_{t}^{a},\pi_{(t+1):T};\pi_{t}^{b},\pi_{(t+1):T}) - \widetilde{\Delta}_{t}(\pi_{t}^{a},\pi_{(t+1):T};\pi_{t}^{b},\pi_{(t+1):T})| \nonumber \\
  &=\sup_{\pi_{(t+1):T} \in \Pi_{(t+1):T}}\sup_{\pi_{t}^{a},\pi_{t}^{b} \in \Pi_{t}} |\widetilde{\Delta}^{\pi_{(t+1):T}}(\pi_{t}^{a},\pi_{(t+1):T};\pi_{t}^{b},\pi_{(t+1):T})  - \Delta^{\pi_{(t+1):T}}(\pi_{t}^{a},\pi_{(t+1):T};\pi_{t}^{b},\pi_{(t+1):T}) | \nonumber \\
  &\leq \sup_{\pi_{(t+1):T} \in \Pi_{(t+1):T}} 
  \sup_{\pi_{t:T}^{a},\pi_{t:T}^{b} \in \Pi_{t:T}} 
  |\widetilde{\Delta}^{\pi_{(t+1):T}}(\pi_{t:T}^{a};\pi_{t:T}^{b})  - \Delta^{\pi_{(t+1):T}}(\pi_{t:T}^{a};\pi_{t:T}^{b})|. \label{eq:tilde_delta_inequalty}
\end{align}

Fix $\pi_{(t+1):T} \in \Pi_{(t+1):T}$. Note that $\left\{\widetilde{\Gamma}_{i,t}^{\pi_{(t+1):T}}(\underline{a}_{t:T})\right\}_{i=1,\ldots,n}$ are i.i.d. random variables with bounded support under Assumptions~\ref{asm:bounded outcome} and \ref{asm:overlap} and that $\kappa(\Pi_{t:T})$ is finite from Lemma~\ref{lem:entropy_integral_bound} and $\kappa(\Pi_{t:T}) \leq \kappa(\Pi)$. 
Therefore, fixing $\pi_{(t+1):T}$ and applying Lemma~\ref{lem:concentration inequality_influence difference function} with $\Gamma_{i}^{\dag}(\underline{a}_{t:T})=\widetilde{\Gamma}_{i,t}^{\pi_{(t+1):T}}(\underline{a}_{t:T})$ leads to the following result:
For any $\delta \in (0,1)$, with probability at least $1-2\delta$,
\begin{align}
    &\sup_{\pi_{t:T}^{a},\pi_{t:T}^{b} \in \Pi_{t:T}} 
  |\widetilde{\Delta}^{\pi_{(t+1):T}}(\pi_{t:T}^{a};\pi_{t:T}^{b})  - \Delta^{\pi_{(t+1):T}}(\pi_{t:T}^{a};\pi_{t:T}^{b})| \nonumber\\
    &\leq \left(54.4 \sqrt{2}\kappa(\Pi_{t:T}) + 435.2 + \sqrt{2 \log \frac{1}{\delta}}\right)\sqrt{\frac{V_{t:T}^{\pi_{(t+1):T},\ast}}{n}} + o\left(\frac{1}{\sqrt{n}}\right), \label{eq:bound_influence_difference_function_2}
\end{align}
with $V_{t:T}^{\pi_{(t+1):T},\ast}  \equiv  \sup_{\pi_{t:T}^{a},\pi_{t:T}^{b} \in \Pi_{t:T}}\E
    \left[\left(\widetilde{\Gamma}_{i,t}^{\pi_{(t+1):T}}(\pi_{t:T}^{a}) - \widetilde{\Gamma}_{i,t}^{\pi_{(t+1):T}}(\pi_{t:T}^{b}) \right)^2\right]$.

Finally, by Assumptions~\ref{asm:bounded outcome} and
\ref{asm:overlap} and the recursive definition of
$\widetilde{\Gamma}_{i,t}$,
\[
V_{t:T}^{\pi_{(t+1):T},\ast}
\le
M^2\left(
1+\frac{2}{\eta_{t:T}}
+3\sum_{s=t}^{T-1}\frac{1}{\eta_{t:s}}
\right) =  M^2
\mathcal V_{t:T}(\boldsymbol{\eta}).
\]
This proves the result.
\end{proof}


\subsection{Proof of Lemma~\ref{lem:helpful_lemma}}\label{app:proofs_of_main_lemmas}

We present the proof of Lemma~\ref{lem:helpful_lemma} in this section. 
The following lemma is a general version of Lemma~\ref{lem:helpful_lemma}.

\medskip

\begin{lemma}\label{lem:helpful_lemma_general}
Fix $\pi=(\pi_1,\ldots,\pi_T)\in \Pi$. Let  $R_{t}^{\pi_{t:T}}(\tilde{\pi}_t)  \equiv  V_{t}(\tilde{\pi}_t,\pi_{(t+1):T}) - V_{t}(\pi_{t:T})$ for any $\tilde{\pi}_t \in \Pi_t$. Then, under Assumptions~\ref{asm:sequential independence}, \ref{asm:overlap}, and \ref{asm:first-best}, the regret of $\pi$ is bounded from above as
\begin{align*}
    R(\pi) \leq R_{1}^{\pi_{1:T}}(\pi_{1}^{\ast,B}) + \sum_{t=2}^{T} \frac{2^{t-2}}{\eta_{1:(t-1)}} R_{t}^{\pi_{t:T}}(\pi_{t}^{\ast,B}). 
\end{align*}
\end{lemma}

\begin{proof}
For any $t$, define $R_{t}(\pi_{t:T})  \equiv  V_{t}(\pi_{t:T}^{\ast,B}) - V_{t}(\pi_{t:T})$, which is a partial regret of $\pi_{t:T}$ for stage $t$. Under Assumptions~\ref{asm:overlap} and \ref{asm:first-best}, $\pi^{\ast,B}$ satisfies the condition in Assumption \ref{asm:first-best}.
For any integers $s$ and $t$ such that $1 \leq t < s \leq T$,
\begin{align}
    &V_{t}(\pi_{t}^{\ast,B},\ldots,\pi_{T}^{\ast,B}) - V_{t}(\pi_{t}^{\ast,B},\ldots,\pi_{s-1}^{\ast,B},\pi_{s},\ldots,\pi_{T}) \nonumber \\
    &= \E\left[\frac{\prod_{\ell=t}^{s-1}\mathbf{1}\{A_{\ell}=\pi_{\ell}^{\ast,B}(H_{\ell})\}}{\prod_{\ell=t}^{s-1}e_{\ell}(H_{\ell},A_{\ell})}\cdot \left(Q_{s}^{\pi_{(s+1):T}^{\ast,B}}\left(H_{s},\pi_{s}^{\ast,B}\right) - Q_{s}^{\pi_{(s+1):T}}\left(H_{s},\pi_{s}\right)\right)\right] \nonumber \\
      &\leq \frac{1}{\eta_{t:(s-1)}} \E\left[\left(Q_{s}^{\pi_{(s+1):T}^{\ast,B}}\left(H_{s},\pi_{s}^{\ast,B}\right) - Q_{s}^{\pi_{(s+1):T}}\left(H_{s},\pi_{s}\right)\right)\right]  \nonumber \\
            &= \frac{1}{\eta_{t:(s-1)}} \left(V_{s}(\pi_{s:T}^{\ast,B}) - V_{s}(\pi_{s:T})\right) \nonumber \\
    &= \frac{1}{\eta_{t:(s-1)}}R_{s}\left(\pi_{s:T}\right), \label{eq:bound_delta}
\end{align}
where the first equality follows from Lemma~\ref{lemma:identification} (i) and Assumption \ref{asm:sequential independence}; the inequality follows from Assumptions~\ref{asm:overlap} and \ref{asm:first-best}; the second equality follows from Lemma~\ref{lemma:identification} (i).

For $t=T$ and $T-1$, we have
\begin{align*}
R_{T}(\pi_T) &= V_{T}(\pi_{T}^{\ast,B}) - V_{T}(\pi_{T}) = R_{T}^{\pi_{T}}(\pi_{T}^{\ast,B});  \\
   R_{T-1}\left(\pi_{(T-1):T}\right) &= \left[V_{T-1}\left(\pi_{T-1}^{\ast,B},\pi_{T}^{\ast,B}\right) - V_{T-1}\left(\pi_{T-1}^{\ast,B},\pi_{T}\right) \right]
   + \left[V_{T-1}\left(\pi_{T-1}^{\ast,B},\pi_{T}\right) - V_{T-1}\left(\pi_{T-1},\pi_{T}\right)\right] \\
   &\leq \frac{1}{\eta_{T-1}}R_{T}^{\pi_{T}}(\pi_{T}^{\ast,B}) + R_{T-1}^{\pi_{(T-1):T}}\left(\pi_{T-1}^{\ast,B}\right) \\
   &= \frac{1}{\eta_{T-1}}R_T(\pi_{T}) + R_{T-1}^{\pi_{(T-1):T}}\left(\pi_{T-1}^{\ast,B}\right),
\end{align*}
where the inequality follows from (\ref{eq:bound_delta}).

Generally, for $k=2,\ldots,T-1$, it follows that
\begin{align*}
    &R_{T-k}\left(\pi_{(T-k):T}\right) \\
    &= V_{T-k}\left(\pi_{T-k}^{\ast,B},\ldots,\pi_{T}^{\ast,B}\right) - V_{T-k}\left(\pi_{T-k},\ldots,\pi_{T}\right) \\ &=\sum_{s=T-k}^{T}\left[V_{T-k}\left(\pi_{T-k}^{\ast,B},\ldots,\pi_{s}^{\ast,B},\pi_{s+1},\ldots,\pi_{T}\right) - V_{T-k}\left(\pi_{T-k}^{\ast,B},\ldots,\pi_{s-1}^{\ast,B},\pi_{s},\ldots,\pi_{T}\right)\right] \\
    &=\sum_{s=T-k+1}^{T}\left[V_{T-k}\left(\pi_{T-k}^{\ast,B},\ldots,\pi_{s}^{\ast,B},\pi_{s+1},\ldots,\pi_{T}\right) - V_{T-k}\left(\pi_{T-k}^{\ast,B},\ldots,\pi_{s-1}^{\ast,B},\pi_{s},\ldots,\pi_{T}\right)\right]  \\
    &+ R_{T-k}^{\pi_{(T-k):T}}(\pi_{T-k}^{\ast,B})\\
    &\leq \sum_{s=T-k+1}^{T}\left[V_{T-k}\left(\pi_{T-k}^{\ast,B},\ldots,\pi_{T}^{\ast,B}\right) - V_{T-k}\left(\pi_{T-k}^{\ast,B},\ldots,\pi_{s-1}^{\ast,B},\pi_{s},\ldots,\pi_{T}\right)\right] + R_{T-k}^{\pi_{(T-k):T}}(\pi_{T-k}^{\ast,B})\\
    &\leq \sum_{s=T-k+1}^{T}\frac{1}{\eta_{(T-k):(s-1)}} R_{s}\left(\pi_{t:T}\right)+ R_{T-k}^{\pi_{(T-k):T}}(\pi_{T-k}^{\ast,B}),
\end{align*}
where the second equality follows from the telescoping sum; the third equality follows from the definition of $R_{T-k}^{\pi_{(T-k):T}}(\pi_{T-k}^{\ast,B})$; the first inequality follows from Lemma~\ref{lemma:identification} and Assumption \ref{asm:first-best}; the last line follows from (\ref{eq:bound_delta}).

Then, recursively, the following hold:
\begin{align*}
    R_{T-1}\left(\pi_{(T-1):T}\right) &\leq \frac{1}{\eta_{T-1}}R_{T}\left(\pi_{T}\right)+ R_{T-1}^{\pi_{(T-1):T}}(\pi_{T-1}^{\ast,B}) \\
    &= \frac{1}{\eta_{T-1}}R_{T}^{\pi_{T}}\left(\pi_{T}^{\ast,B}\right)+ R_{T-1}^{\pi_{(T-1):T}}(\pi_{T-1}^{\ast,B}), \\
    R_{T-2}\left(\pi_{(T-2):T}\right) &\leq \frac{1}{\eta_{T-2}}R_{T-1}\left(\pi_{(T-1):T}\right)+
    \frac{1}{\eta_{(T-2):(T-1)}}R_{T}\left(\pi_{T}\right)+R_{T-2}^{\pi_{(T-2):T}}(\pi_{T-2}^{\ast,B}) \\
    &\leq \frac{2}{\eta_{(T-2):(T-1)}}R_{T}^{\pi_{T}}\left(\pi_{T}^{\ast,B}\right) + \frac{1}{\eta_{T-2}}R_{T-1}^{\pi_{(T-1):T}}\left(\pi_{T-1}^{\ast,B}\right) +R_{T-2}^{\pi_{(T-2):T}}(\pi_{T-2}^{\ast,B}),\\
    & \ \ \vdots \\
R_{T-k}\left(\pi_{(T-k):T}\right) & \leq \sum_{s=1}^{k} \frac{2^{k-s}}{\eta_{(T-k):(T-s)}} R_{T-s+1}^{\pi_{(T-s+1):T}}(\pi_{T-s+1}^{\ast,B}) + R_{T-k}^{\pi_{(T-k):T}}(\pi_{T-k}^{\ast,B}).
\end{align*}
Therefore, setting $k=T-1$ and noting that $R_{1}\left(\pi_{1:T}\right) = R(\pi)$, we obtain
\begin{align*}
    R(\pi) &\leq  \sum_{s=1}^{T-1} \frac{2^{T-1-s}}{\eta_{1:(T-s)}} R_{T-s+1}^{\pi_{(T-s+1):T}}(\pi_{T-s+1}^{\ast,B}) + R_{1}^{\pi_{1:T}}(\pi_{1}^{\ast,B}) \\
    & = R_{1}^{\pi_{1:T}}(\pi_{1}^{\ast,B}) + \sum_{s=1}^{T-1} \frac{2^{s-1}}{\eta_{1:s}} R_{s+1}^{\pi_{(s+1):T}}(\pi_{s+1}^{\ast,B}).
\end{align*}
Setting $t=s+1$ in the above equation leads to the result.
\end{proof}

\medskip


\begin{proof}[Proof of Lemma~\ref{lem:helpful_lemma}.]
   Lemma~\ref{lem:helpful_lemma} follows immediately from Lemma~\ref{lem:helpful_lemma_general} by setting $\pi=\hat{\pi}$. 
\end{proof}


\subsection{Proof of Lemma~\ref{lem:asymptotic_estimated_policy_difference_function}}
\label{app:proof_of_main_lemma_2}

This section presents the proof of Lemma~\ref{lem:asymptotic_estimated_policy_difference_function}. We here extend the analytical strategy of \cite{Athey_Wager_2020} and \cite{zhou2022offline}, 
which leverages orthogonal moments and cross-fitting, to the sequential multi-stage setting.

We begin by introducing several notational definitions. For any $\underline{a}_{t:s} \in \underline{\MA}_{t:s}$ and $\pi_{t:s}^{a},\pi_{t:s}^{b} \in \Pi_{t:s}$ ($t\leq s$), let $G_{i,\pi_{t:s}^{a},\pi_{t:s}^{b}}^{\underline{a}_{t:s}}  \equiv  \prod_{\ell=t}^{s}\mathbf{1}\{\pi_{\ell}^{a}(H_{i,\ell})=a_\ell\} - \prod_{\ell=t}^{s}\mathbf{1}\{\pi_{\ell}^{b}(H_{i,\ell})=a_\ell\}$.

Given a fixed DTR $\pi=(\pi_1,\ldots,\pi_T)$, with some abuse of notation, we define
\begin{align*}
\widetilde{V}_{i,T}(a_{T}) &\equiv \frac{ Y_{i,T} - Q_{T}(H_{i,T},a_{T}) }{e_{T}(H_{i,T},a_{T})}\cdot \mathbf{1}\{A_{i,T}=a_{T}\} + Q_{T}(H_{i,T},a_{T}),\\
\check{V}_{i,T}(a_{T}) &\equiv \frac{ Y_{i,T} - \widehat{Q}_{T}^{-k(i)}(H_{i,T},a_{T}) }{\hat{e}_{T}^{-k(i)}(H_{i,T},a_{T})}\cdot \mathbf{1}\{A_{i,T}=a_{T}\} + \widehat{Q}_{T}^{-k(i)}(H_{i,T},a_{T}),
\end{align*}
and, recursively, for $t=T-1,\ldots,1$,
\begin{align*}
\widetilde{V}_{i,t}^{\pi_{(t+1):T}}(a_{t}) &\equiv \frac{ Y_{i,t} + \widetilde{V}_{i,t+1}^{\pi_{(t+2):T}}(\pi_{t+1}(H_{i,t+1})) - Q_{t}^{\pi_{(t+1):T}}(H_{i,t},a_{t}) }{e_{t}(H_{i,t},a_{t})}\cdot \mathbf{1}\{A_{i,t} = a_{t}\} \\
&+ Q_{t}^{\pi_{(t+1):T}}(H_{i,t},a_{t}),\\
\check{V}_{i,t}^{\pi_{(t+1):T}}(a_t) &\equiv \frac{ Y_{i,t} + \widetilde{V}_{i,t+1}^{\pi_{(t+2):T}}(\pi_{t+1}(H_{i,t+1})) - \widehat{Q}_{t}^{\pi_{(t+1):T},-k(i)}(H_{i,t},a_{t}) }{\hat{e}_{t}^{-k(i)}(H_{i,t},a_{t})}\cdot \mathbf{1}\{A_{i,t} = a_{t}\} \\
&+ \widehat{Q}_{t}^{\pi_{(t+1):T},-k(i)}(H_{i,t},a_{t}),
\end{align*}
where, for $t=T$, we denote $\widetilde{V}_{i,T}^{\pi_{(T+1):T}}(a_{t}) = \widetilde{V}_{i,T}(a_{T})$ and $\check{V}_{i,T}^{\pi_{(T+1):T}}(a_{t}) = \check{V}_{i,T}(a_{T})$.

For integers $s$ and $t$ such that $1 \leq t \leq s \leq T$ and $\underline{a}_{t:s} \in \underline{\MA}_{t:s}$, let
\begin{align*}
    &\widetilde{S}_{t:s}^{\underline{a}_{t:s}}(\pi_{t:s}^{a},\pi_{t:s}^{b},\pi_{(s+1):T}) \\
    &\equiv  \frac{1}{n}\sum_{i=1}^{n} G_{i,\pi_{t:s}^{a},\pi_{t:s}^{b}}^{\underline{a}_{t:s}}\cdot \left(\frac{\prod_{\ell = t}^{s-1}\mathbf{1}\{ A_{i,\ell} = a_{\ell}\} }{\prod_{\ell = t}^{s-1}\hat{e}_{\ell}^{-k(i)}(H_{i,\ell},a_{\ell}) } \right) 
    \cdot \left(\check{V}_{i,s}^{\pi_{(s+1):T}}(a_s) - \widetilde{V}_{i,s}^{\pi_{(s+1):T}}(a_s)\right).
\end{align*}
For the case $s=t$, we denote $ \widetilde{S}_{t:t}^{\underline{a}_{t:t}}(\pi_{t:t}^{a},\pi_{t:t}^{b},\pi_{(t+1):T}) = \widetilde{S}_{t}^{a_{t}}(\pi_{t}^{a},\pi_{t}^{b},\pi_{(t+1):T})$.
When $s=T$, we also denote $\widetilde{S}_{t:T}^{\underline{a}_{t:T}}(\pi_{t:T}^{a},\pi_{t:T}^{b}) = \widetilde{S}_{t:T}^{\underline{a}_{t:T}}(\pi_{t:T}^{a},\pi_{t:T}^{b},\pi_{(T+1):T})$ where
\begin{align*}
    \widetilde{S}_{t:T}^{\underline{a}_{t:T}}(\pi_{t:T}^{a},\pi_{t:T}^{b}) = \frac{1}{n}\sum_{i=1}^{n} G_{i,\pi_{t:T}^{a},\pi_{t:T}^{b}}^{\underline{a}_{t:T}}\cdot \left(\frac{\prod_{\ell = t}^{T-1}\mathbf{1}\{ A_{i,\ell} = a_{\ell}\} }{\prod_{\ell = t}^{T-1}\hat{e}_{\ell}^{-k(i)}(H_{i,\ell},a_{\ell}) } \right) 
    \cdot \left(\check{V}_{i,T}(a_T) - \widetilde{V}_{i,T}(a_T)\right).
\end{align*}

Note that each term in (\ref{eq:latter_2}) can be expressed as
\begin{align*}
    &\check{\Delta}_{t,s}^{\dagger}(\pi_{t}^{a},\pi_{(t+1):T};\pi_{t}^{b},\pi_{(t+1):T}) - \widetilde{\Delta}_{t,s}^{\dagger}(\pi_{t}^{a},\pi_{(t+1):T};\pi_{t}^{b},\pi_{(t+1):T})\\
    &=\sum_{\underline{a}_{t:s} \in \underline{\MA}_{t:s}} \widetilde{S}_{t:s}^{\underline{a}_{t:s}}(\pi_{t}^{a},\pi_{(t+1):s},\pi_{t}^{b},\pi_{(t+1):s},\pi_{(s+1):T}). 
\end{align*}
Hence, regarding (\ref{eq:latter_2}), it follows that
\begin{align}
     &\sup_{\pi_{(t+1):T} \in \Pi_{(t+1):T}} \sup_{\pi_{t}^{a},\pi_{t}^{b} \in \Pi_{t}} |\check{\Delta}_{t,s}^{\dagger}(\pi_{t}^{a},\pi_{(t+1):T};\pi_{t}^{b},\pi_{(t+1):T}) - \widetilde{\Delta}_{t,s}^{\dagger}(\pi_{t}^{a},\pi_{(t+1):T};\pi_{t}^{b},\pi_{(t+1):T})| \nonumber \\
     &\leq \sum_{\underline{a}_{t:s} \in \underline{\MA}_{t:s}} \sup_{\pi_{(t+1):T} \in \Pi_{(t+1):T}} \sup_{\pi_{t}^{a},\pi_{t}^{b} \in \Pi_{t}} |\widetilde{S}_{t:s}^{\underline{a}_{t:s}}(\pi_{t}^{a},\pi_{(t+1):s},\pi_{t}^{b},\pi_{(t+1):s},\pi_{(s+1):T})|. \label{eq:latter_2_bound}
\end{align}
This result enables us to evaluate (\ref{eq:latter_2}) through evaluating $$\sup_{\pi_{(t+1):T} \in \Pi_{(t+1):T}} \sup_{\pi_{t}^{a},\pi_{t}^{b} \in \Pi_{t}} |\widetilde{S}_{t:s}^{\underline{a}_{t:s}}(\pi_{t}^{a},\pi_{(t+1):s},\pi_{t}^{b},\pi_{(t+1):s},\pi_{(s+1):T})|$$
for each $\underline{a}_{t:s} \in \underline{\MA}_{t:s}$.

\medskip

\begin{lemma} \label{lem:convergence_rate_Stilde}
Suppose that Assumptions~\ref{asm:sequential independence}, \ref{asm:bounded outcome}, \ref{asm:overlap}, \ref{asm:rate_of_convergence_backward}, and \ref{asm:bounded entropy} hold. 
Then, for any integers $s$ and $t$ such that $1\leq s \leq t \leq T$ and any
$\underline a_{s:t}\in\underline{\mathcal A}_{s:t}$,
\begin{align}
        &\sup_{\pi_{(t+1):T} \in  \Pi_{(t+1):T}}\sup_{\pi_{s:t}^{a},\pi_{s:t}^{b}\in \Pi_{s:t}} \left| \widetilde{S}_{s:t}^{\underline{a}_{s:t}}(\pi_{s:t}^{a},\pi_{s:t}^{b},\pi_{(t+1):T})\right|  = O_{p}(n^{-\min\{1/2,\tau/2\}}). \label{eq:sequential_bound}
\end{align}
\end{lemma}

\medskip

\begin{proof}
We first consider the case that $t <T$. For any integers $s$ and $t$ such that $1\leq s\leq t < T$, we define
\begin{align*}
\widetilde{S}_{s:t,(A)}^{\underline{a}_{s:t}}(\pi_{s:t}^{a},\pi_{s:t}^{b},\pi_{(t+1):T})& \equiv \frac{1}{n}\sum_{i=1}^{n}
G_{i,\pi_{s:t}^{a},\pi_{s:t}^{b}}^{\underline{a}_{s:t}}
\frac{\prod_{\ell=s}^{t-1}\mathbf{1}\{A_{i,\ell}=a_{\ell}\}}{\prod_{\ell=s}^{t-1}\hat{e}_{\ell}^{-k(i)}(H_{i,\ell},a_{\ell})} \nonumber \\
&\times\left(\widehat{Q}_{t}^{\pi_{(t+1):T},-k(i)}\left(H_{i,t},a_{t}\right)-Q_{t}^{\pi_{(t+1):T}}\left(H_{i,t},a_{t}\right)\right) \nonumber\\
&\times \left(1-\frac{\mathbf{1}\left\{ A_{i,t}=a_{t}\right\} }{e_{t}\left(H_{i,t},a_{t}\right)}\right); \\
\widetilde{S}_{s:t,(B)}^{\underline{a}_{s:t}}(\pi_{s:t}^{a},\pi_{s:t}^{b},\pi_{(t+1):T})& \equiv \frac{1}{n}\sum_{i=1}^{n}G_{i,\pi_{s:t}^{a},\pi_{s:t}^{b}}^{\underline{a}_{s:t}}\frac{\prod_{\ell=s}^{t-1}\mathbf{1}\{A_{i,\ell}=a_{\ell}\}}{\prod_{\ell=s}^{t-1}\hat{e}_{\ell}^{-k(i)}(H_{i,\ell},a_{\ell})} \nonumber \\
	&\times\left(\widetilde{V}_{i,t+1}^{\pi_{(t+2):T}}(\pi_{t+1}(H_{i,t+1}))-Q_{t}^{\pi_{(t+1):T}}\left(H_{i,t},a_{t}\right)\right) \nonumber \\
 &\times \left(\frac{\mathbf{1}\left\{ A_{i,t}=a_{t}\right\} }{\hat{e}_{t}^{-k(i)}\left(H_{i,t},a_{t}\right)}-\frac{\mathbf{1}\left\{ A_{i,t}=a_{t}\right\} }{e_{t}\left(H_{i,t},a_{t}\right)}\right);  \\
\widetilde{S}_{s:t,(C)}^{\underline{a}_{s:t}}(\pi_{s:t}^{a},\pi_{s:t}^{b},\pi_{(t+1):T})& \equiv \frac{1}{n}\sum_{i=1}^{n}G_{i,\pi_{s:t}^{a},\pi_{s:t}^{b}}^{\underline{a}_{s:t}}\frac{\prod_{\ell=s}^{t-1}\mathbf{1}\{A_{i,\ell}=a_{\ell}\}}{\prod_{\ell=s}^{t-1}\hat{e}_{\ell}^{-k(i)}(H_{i,\ell},a_{\ell})} \nonumber \\
&\times\left(Q_{t}^{\pi_{(t+1):T}}\left(H_{i,t},a_{t}\right)-\widehat{Q}_{t}^{\pi_{(t+1):T},-k(i)}\left(H_{i,t},a_{t}\right)\right) \notag \\ 
&\times \left(\frac{\mathbf{1}\left\{ A_{i,t}=a_{t}\right\} }{\hat{e}_{t}^{-k(i)}\left(H_{i,t},a_{t}\right)}-\frac{\mathbf{1}\left\{ A_{i,t}=a_{t}\right\} }{e_{t}\left(H_{i,t},a_{t}\right)}\right).
\end{align*}
Using these definitions, we can decompose $\widetilde{S}_{s:t}^{\underline{a}_{s:t}}(\pi_{s:t}^{a},\pi_{s:t}^{b},\pi_{(t+1):T})$ as follows:
\begin{align*}
    &\widetilde{S}_{s:t}^{\underline{a}_{s:t}}(\pi_{s:t}^{a},\pi_{s:t}^{b},\pi_{(t+1):T}) \\
    &= \widetilde{S}_{s:t,(A)}^{\underline{a}_{s:t}}(\pi_{s:t}^{a},\pi_{s:t}^{b},\pi_{(t+1):T}) 
    + \widetilde{S}_{s:t,(B)}^{\underline{a}_{s:t}}(\pi_{s:t}^{a},\pi_{s:t}^{b},\pi_{(t+1):T})  
    + \widetilde{S}_{s:t,(C)}^{\underline{a}_{s:t}}(\pi_{s:t}^{a},\pi_{s:t}^{b},\pi_{(t+1):T}).
\end{align*}
We hence have 
\begin{align}
        &\sup_{\pi_{(t+1):T} \in  \Pi_{(t+1):T}}\sup_{\pi_{s:t}^{a},\pi_{s:t}^{b}\in \Pi_{s:t}} \left| \widetilde{S}_{s:t}^{\underline{a}_{s:t}}(\pi_{s:t}^{a},\pi_{s:t}^{b},\pi_{(t+1):T})\right| \nonumber \\
        & \leq  \sup_{\pi_{(t+1):T} \in  \Pi_{(t+1):T}}\sup_{\pi_{s:t}^{a},\pi_{s:t}^{b}\in \Pi_{s:t}} \left| \widetilde{S}_{s:t,(A)}^{\underline{a}_{s:t}}(\pi_{s:t}^{a},\pi_{s:t}^{b},\pi_{(t+1):T})\right| \nonumber \\
        &+ \sup_{\pi_{(t+1):T} \in  \Pi_{(t+1):T}}\sup_{\pi_{s:t}^{a},\pi_{s:t}^{b}\in \Pi_{s:t}} \left| \widetilde{S}_{s:t,(B)}^{\underline{a}_{s:t}}(\pi_{s:t}^{a},\pi_{s:t}^{b},\pi_{(t+1):T})\right| \nonumber \\
        &+ \sup_{\pi_{(t+1):T} \in  \Pi_{(t+1):T}}\sup_{\pi_{s:t}^{a},\pi_{s:t}^{b}\in \Pi_{s:t}} \left| \widetilde{S}_{s:t,(C)}^{\underline{a}_{s:t}}(\pi_{s:t}^{a},\pi_{s:t}^{b},\pi_{(t+1):T})\right|  \label{eq:Stilde_bound}
\end{align}

In what follows, we will prove that:
\begin{align}
    \sup_{\pi_{(t+1):T} \in  \Pi_{(t+1):T}}\sup_{\pi_{s:t}^{a},\pi_{s:t}^{b}\in \Pi_{s:t}} \left| \widetilde{S}_{s:t,(A)}^{\underline{a}_{s:t}}(\pi_{s:t}^{a},\pi_{s:t}^{b},\pi_{(t+1):T})\right| &= O_{p}\left(n^{-1/2}\right); \label{eq:S_tilde_A_bound}\\
    \sup_{\pi_{(t+1):T} \in  \Pi_{(t+1):T}}\sup_{\pi_{s:t}^{a},\pi_{s:t}^{b}\in \Pi_{s:t}} \left| \widetilde{S}_{s:t,(B)}^{\underline{a}_{s:t}}(\pi_{s:t}^{a},\pi_{s:t}^{b},\pi_{(t+1):T})\right| &= O_{p}\left(n^{-1/2}\right); \label{eq:S_tilde_B_bound}\\
    \sup_{\pi_{(t+1):T} \in  \Pi_{(t+1):T}}\sup_{\pi_{s:t}^{a},\pi_{s:t}^{b}\in \Pi_{s:t}} \left| \widetilde{S}_{s:t,(C)}^{\underline{a}_{s:t}}(\pi_{s:t}^{a},\pi_{s:t}^{b},\pi_{(t+1):T})\right| &= O_{p}\left(n^{-\min\{1/2,\tau/2\}}\right). \label{eq:S_tilde_C_bound}
\end{align}
Then we can obtain the result (\ref{eq:sequential_bound}) from equation (\ref{eq:Stilde_bound}).

Throughout the proof, without loss of generality, we assume that $n>n_0$, where $n_0$ is defined in Assumption \ref{asm:rate_of_convergence_backward}.
We begin by examining $\widetilde{S}_{s:t,(A)}^{\underline{a}_{s:t}}(\pi_{s:t}^{a},\pi_{s:t}^{b},\pi_{(t+1):T})$, which we further decompose as follows:
\begin{align*}
    &\widetilde{S}_{s:t,(A)}^{\underline{a}_{s:t}}(\pi_{s:t}^{a},\pi_{s:t}^{b},\pi_{(t+1):T}) = \widetilde{S}_{s:t,(A1)}^{\underline{a}_{s:t}}(\pi_{s:t}^{a},\pi_{s:t}^{b},\pi_{(t+1):T}) + \widetilde{S}_{s:t,(A2)}^{\underline{a}_{s:t}}(\pi_{s:t}^{a},\pi_{s:t}^{b},\pi_{(t+1):T}),
\end{align*}
where
\begin{align*}
\widetilde{S}_{s:t,(A1)}^{\underline{a}_{s:t}}(\pi_{s:t}^{a},\pi_{s:t}^{b},\pi_{(t+1):T})& \equiv \frac{1}{n}\sum_{i=1}^{n}G_{i,\pi_{s:t}^{a},\pi_{s:t}^{b}}^{\underline{a}_{s:t}}\cdot\frac{\prod_{\ell=s}^{t-1}\mathbf{1}\{A_{i,\ell}=a_{\ell}\}}{\prod_{\ell=s}^{t-1}e_{\ell}(H_{i,\ell},a_{\ell})}\\
      &\times\left(\widehat{Q}_{t}^{\pi_{(t+1):T},-k(i)}\left(H_{i,t},a_{t}\right)-Q_{t}^{\pi_{(t+1):T}}\left(H_{i,t},a_{t}\right)\right)\left(1-\frac{\mathbf{1}\left\{ A_{i,t}=a_{t}\right\} }{e_{t}\left(H_{i,t},a_{t}\right)}\right);\\
\widetilde{S}_{s:t,(A2)}^{\underline{a}_{s:t}}(\pi_{s:t}^{a},\pi_{s:t}^{b},\pi_{(t+1):T})& \equiv \frac{1}{n}\sum_{i=1}^{n}G_{i,\pi_{s:t}^{a},\pi_{s:t}^{b}}^{\underline{a}_{s:t}}\cdot\left(\frac{\prod_{\ell=s}^{t-1}\mathbf{1}\{A_{i,\ell}=a_{\ell}\}}{\prod_{\ell=s}^{t-1}\hat{e}_{\ell}^{-k(i)}(H_{i,\ell},a_{\ell})}-\frac{\prod_{\ell=s}^{t-1}\mathbf{1}\{A_{i,\ell}=a_{\ell}\}}{\prod_{\ell=s}^{t-1}e_{\ell}(H_{i,\ell},a_{\ell})}\right)\\
 &\times\left(\widehat{Q}_{t}^{\pi_{(t+1):T},-k(i)}\left(H_{i,t},a_{t}\right)-Q_{t}^{\pi_{(t+1):T}}\left(H_{i,t},a_{t}\right)\right)\left(1-\frac{\mathbf{1}\left\{ A_{i,t}=a_{t}\right\} }{e_{t}\left(H_{i,t},a_{t}\right)}\right).
\end{align*}
For each fold index $k$, we define
\begin{align*}
\widetilde{S}_{s:t,(A1)}^{\underline{a}_{s:t},k}(\pi_{s:t}^{a},\pi_{s:t}^{b},\pi_{(t+1):T})& \equiv \frac{1}{n}\sum_{i \in I_{k}}G_{i,\pi_{s:t}^{a},\pi_{s:t}^{b}}^{\underline{a}_{s:t}}\cdot\frac{\prod_{\ell=s}^{t-1}\mathbf{1}\{A_{i,\ell}=a_{\ell}\}}{\prod_{\ell=s}^{t-1}e_{\ell}(H_{i,\ell},a_{\ell})}\\
      &\times\left(\widehat{Q}_{t}^{\pi_{(t+1):T},-k}\left(H_{i,t},a_{t}\right)-Q_{t}^{\pi_{(t+1):T}}\left(H_{i,t},a_{t}\right)\right)\left(1-\frac{\mathbf{1}\left\{ A_{i,t}=a_{t}\right\} }{e_{t}\left(H_{i,t},a_{t}\right)}\right).
\end{align*}
For notational simplicity, let $\eta := \min_{t} \eta_t$.

Fix $k \in \{1,\ldots,K\}$. We now consider $\widetilde{S}_{s:t,(A1)}^{\underline{a}_{s:t},k}(\pi_{s:t}^{a},\pi_{s:t}^{b},\pi_{(t+1):T})$. Since $\widehat{Q}_{t}^{\pi_{(t+1):T},-k}(\cdot,a_{t})$ is computed using the data in the rest $K-1$ folds, when the data $\MS_{-k} \equiv \{Z_i : i \notin I_{k}\}$ in the rest $K-1$ folds is conditioned,  $\widehat{Q}_{t}^{\pi_{(t+1):T},-k}(\cdot,a_t)$ is fixed; hence, $\widetilde{S}_{s:t,(A1)}^{\underline{a}_{s:t},k}(\pi_{s:t}^{a},\pi_{s:t}^{b},\pi_{(t+1):T})$ is a sum of i.i.d. bounded random variables under Assumptions~\ref{asm:bounded outcome}, \ref{asm:overlap}, and \ref{asm:rate_of_convergence_backward} (ii).

It follows that 
\begin{align*}
&\E\left[G_{i,\pi_{s:t}^{a},\pi_{s:t}^{b}}^{\underline{a}_{s:t}}\cdot\frac{\prod_{\ell=s}^{t-1}\mathbf{1}\{A_{i,\ell}=a_{\ell}\}}{\prod_{\ell=s}^{t-1}e_{\ell}(H_{i,\ell},a_{\ell})}\cdot\left(\widehat{Q}_{t}^{\pi_{(t+1):T},-k}\left(H_{i,t},a_{t}\right)-Q_{t}^{\pi_{(t+1):T}}\left(H_{i,t},a_{t}\right)\right) \right. \\ & \left. \  \times \left(1-\frac{\mathbf{1}\left\{ A_{i,t}=a_{t}\right\} }{e_{t}\left(H_{i,t},a_{t}\right)}\right)\middle| \MS_{-k}\right]\\
&=	\E\left[G_{i,\pi_{s:t}^{a},\pi_{s:t}^{b}}^{\underline{a}_{s:t}}\cdot\frac{\prod_{\ell=s}^{t-1}\mathbf{1}\{A_{i,\ell}=a_{\ell}\}}{\prod_{\ell=s}^{t-1}e_{\ell}(H_{i,\ell},a_{\ell})} \cdot \left(\widehat{Q}_{t}^{\pi_{(t+1):T},-k}\left(H_{i,t},a_{t}\right)-Q_{t}^{\pi_{(t+1):T}}\left(H_{i,t},a_{t}\right)\right)\right. \\ 
& \left. \times \E\left[\left(1-\frac{\mathbf{1}\left\{ A_{i,t}=a_{t}\right\} }{e_{t}\left(H_{i,t},a_{t}\right)}\right)\middle| H_{i,t}\right]\middle|\MS_{-k} \right]\\
&=	\E\left[G_{i,\pi_{s:t}^{a},\pi_{s:t}^{b}}^{\underline{a}_{s:t}}\cdot\frac{\prod_{\ell=s}^{t-1}\mathbf{1}\{A_{i,\ell}=a_{\ell}\}}{\prod_{\ell=s}^{t-1}e_{\ell}(H_{i,\ell},a_{\ell})} \cdot \left(\widehat{Q}_{t}^{\pi_{(t+1):T},-k}\left(H_{i,t},a_{t}\right)-Q_{t}^{\pi_{(t+1):T}}\left(H_{i,t},a_{t}\right)\right)\right. \\ 
& \left. \times \left(1-\frac{e_{t}\left(H_{i,t},a_{t}\right)}{e_{t}\left(H_{i,t},a_{t}\right)}\right)\middle|\MS_{-k} \right]\\
&=	0.
\end{align*}
Hence, fixing $\pi_{(t+1):T}$, $\sup_{\pi_{s:t}^{a},\pi_{s:t}^{b}\in \Pi_{s:t}} \left| \widetilde{S}_{s:t,(A1)}^{\underline{a}_{s:t},k}(\pi_{s:t}^{a},\pi_{s:t}^{b},\pi_{(t+1):T})\right|$ can be written as
\begin{align*}
     	&\sup_{\pi_{s:t}^{a},\pi_{s:t}^{b}\in \Pi_{s:t}}\left|\widetilde{S}_{s:t,(A1)}^{\underline{a}_{s:t},k}(\pi_{s:t}^{a},\pi_{s:t}^{b},\pi_{(t+1):T})\right|\\
&=	\frac{1}{K}\sup_{\pi_{s:t}^{a},\pi_{s:t}^{b}\in \Pi_{s:t}}\left|\frac{1}{n/K}\sum_{i \in I_k }
G_{i,\pi_{s:t}^{a},\pi_{s:t}^{b}}^{\underline{a}_{s:t}}\cdot\frac{\prod_{\ell=s}^{t-1}\mathbf{1}\{A_{i,\ell}=a_{\ell}\}}{\prod_{\ell=s}^{t-1}e_{\ell}(H_{i,\ell},a_{\ell})}\right.\\
	&\times \left(\widehat{Q}_{t}^{\pi_{(t+1):T},-k}\left(H_{i,t},a_{t}\right)-Q_{t}^{\pi_{(t+1):T}}\left(H_{i,t},a_{t}\right)\right)\left(1-\frac{\mathbf{1}\left\{ A_{i,t}=a_{t}\right\} }{e_{t}\left(H_{i,t},a_{t}\right)}\right) \\
	&-\E\left[\frac{1}{n/K}\sum_{i \in I_k }G_{i,\pi_{s:t}^{a},\pi_{s:t}^{b}}^{\underline{a}_{s:t}}\cdot\frac{\prod_{\ell=s}^{t-1}\mathbf{1}\{A_{i,\ell}=a_{\ell}\}}{\prod_{\ell=s}^{t-1}e_{\ell}(H_{i,\ell},a_{\ell})}\right.\\
	&\left.\left.\times\left(\widehat{Q}_{t}^{\pi_{(t+1):T},-k}\left(H_{i,t},a_{t}\right)-Q_{t}^{\pi_{(t+1):T}}\left(H_{i,t},a_{t}\right)\right)\left(1-\frac{\mathbf{1}\left\{ A_{i,t}=a_{t}\right\} }{e_{t}\left(H_{i,t},a_{t}\right)}\right)\middle|\MS_{-k} \right] \right|.
\end{align*}

By applying Lemma~\ref{lem:concentration inequality_influence difference function} while fixing $\MS_{-k}$ and setting $i \in I_k$ and 
\begin{align*}
    \Gamma_{i}^{\dag}(\underline{a}_{s:t})= \frac{\prod_{\ell=s}^{t-1}\mathbf{1}\{A_{i,\ell}=a_{\ell}\}}{\prod_{\ell=s}^{t-1}e_{\ell}(H_{i,\ell},a_{\ell})}\left(\widehat{Q}_{t}^{\pi_{(t+1):T},-k}\left(H_{i,t},a_{t}\right)-Q_{t}^{\pi_{(t+1):T}}\left(H_{i,t},a_{t}\right)\right)\left(1-\frac{\mathbf{1}\left\{ A_{i,t}=a_{t}\right\} }{e_{t}\left(H_{i,t},a_{t}\right)}\right),
\end{align*}
the following holds: $\forall \delta > 0$, with probability at least $1-2\delta$,
\begin{align*}
&\sup_{\pi_{s:t}^{a},\pi_{s:t}^{b}\in \Pi_{s:t}}\left|\widetilde{S}_{s:t,(A1)}^{\underline{a}_{s:t},k}(\pi_{s:t}^{a},\pi_{s:t}^{b},\pi_{(t+1):T})\right|\\
&\leq	o\left(n^{-1/2}\right)+\left(54.4\kappa\left(\Pi_{s:t}\right)+435.2+\sqrt{2\log(1/\delta)}\right) \\
	&\times\left[\sup_{\pi_{s:T}\in \Pi_{s:T}}\E\left[\left(G_{i,\pi_{s:t}^{a},\pi_{s:t}^{b}}^{\underline{a}_{s:t}}\right)^{2}\cdot\left(\frac{\prod_{\ell=s}^{t-1}\mathbf{1}\{A_{i,\ell}=a_{\ell}\}}{\prod_{\ell=s}^{t-1}e_{\ell}(H_{i,\ell},a_{\ell})}\right)^{2}\right.\right.\\
&	\left.\left.\left.\times\left(\widehat{Q}_{t}^{\pi_{(t+1):T},-k}\left(H_{i,t},a_{t}\right)-Q_{t}^{\pi_{(t+1):T}}\left(H_{i,t},a_{t}\right)\right)^{2}\left(1-\frac{\mathbf{1}\left\{ A_{i,t}=a_{t}\right\} }{e_{t}\left(H_{i,t},a_{t}\right)}\right)^{2}\right|\MS_{-k} \right] \middle/ \left(\frac{n}{K}\right) \right]^{1/2} \\
&\leq	o\left(n^{-1/2}\right)+\sqrt{K} \cdot \left(54.4\kappa\left(\Pi_{s:t}\right)+435.2+\sqrt{2\log(1/\delta)}\right)
	\cdot \left(\frac{1}{\eta}\right)^{t-s+1}\\
	&\times \sqrt{\frac{\E\left[\left.\left(\widehat{Q}_{t}^{\pi_{(t+1):T},-k}\left(H_{i,t},a_{t}\right)-Q_{t}^{\pi_{(t+1):T}}\left(H_{i,t},a_{t}\right)\right)^2\right|\MS_{-k} \right]}{n}},
\end{align*}
where the last inequality follows from $\left(G_{i,\pi_{s:t}^{a},\pi_{s:t}^{b}}^{\underline{a}_{s:t}}\right)^{2}\leq 1$ a.s. and Assumption \ref{asm:overlap} (overlap condition). 
Note that $\kappa\left(\Pi_{s:t}\right)$ is finite by Lemma~\ref{lem:entropy_integral_bound} and the inequality $\kappa\left(\Pi_{s:t}\right) \leq \kappa\left(\Pi \right)$.
From Assumptions~\ref{asm:bounded outcome} and \ref{asm:rate_of_convergence_backward} (ii), we have 
\begin{align*}
    \sup_{\pi_{(t+1):T} \in \Pi_{(t+1):T}}\E\left[\left(\widehat{Q}_{t}^{\pi_{(t+1):T},-k}\left(H_{i,t},a_{t}\right)-Q_{t}^{\pi_{(t+1):T}}\left(H_{i,t},a_{t}\right)\right)^2\right]<\infty.
\end{align*}
Hence, Markov's inequality yields
\begin{align*}
    \sup_{\pi_{(t+1):T} \in \Pi_{(t+1):T}} \E\left[\left.\left(\widehat{Q}_{t}^{\pi_{(t+1):T},-k}\left(H_{i,t},a_{t}\right)-Q_{t}^{\pi_{(t+1):T}}\left(H_{i,t},a_{t}\right)\right)^2\right|\MS_{-k} \right] = O_p(1).
\end{align*}
Combining these results, we obtain
\begin{align}
    \sup_{\pi_{(t+1):T} \in \Pi_{(t+1):T}}\sup_{\pi_{s:t}^{a},\pi_{s:t}^{b}\in \Pi_{s:t}}\left|\widetilde{S}_{s:t,(A1)}^{\underline{a}_{s:t},k}(\pi_{s:t}^{a},\pi_{s:t}^{b},\pi_{(t+1):T})\right| = O_p \left(\frac{1}{\sqrt{n}}\right). \label{eq:A1_convergence}
\end{align}
Consequently, 
\begin{align*}
    &\sup_{\pi_{(t+1):T} \in \Pi_{(t+1):T}}\sup_{\pi_{s:t}^{a},\pi_{s:t}^{b}\in \Pi_{s:t}}\left|\widetilde{S}_{s:t,(A1)}^{\underline{a}_{s:t}}(\pi_{s:t}^{a},\pi_{s:t}^{b},\pi_{(t+1):T})\right| \\
    &\leq  \sum_{k=1}^{K}\sup_{\pi_{(t+1):T} \in \Pi_{(t+1):T}}\sup_{\pi_{s:t}^{a},\pi_{s:t}^{b}\in \Pi_{s:t}}\left|\widetilde{S}_{s:t,(A1)}^{\underline{a}_{s:t},k}(\pi_{s:t}^{a},\pi_{s:t}^{b},\pi_{(t+1):T})\right| = O_p \left(\frac{1}{\sqrt{n}}\right).
\end{align*}

We next consider $\widetilde{S}_{s:t,(B)}^{\underline{a}_{s:t}}(\pi_{s:t}^{a},\pi_{s:t}^{b},\pi_{(t+1):T})$ (we will consider $\widetilde{S}_{s:t,(A2)}^{\underline{a}_{s:t}}(\pi_{s:t}^{a},\pi_{s:t}^{b},\pi_{(t+1):T})$ later). We begin by decomposing $\widetilde{S}_{s:t,(B)}^{\underline{a}_{s:t}}(\pi_{s:t}^{a},\pi_{s:t}^{b},\pi_{(t+1):T})$ as follows:
\begin{align*}
    \widetilde{S}_{s:t,(B)}^{\underline{a}_{s:t}}(\pi_{s:t}^{a},\pi_{s:t}^{b},\pi_{(t+1):T})=\sum_{k=1}^{K}\left(\widetilde{S}_{s:t,(B1)}^{\underline{a}_{s:t},k}(\pi_{s:t}^{a},\pi_{s:t}^{b},\pi_{(t+1):T})+\widetilde{S}_{s:t,(B2)}^{\underline{a}_{s:t},k}(\pi_{s:t}^{a},\pi_{s:t}^{b},\pi_{(t+1):T})\right),
\end{align*}
where
\begin{align*}
    \widetilde{S}_{s:t,(B1)}^{\underline{a}_{s:t},k}(\pi_{s:t}^{a},\pi_{s:t}^{b},\pi_{(t+1):T})& \equiv \frac{1}{n}\sum_{i\in I_k}G_{i,\pi_{s:t}^{a},\pi_{s:t}^{b}}^{\underline{a}_{s:t}}\left(\widetilde{V}_{i,t+1}^{\pi_{(t+2):T}}(\pi_{t+1}(H_{i,t+1}))-Q_{t}^{\pi_{(t+1):T}}\left(H_{i,t},a_{t}\right)\right)\\
    &\times \left(\frac{\prod_{\ell=s}^{t-1}\mathbf{1}\{A_{i,\ell}=a_{\ell}\}}{\prod_{\ell=s}^{t-1}e_{\ell}(H_{i,\ell},a_{\ell})}-\frac{\prod_{\ell=s}^{t-1}\mathbf{1}\{A_{i,\ell}=a_{\ell}\}}{\prod_{\ell=s}^{t-1}\hat{e}_{\ell}^{-k}(H_{i,\ell},a_{\ell})}\right)\frac{\mathbf{1}\left\{ A_{i,t}=a_{t}\right\} }{e_{t}\left(H_{i,t},a_{t}\right)};\\
    \widetilde{S}_{s:t,(B2)}^{\underline{a}_{s:t},k}(\pi_{s:t}^{a},\pi_{s:t}^{b},\pi_{(t+1):T})& \equiv \frac{1}{n}\sum_{i \in I_k}G_{i,\pi_{s:t}^{a},\pi_{s:t}^{b}}^{\underline{a}_{s:t}}\left(\widetilde{V}_{i,t+1}^{\pi_{(t+2):T}}(\pi_{t+1}(H_{i,t+1}))-Q_{t}^{\pi_{(t+1):T}}\left(H_{i,t},a_{t}\right)\right)\\
    &\times \left(\frac{\prod_{\ell=s}^{t}\mathbf{1}\{A_{i,\ell}=a_{\ell}\}}{\prod_{\ell=s}^{t}\hat{e}_{\ell}^{-k(i)}(H_{i,\ell},a_{\ell})}-\frac{\prod_{\ell=s}^{t}\mathbf{1}\{A_{i,\ell}=a_{\ell}\}}{\prod_{\ell=s}^{t}e_{\ell}(H_{i,\ell},a_{\ell})}\right).
\end{align*}

Fix $k \in \{1,\ldots,K\}$. As for $  \widetilde{S}_{s:t,(B1)}^{\underline{a}_{s:t},k}(\pi_{s:t}^{a},\pi_{s:t}^{b},\pi_{(t+1):T})$, taking the conditional expectation given the date $\MS_{-k}$ in the rest $k-1$ folds leads to 
\begin{align}
&\E\left[ \widetilde{S}_{s:t,(B1)}^{\underline{a}_{s:t},k}(\pi_{s:t}^{a},\pi_{s:t}^{b},\pi_{(t+1):T})\middle|\MS_{-k}\right] \nonumber \\
 	&=\E\left[\frac{1}{n/K}\sum_{i \in I_k}G_{i,\pi_{s:t}^{a},\pi_{s:t}^{b}}^{\underline{a}_{s:t}}\left(\widetilde{V}_{i,t+1}^{\pi_{(t+2):T}}(\pi_{t+1}(H_{i,t+1}))-Q_{t}^{\pi_{(t+1):T}}\left(H_{i,t},a_{t}\right)\right)\right. \nonumber \\
 	&\left.\times\left(\frac{\prod_{\ell=s}^{t-1}\mathbf{1}\{A_{i,\ell}=a_{\ell}\}}{\prod_{\ell=s}^{t-1}e_{\ell}(H_{i,\ell},a_{\ell})}-\frac{\prod_{\ell=s}^{t-1}\mathbf{1}\{A_{i,\ell}=a_{\ell}\}}{\prod_{\ell=s}^{t-1}\hat{e}_{\ell}^{-k}(H_{i,\ell},a_{\ell})}\right)\frac{\mathbf{1}\left\{ A_{i,t}=a_{t}\right\} }{e_{t}\left(H_{i,t},a_{t}\right)}\middle|\MS_{-k}\right] \nonumber \\
&=	\E\left[\frac{1}{n/K}\sum_{i \in I_k}G_{i,\pi_{s:t}^{a},\pi_{s:t}^{b}}^{\underline{a}_{s:t}}\E\left[\widetilde{V}_{i,t+1}^{\pi_{(t+2):T}}(\pi_{t+1}(H_{i,t+1}))-Q_{t}^{\pi_{(t+1):T}}\left(H_{i,t},a_{t}\right)\middle|H_{i,t},A_{i,t}=a_{t}\right]\right. \nonumber \\
&\left. \times \left(\frac{\prod_{\ell=s}^{t-1}\mathbf{1}\{A_{i,\ell}=a_{\ell}\}}{\prod_{\ell=s}^{t-1}e_{\ell}(H_{i,\ell},a_{\ell})}-\frac{\prod_{\ell=s}^{t-1}\mathbf{1}\{A_{i,\ell}=a_{\ell}\}}{\prod_{\ell=s}^{t-1}\hat{e}_{\ell}^{-k}(H_{i,\ell},a_{\ell})}\right)\frac{\mathbf{1}\left\{ A_{i,t}=a_{t}\right\} }{e_{t}\left(H_{i,t},a_{t}\right)}\middle|\MS_{-k}\right] \nonumber \\
&=	\E\left[\frac{1}{n/K}\sum_{i \in I_k}G_{i,\pi_{s:t}^{a},\pi_{s:t}^{b}}^{\underline{a}_{s:t}}\left(Q_{t}^{\pi_{(t+1):T}}\left(H_{i,t},a_{t}\right)-Q_{t}^{\pi_{(t+1):T}}\left(H_{i,t},a_{t}\right)\right)\right. \nonumber \\
&\left.\times \left(\frac{\prod_{\ell=s}^{t-1}\mathbf{1}\{A_{i,\ell}=a_{\ell}\}}{\prod_{\ell=s}^{t-1}e_{\ell}(H_{i,\ell},a_{\ell})}-\frac{\prod_{\ell=s}^{t-1}\mathbf{1}\{A_{i,\ell}=a_{\ell}\}}{\prod_{\ell=s}^{t-1}\hat{e}_{\ell}^{-k}(H_{i,\ell},a_{\ell})}\right)\frac{\mathbf{1}\left\{ A_{i,t}=a_{t}\right\} }{e_{t}\left(H_{i,t},a_{t}\right)}\middle|\MS_{-k}\right] \nonumber \\
&=	0, \nonumber
\end{align}
where the third equality follows from Lemma~\ref{lemma:identification} (i). Note that conditional on $\MS_{-k}$, $\widetilde{S}_{s:t,(B1)}^{\underline{a}_{s:t},k}(\pi_{s:t}^{a},\pi_{s:t}^{b},\pi_{(t+1):T})$ is a sum of i.i.d. bounded random variables under Assumptions~\ref{asm:bounded outcome}, \ref{asm:overlap}, and \ref{asm:rate_of_convergence_backward} (ii), and its conditional mean is zero.
Hence, fixing $\pi_{(t+1):T}$ and conditioning on $\MS_{-k}$, we can apply Lemma~\ref{lem:concentration inequality_influence difference function} with setting $i \in I_k$ and 
\begin{align*}
    \Gamma_{i}^{\dag}(\underline{a}_{s:t})&=	\left(\widetilde{V}_{i,t+1}^{\pi_{(t+2):T}}(\pi_{t+1}(H_{i,t+1}))-Q_{t}^{\pi_{(t+1):T}}\left(H_{i,t},a_{t}\right)\right) \\
&\times \left(\frac{\prod_{\ell=s}^{t-1}\mathbf{1}\{A_{i,\ell}=a_{\ell}\}}{\prod_{\ell=s}^{t-1}e_{\ell}(H_{i,\ell},a_{\ell})}-\frac{\prod_{\ell=s}^{t-1}\mathbf{1}\{A_{i,\ell}=a_{\ell}\}}{\prod_{\ell=s}^{t-1}\hat{e}_{\ell}^{-k}(H_{i,\ell},a_{\ell})}\right)\frac{\mathbf{1}\left\{ A_{i,t}=a_{t}\right\} }{e_{t}\left(H_{i,t},a_{t}\right)}
\end{align*}
to obtain the following: $\forall \delta > 0$, with probability at least $1-2\delta$,
\begin{align*}
    	&\sup_{\pi_{s:t}^{a},\pi_{s:t}^{b}\in\Pi_{s:t}}\left|\widetilde{S}_{s:t,(B1)}^{\underline{a}_{s:t},k}(\pi_{s:t}^{a},\pi_{s:t}^{b},\pi_{(t+1):T})\right| \\
&\leq	o\left(n^{-1/2}\right)+\left(54.4\kappa\left(\Pi_{s:t}\right)+435.2+\sqrt{2\log(1/\delta)}\right) \\
&\times	\left[\sup_{\pi_{s:T}\in\Pi_{t:T}}\E\left[\left(G_{i,\pi_{s:t}^{a},\pi_{s:t}^{b}}^{\underline{a}_{s:t}}\right)^{2}\cdot\left(\widetilde{V}_{i,t+1}^{\pi_{(t+2):T}}(\pi_{t+1}(H_{i,t+1}))-Q_{t}^{\pi_{(t+1):T}}\left(H_{i,t},a_{t}\right)\right)^{2}\right.\right.\\
&\left.\left. \left(\frac{\prod_{\ell=s}^{t-1}\mathbf{1}\{A_{i,\ell}=a_{\ell}\}}{\prod_{\ell=s}^{t-1}e_{\ell}(H_{i,\ell},a_{\ell})}-\frac{\prod_{\ell=s}^{t-1}\mathbf{1}\{A_{i,\ell}=a_{\ell}\}}{\prod_{\ell=s}^{t-1}\hat{e}_{\ell}^{-k}(H_{i,\ell},a_{\ell})}\right)^{2}\left(\frac{\mathbf{1}\left\{ A_{i,t}=a_{t}\right\} }{e_{t}\left(H_{i,t},a_{t}\right)}\right)^{2}\middle| \MS_{-k} \right]\middle/ \left(\frac{n}{K}\right)\right]^{1/2} \\
&\leq	o\left(n^{-1/2}\right)+\sqrt{K}\cdot\left(54.4\kappa\left(\Pi_{s:t}\right)+435.2+\sqrt{2\log(1/\delta)}\right)\cdot\left(\sum_{j=0}^{T-t}\frac{3M}{\eta^j}\right) \\
&\times	\sqrt{\frac{\E\left[\left(\frac{1}{\prod_{\ell=s}^{t-1}e_{\ell}(H_{i,\ell},a_{\ell})}-\frac{1}{\prod_{\ell=s}^{t-1}\hat{e}_{\ell}^{-k}(H_{i,\ell},a_{\ell})}\right)^{2}\middle|\MS_{-k} \right]}{n}},
\end{align*}
where the last inequality follows from $\left(G_{i,\pi_{s:t}^{a},\pi_{s:t}^{b}}^{\underline{a}_{s:t}}\right)^{2}\leq 1$ a.s. and Assumptions~\ref{asm:bounded outcome} and \ref{asm:overlap}. From Assumptions~\ref{asm:overlap} and \ref{asm:rate_of_convergence_backward} (ii), we have
\begin{align*}
    \E\left[\left(\frac{1}{\prod_{\ell=s}^{t-1}e_{\ell}(H_{i,\ell},a_{\ell})}-\frac{1}{\prod_{\ell=s}^{t-1}\hat{e}_{\ell}^{-k}(H_{i,\ell},a_{\ell})}\right)^{2}\right] < \infty.
\end{align*}
Hence, Markov's inequality leads to
\begin{align*}
    \E\left[\left(\frac{1}{\prod_{\ell=s}^{t-1}e_{\ell}(H_{i,\ell},a_{\ell})}-\frac{1}{\prod_{\ell=s}^{t-1}\hat{e}_{\ell}^{-k}(H_{i,\ell},a_{\ell})}\right)^{2}\middle|\MS_{-k} \right] = O_p(1).
\end{align*}
Note also that $\kappa(\Pi_{s:t})<\infty$ by Lemma~\ref{lem:entropy_integral_bound}. Combining these results, we have
\begin{align}
   \sup_{\pi_{(t+1):T}\in\Pi_{(t+1):T}}
    \sup_{\pi_{s:t}^{a},\pi_{s:t}^{b}\in\Pi_{s:t}}\left|\widetilde{S}_{s:t,(B1)}^{\underline{a}_{s:t},k}(\pi_{s:t}^{a},\pi_{s:t}^{b},\pi_{(t+1):T})\right| = O_p \left(\frac{1}{\sqrt{n}}\right).  \label{eq:B1_convergence}
\end{align}
By applying the same argument to derive (\ref{eq:B1_convergence}), we also obtain 
\begin{align*}
   \sup_{\pi_{(t+1):T}\in\Pi_{(t+1):T}}
    \sup_{\pi_{s:t}^{a},\pi_{s:t}^{b}\in\Pi_{s:t}}\left|\widetilde{S}_{s:t,(B2)}^{\underline{a}_{s:t},k}(\pi_{s:t}^{a},\pi_{s:t}^{b},\pi_{(t+1):T})\right| = O_p \left(\frac{1}{\sqrt{n}}\right).
\end{align*}
Consequently, 
\begin{align*}
    &\sup_{\pi_{(t+1):T}\in\Pi_{(t+1):T}}
    \sup_{\pi_{s:t}^{a},\pi_{s:t}^{b}\in\Pi_{s:t}}\left|\widetilde{S}_{s:t,(B)}^{\underline{a}_{s:t},k}(\pi_{s:t}^{a},\pi_{s:t}^{b},\pi_{(t+1):T})\right|\\
    &\leq  \sum_{k=1}^{K}\sup_{\pi_{(t+1):T}\in\Pi_{(t+1):T}}
    \sup_{\pi_{s:t}^{a},\pi_{s:t}^{b}\in\Pi_{s:t}}\left|\widetilde{S}_{s:t,(B1)}^{\underline{a}_{s:t},k}(\pi_{s:t}^{a},\pi_{s:t}^{b},\pi_{(t+1):T})\right|\\
    & + \sum_{k=1}^{K}\sup_{\pi_{(t+1):T}\in\Pi_{(t+1):T}}
    \sup_{\pi_{s:t}^{a},\pi_{s:t}^{b}\in\Pi_{s:t}}\left|\widetilde{S}_{s:t,(B2)}^{\underline{a}_{s:t},k}(\pi_{s:t}^{a},\pi_{s:t}^{b},\pi_{(t+1):T})\right|\\
    &= O_p \left(\frac{1}{\sqrt{n}}\right),
\end{align*}
which proves equation (\ref{eq:S_tilde_B_bound}).

We next bound $\sup_{\pi_{(t+1):T} \in  \Pi_{(t+1):T}}\sup_{\pi_{s:t}^{a},\pi_{s:t}^{b}\in \Pi_{s:t}} \left| \widetilde{S}_{s:t,(C)}^{\underline{a}_{s:t}}(\pi_{s:t}^{a},\pi_{s:t}^{b},\pi_{(t+1):T})\right|$ from above. 
It follows that 
\begin{align*}
&\sup_{\pi_{s:t}^{a},\pi_{s:t}^{b}\in \Pi_{s:t}} \left| \widetilde{S}_{s:t,(C)}^{\underline{a}_{s:t}}(\pi_{s:t}^{a},\pi_{s:t}^{b},\pi_{(t+1):T})\right|\\
	&=\sup_{\pi_{s:t}^{a},\pi_{s:t}^{b}\in \Pi_{s:t}} \left| \frac{1}{n}\sum_{i=1}^{n}G_{i,\pi_{s:t}^{a},\pi_{s:t}^{b}}^{\underline{a}_{s:t}}\frac{\prod_{\ell=s}^{t-1}\mathbf{1}\{A_{i,\ell}=a_{\ell}\}}{\prod_{\ell=s}^{t-1}\hat{e}_{\ell}^{-k(i)}(H_{i,\ell},a_{\ell})} 
\left(Q_{t}^{\pi_{(t+1):T}}\left(H_{i,t},a_{t}\right)-\widehat{Q}_{t}^{\pi_{(t+1):T},-k(i)}\left(H_{i,t},a_{t}\right)\right) \right. \notag \\ 
&\left. \times \left(\frac{\mathbf{1}\left\{ A_{i,t}=a_{t}\right\} }{\hat{e}_{t}^{-k(i)}\left(H_{i,t},a_{t}\right)}-\frac{\mathbf{1}\left\{ A_{i,t}=a_{t}\right\} }{e_{t}\left(H_{i,t},a_{t}\right)}\right) \right| \\
	&\leq\frac{1}{n}\sum_{i=1}^{n}\left|Q_{t}^{\pi_{(t+1):T}}\left(H_{i,t},a_{t}\right)-\widehat{Q}_{t}^{\pi_{(t+1):T},-k(i)}\left(H_{i,t},a_{t}\right)\right| \\
 &\times \left|\frac{1}{\prod_{\ell=s}^{t}\hat{e}_{\ell}^{-k(i)}(H_{i,\ell},a_{\ell})}-\frac{1}{e_{t}\left(H_{i,t},a_{t}\right)\cdot\prod_{\ell=s}^{t-1}\hat{e}_{\ell}^{-k(i)}(H_{i,\ell},a_{\ell})}\right|\\
	&\leq\frac{1}{n}\sum_{i=1}^{n}\left|Q_{t}^{\pi_{(t+1):T}}\left(H_{i,t},a_{t}\right)-\widehat{Q}_{t}^{\pi_{(t+1):T},-k(i)}\left(H_{i,t},a_{t}\right)\right| \cdot \left|\frac{1}{\prod_{\ell=s}^{t}\hat{e}_{\ell}^{-k(i)}(H_{i,\ell},a_{\ell})}-\frac{1}{\prod_{\ell=s}^{t}e_{\ell}(H_{i,\ell},a_{\ell})}\right|\\
	&+\frac{1}{n}\sum_{i=1}^{n}\left|Q_{t}^{\pi_{(t+1):T}}\left(H_{i,t},a_{t}\right)-\widehat{Q}_{t}^{\pi_{(t+1):T},-k(i)}\left(H_{i,t},a_{t}\right)\right| \\
 &\times \left|\frac{1}{\prod_{\ell=s}^{t-1}\hat{e}_{\ell}^{-k(i)}(H_{i,\ell},a_{\ell})}-\frac{1}{\prod_{\ell=s}^{t-1}e_{\ell}(H_{i,\ell},a_{\ell})}\right|\left(\frac{1}{e_t(H_{i,t},a_t)}\right)\\
	&\leq\sqrt{\frac{1}{n}\sum_{i=1}^{n}\left(Q_{t}^{\pi_{(t+1):T}}\left(H_{i,t},a_{t}\right)-\widehat{Q}_{t}^{\pi_{(t+1):T},-k(i)}\left(H_{i,t},a_{t}\right)\right)^{2}} \\
 &\times \sqrt{\frac{1}{n}\sum_{i=1}^{n}\left(\frac{1}{\prod_{\ell=s}^{t}\hat{e}_{\ell}^{-k(i)}(H_{i,\ell},a_{\ell})}-\frac{1}{\prod_{\ell=s}^{t}e_{\ell}(H_{i,\ell},a_{\ell})}\right)^{2}}\\
	&+\left(\frac{1}{\eta}\right)\sqrt{\frac{1}{n}\sum_{i=1}^{n}\left(Q_{t}^{\pi_{(t+1):T}}\left(H_{i,t},a_{t}\right)-\widehat{Q}_{t}^{\pi_{(t+1):T},-k(i)}\left(H_{i,t},a_{t}\right)\right)^{2}}\\ &\times \sqrt{\frac{1}{n}\sum_{i=1}^{n}\left(\frac{1}{\prod_{\ell=s}^{t-1}\hat{e}_{\ell}^{-k(i)}(H_{i,\ell},a_{\ell})}-\frac{1}{\prod_{\ell=s}^{t-1}e_{\ell}(H_{i,\ell},a_{\ell})}\right)^{2}}
\end{align*}
where the last inequality follows from the Cauchy--Schwarz inequality and Assumption \ref{asm:overlap} (overlap condition). Maximizing over $\Pi_{(t+1):T}$ and taking the expectation of both sides yields:
\begin{align*}
 & \E\left[\sup_{\pi_{(t+1):T}\in \Pi_{(t+1):T}}\sup_{\pi_{s:t}^{a},\pi_{s:t}^{b}\in \Pi_{s:t}} \left| \widetilde{S}_{s:t,(C)}^{\underline{a}_{s:t}}(\pi_{s:t}^{a},\pi_{s:t}^{b},\pi_{(t+1):T})\right|\right] \\
 &\leq	\E\left[\sup_{\pi_{(t+1):T}\in \Pi_{(t+1):T}}\sqrt{\frac{1}{n}\sum_{i=1}^{n}\left(Q_{t}^{\pi_{(t+1):T}}\left(H_{i,t},a_{t}\right)-\widehat{Q}_{t}^{\pi_{(t+1):T},-k(i)}\left(H_{i,t},a_{t}\right)\right)^{2}}\right. \\
&	\left.\times\sqrt{\frac{1}{n}\sum_{i=1}^{n}\left(\frac{1}{\prod_{\ell=s}^{t}\hat{e}_{\ell}^{-k(i)}\left(H_{i,\ell},a_{\ell}\right)}-\frac{1}{\prod_{\ell=s}^{t}e_{\ell}\left(H_{i,\ell},a_{\ell}\right)}\right)^{2}}\right] \\
& + \eta^{-1}\E\left[\sup_{\pi_{(t+1):T}\in \Pi_{(t+1):T}}\sqrt{\frac{1}{n}\sum_{i=1}^{n}\left(Q_{t}^{\pi_{(t+1):T}}\left(H_{i,t},a_{t}\right)-\widehat{Q}_{t}^{\pi_{(t+1):T},-k(i)}\left(H_{i,t},a_{t}\right)\right)^{2}}\right. \\
&	\left.\times\sqrt{\frac{1}{n}\sum_{i=1}^{n}\left(\frac{1}{\prod_{\ell=s}^{t-1}\hat{e}_{\ell}^{-k(i)}\left(H_{i,\ell},a_{\ell}\right)}-\frac{1}{\prod_{\ell=s}^{t-1}e_{\ell}\left(H_{i,\ell},a_{\ell}\right)}\right)^{2}}\right] \\
 &\leq	\sqrt{\frac{1}{n}\sum_{i=1}^{n}\E\left[\sup_{\pi_{(t+1):T}\in \Pi_{(t+1):T}}\left(Q_{t}^{\pi_{(t+1):T}}\left(H_{i,t},a_{t}\right)-\widehat{Q}_{t}^{\pi_{(t+1):T},-k(i)}\left(H_{i,t},a_{t}\right)\right)^{2}\right]} \\
&	\times\sqrt{\frac{1}{n}\sum_{i=1}^{n}\E\left[\left(\frac{1}{\prod_{\ell=s}^{t}\hat{e}_{\ell}^{-k(i)}\left(H_{i,\ell},a_{\ell}\right)}-\frac{1}{\prod_{\ell=s}^{t}e_{\ell}\left(H_{i,\ell},a_{\ell}\right)}\right)^{2}\right]} \\
&+ \eta^{-1}\sqrt{\frac{1}{n}\sum_{i=1}^{n}\E\left[\sup_{\pi_{(t+1):T}\in \Pi_{(t+1):T}}\left(Q_{t}^{\pi_{(t+1):T}}\left(H_{i,t},a_{t}\right)-\widehat{Q}_{t}^{\pi_{(t+1):T},-k(i)}\left(H_{i,t},a_{t}\right)\right)^{2}\right]} \\
&	\times\sqrt{\frac{1}{n}\sum_{i=1}^{n}\E\left[\left(\frac{1}{\prod_{\ell=s}^{t-1}\hat{e}_{\ell}^{-k(i)}\left(H_{i,\ell},a_{\ell}\right)}-\frac{1}{\prod_{\ell=s}^{t-1}e_{\ell}\left(H_{i,\ell},a_{\ell}\right)}\right)^{2}\right]} \\
&= \sum_{k=1}^{K}	\sqrt{\E\left[\sup_{\pi_{(t+1):T}\in \Pi_{(t+1):T}}\left(Q_{t}^{\pi_{(t+1):T}}\left(H_{t},a_{t}\right)-\widehat{Q}_{t}^{\pi_{(t+1):T},-k}\left(H_{i,t},a_{t}\right)\right)^{2}\right]} \\
&	\times\sqrt{\E\left[\left(\frac{1}{\prod_{\ell=s}^{t}\hat{e}_{\ell}^{-k}\left(H_{\ell},a_{\ell}\right)}-\frac{1}{\prod_{\ell=s}^{t}e_{\ell}\left(H_{\ell},a_{\ell}\right)}\right)^{2}\right]} \\
&+ \eta^{-1}\sum_{k=1}^{K}\sqrt{\E\left[\sup_{\pi_{(t+1):T}\in \Pi_{(t+1):T}}\left(Q_{t}^{\pi_{(t+1):T}}\left(H_{t},a_{t}\right)-\widehat{Q}_{t}^{\pi_{(t+1):T},-k}\left(H_{t},a_{t}\right)\right)^{2}\right]} \\
&	\times\sqrt{\E\left[\left(\frac{1}{\prod_{\ell=s}^{t-1}\hat{e}_{\ell}^{-k(i)}\left(H_{\ell},a_{\ell}\right)}-\frac{1}{\prod_{\ell=s}^{t-1}e_{\ell}\left(H_{\ell},a_{\ell}\right)}\right)^{2}\right]} \\
 &=	O\left(n^{-\tau/2}\right),
\end{align*}
where the second inequality follows from Cauchy-Schwartz inequality and the last line follows from Assumption \ref{asm:rate_of_convergence_backward} (i). Then applying Markov's inequality leads to
\begin{align}
   \sup_{\pi_{(t+1):T} \in \Pi_{(t+1):T}}\sup_{\pi_{s:t}^{a},\pi_{s:t}^{b}\in \Pi_{s:t}} \left| \widetilde{S}_{s:t,(C)}^{\underline{a}_{s:t}}(\pi_{s:t}^{a},\pi_{s:t}^{b},\pi_{(t+1):T})\right| &= O_{p}\left(n^{-\tau/2}\right), \label{eq:S_C_convergence}
\end{align}
which proves equation (\ref{eq:S_tilde_C_bound}).

Now let us consider $\widetilde{S}_{s:t,(A2)}^{\underline{a}_{s:t}}(\pi_{s:t}^{a},\pi_{s:t}^{b},\pi_{(t+1):T})$. Note that 
\begin{align*}
    	&\sup_{\pi_{(t+1):T}\in \Pi_{(t+1):T}}\sup_{\pi_{s:t}^{a},\pi_{s:t}^{b}\in \Pi_{s:t}}\left|\widetilde{S}_{s:t,(A2)}^{\underline{a}_{s:t}}(\pi_{s:t}^{a},\pi_{s:t}^{b},\pi_{(t+1):T})\right|\\
    	&\leq\sup_{\pi_{(t+1):T}\in \Pi_{(t+1):T}}\frac{1}{n}\sum_{i=1}^{n}\left|\frac{1}{\prod_{\ell=s}^{t-1}\hat{e}_{\ell}^{-k(i)}(H_{i,\ell},a_{\ell})}-\frac{1}{\prod_{\ell=s}^{t-1}e_{\ell}(H_{i,\ell},a_{\ell})}\right|\\
     &\times\left|\widehat{Q}_{t}^{\pi_{(t+1):T},-k(i)}\left(H_{i,t},a_{t}\right)-Q_{t}^{\pi_{(t+1):T}}\left(H_{i,t},a_{t}\right)\right|
    	\cdot \left|1-\frac{\mathbf{1}\left\{ A_{i,t}=a_{t}\right\} }{e_{t}\left(H_{i,t},a_{t}\right)}\right|\\
	&\leq\left(\frac{1}{\eta}\right)\sqrt{\frac{1}{n}\sum_{i=1}^{n}\left(\frac{1}{\prod_{\ell=s}^{t-1}\hat{e}_{\ell}^{-k(i)}(H_{i,\ell},a_{\ell})}-\frac{1}{\prod_{\ell=s}^{t-1}e_{\ell}(H_{i,\ell},a_{\ell})}\right)^{2}}\\
	&\times \sqrt{\frac{1}{n}\sum_{i=1}^{n}\sup_{\pi_{(t+1):T}\in \Pi_{(t+1):T}}\left(\widehat{Q}_{t}^{\pi_{(t+1):T},-k(i)}\left(H_{i,t},a_{t}\right)-Q_{t}^{\pi_{(t+1):T}}\left(H_{i,t},a_{t}\right)\right)^{2}},
\end{align*}
where the last inequality follows from Assumption \ref{asm:overlap} (overlap condition) and Cauchy-Schwartz inequality. Then, by applying the same argument to derive (\ref{eq:S_C_convergence}), we obtain
\begin{align*}
    \sup_{\pi_{(t+1):T} \in \Pi_{(t+1):T}}\sup_{\pi_{s:t}^{a},\pi_{s:t}^{b}\in \Pi_{s:t}} \left| \widetilde{S}_{s:t,(A2)}^{\underline{a}_{s:t}}(\pi_{s:t}^{a},\pi_{s:t}^{b},\pi_{(t+1):T})\right| &= O_{p}\left(n^{-\tau/2}\right).
\end{align*}
 Combining this result with (\ref{eq:A1_convergence}) leads to: 
 \begin{align*}
     &\sup_{\pi_{(t+1):T} \in \Pi_{(t+1):T}}\sup_{\pi_{s:t}^{a},\pi_{s:t}^{b}\in \Pi_{s:t}} \left| \widetilde{S}_{s:t,(A)}^{\underline{a}_{s:t}}(\pi_{s:t}^{a},\pi_{s:t}^{b},\pi_{(t+1):T})\right| \\
     &\leq \sum_{k=1}^{K} \sup_{\pi_{s:t}^a , \pi_{s:t}^b \in \Pi_{s:t}} \left| \widetilde{S}_{s:t,(A1)}^{\underline{a}_{s:t},k}\left(\pi_{s:t}\right)\right| 
     + \sup_{\pi_{s:t}^a , \pi_{s:t}^b \in \Pi_{s:t}} \left| \widetilde{S}_{s:t,(A2)}^{\underline{a}_{s:t}}\left(\pi_{s:t}\right)\right| \\
     &= O_{p}\left(n^{-\min\{1/2,\tau/2\}}\right).
 \end{align*}
 This result proves equation (\ref{eq:S_tilde_A_bound}). 

Consequently, combining equations (\ref{eq:Stilde_bound})--(\ref{eq:S_tilde_C_bound}), we obtain the result (\ref{eq:sequential_bound}).

We next consider the case that $t=T$. In this case, $\widetilde{S}_{t:T}^{\underline{a}_{t:T}}(\pi_{t:T}^{a},\pi_{t:T}^{b})$ is decomposed as
\begin{align*}
    &\widetilde{S}_{t:T}^{\underline{a}_{t:T}}(\pi_{t:T}^{a},\pi_{t:T}^{b}) = \widetilde{S}_{t:T,(A)}^{\underline{a}_{t:T}}(\pi_{t:T}^{a},\pi_{t:T}^{b}) 
    + \widetilde{S}_{t:T,(B)}^{\underline{a}_{t:T}}(\pi_{t:T}^{a},\pi_{t:T}^{b})  
    + \widetilde{S}_{t:T,(C)}^{\underline{a}_{t:T}}(\pi_{t:T}^{a},\pi_{t:T}^{b}),
\end{align*}
where 
\begin{align*}
\widetilde{S}_{t:T,(A)}^{\underline{a}_{t:T}}(\pi_{t:T}^{a},\pi_{t:T}^{b})& \equiv \frac{1}{n}\sum_{i=1}^{n}
G_{i,\pi_{t:T}^{a},\pi_{t:T}^{b}}^{\underline{a}_{t:T}}
\frac{\prod_{\ell=s}^{T-1}\mathbf{1}\{A_{i,\ell}=a_{\ell}\}}{\prod_{\ell=s}^{T-1}\hat{e}_{\ell}^{-k(i)}(H_{i,\ell},a_{\ell})} \nonumber \\
&\times\left(\widehat{Q}_{T}^{-k(i)}\left(H_{i,T},a_{T}\right)-Q_{T}\left(H_{i,T},a_{T}\right)\right) \left(1-\frac{\mathbf{1}\left\{ A_{i,T}=a_{T}\right\} }{e_{T}\left(H_{i,T},a_{T}\right)}\right); \ \\
\widetilde{S}_{t:T,(B)}^{\underline{a}_{t:T}}(\pi_{t:T}^{a},\pi_{t:T}^{b})& \equiv \frac{1}{n}\sum_{i=1}^{n}G_{i,\pi_{t:T}^{a},\pi_{t:T}^{b}}^{\underline{a}_{t:T}}\frac{\prod_{\ell=s}^{T-1}\mathbf{1}\{A_{i,\ell}=a_{\ell}\}}{\prod_{\ell=s}^{T-1}\hat{e}_{\ell}^{-k(i)}(H_{i,\ell},a_{\ell})} \nonumber \\
	&\times\left(Y_{i,T}-Q_{T}\left(H_{i,T},a_{T}\right)\right)  \left(\frac{\mathbf{1}\left\{ A_{i,T}=a_{T}\right\} }{\hat{e}_{T}^{-k(i)}\left(H_{i,T},a_{T}\right)}-\frac{\mathbf{1}\left\{ A_{i,T}=a_{T}\right\} }{e_{T}\left(H_{i,T},a_{T}\right)}\right);  \\
\widetilde{S}_{t:T,(C)}^{\underline{a}_{t:T}}(\pi_{t:T}^{a},\pi_{t:T}^{b})& \equiv \frac{1}{n}\sum_{i=1}^{n}G_{i,\pi_{t:T}^{a},\pi_{t:T}^{b}}^{\underline{a}_{t:T}}\frac{\prod_{\ell=s}^{T-1}\mathbf{1}\{A_{i,\ell}=a_{\ell}\}}{\prod_{\ell=s}^{T-1}\hat{e}_{\ell}^{-k(i)}(H_{i,\ell},a_{\ell})} \nonumber \\
&\times\left(Q_{T}\left(H_{i,T},a_{T}\right)-\widehat{Q}_{T}^{-k(i)}\left(H_{i,T},a_{T}\right)\right) 
\left(\frac{\mathbf{1}\left\{ A_{i,T}=a_{T}\right\} }{\hat{e}_{T}^{-k(i)}\left(H_{i,T},a_{T}\right)}-\frac{\mathbf{1}\left\{ A_{i,T}=a_{T}\right\} }{e_{T}\left(H_{i,T},a_{T}\right)}\right).
\end{align*}
The same arguments used to derive the results (\ref{eq:S_tilde_A_bound})--(\ref{eq:S_tilde_C_bound}) also show that
\begin{align*}
    \sup_{\pi_{t:T}^{a},\pi_{t:T}^{b}\in \Pi_{t:T}} \left| \widetilde{S}_{t:T,(A)}^{\underline{a}_{t:T}}(\pi_{t:T}^{a},\pi_{t:T}^{b})\right| &= O_{p}\left(n^{-1/2}\right); \\
    \sup_{\pi_{t:T}^{a},\pi_{t:T}^{b}\in \Pi_{t:T}} \left| \widetilde{S}_{t:T,(B)}^{\underline{a}_{t:T}}(\pi_{t:T}^{a},\pi_{t:T}^{b})\right| &= O_{p}\left(n^{-1/2}\right); \\
    \sup_{\pi_{t:T}^{a},\pi_{t:T}^{b}\in \Pi_{t:T}} \left| \widetilde{S}_{t:T,(C)}^{\underline{a}_{t:T}}(\pi_{t:T}^{a},\pi_{t:T}^{b})\right| &= O_{p}\left(n^{-\min\{1/2,\tau/2\}}\right). 
\end{align*}
Therefore, 
\begin{align*}
    &\sup_{\pi_{t:T}^{a},\pi_{t:T}^{b}\in \Pi_{t:T}} \left| \widetilde{S}_{t:T}^{\underline{a}_{t:T}}(\pi_{t:T}^{a},\pi_{t:T}^{b})\right| \\
    &\leq \sup_{\pi_{t:T}^{a},\pi_{t:T}^{b}\in \Pi_{t:T}} \left| \widetilde{S}_{t:T,(A)}^{\underline{a}_{t:T}}(\pi_{t:T}^{a},\pi_{t:T}^{b})\right| 
    + \sup_{\pi_{t:T}^{a},\pi_{t:T}^{b}\in \Pi_{t:T}} \left| \widetilde{S}_{t:T,(B)}^{\underline{a}_{t:T}}(\pi_{t:T}^{a},\pi_{t:T}^{b})\right| \\
    &+ \sup_{\pi_{t:T}^{a},\pi_{t:T}^{b}\in \Pi_{t:T}} \left| \widetilde{S}_{t:T,(C)}^{\underline{a}_{t:T}}(\pi_{t:T}^{a},\pi_{t:T}^{b})\right| \\
    &= O_{p}\left(n^{-\min\{1/2,\tau/2\}}\right), 
\end{align*}
which leads to the result (\ref{eq:sequential_bound}).
\end{proof}

\medskip

We finally present the proof of Lemma~\ref{lem:asymptotic_estimated_policy_difference_function}.
\medskip

\begin{proof}[Proof of  Lemma~\ref{lem:asymptotic_estimated_policy_difference_function}]
From equation (\ref{eq:latter_2_bound}), we have 
\begin{align*}
    &\sup_{\pi_{(t+1):T} \in \Pi_{(t+1):T}} \sup_{\pi_{t}^{a},\pi_{t}^{b} \in \Pi_{t}} |\check{\Delta}_{t,s}^{\dagger}(\pi_{t}^{a},\pi_{(t+1):T};\pi_{t}^{b},\pi_{(t+1):T}) - \widetilde{\Delta}_{t,s}^{\dagger}(\pi_{t}^{a},\pi_{(t+1):T};\pi_{t}^{b},\pi_{(t+1):T})| \nonumber \\
     &\leq \sum_{\underline{a}_{t:s} \in \underline{\MA}_{t:s}} \sup_{\pi_{(t+1):T} \in \Pi_{(t+1):T}} \sup_{\pi_{t}^{a},\pi_{t}^{b} \in \Pi_{t}} |\widetilde{S}_{t:s}^{\underline{a}_{t:s}}(\pi_{t}^{a},\pi_{(t+1):s},\pi_{t}^{b},\pi_{(t+1):s},\pi_{(s+1):T})|\\
      &\leq \sum_{\underline{a}_{t:s} \in \underline{\MA}_{t:s}} \sup_{\pi_{(s+1):T} \in \Pi_{(s+1):T}} \sup_{\pi_{t:s}^{a},\pi_{t:s}^{b} \in \Pi_{t:s}} |\widetilde{S}_{t:s}^{\underline{a}_{t:s}}(\pi_{t:s}^{a},\pi_{t:s}^{b},\pi_{(s+1):T})|.
\end{align*}
The result then follows from Lemma~\ref{lem:convergence_rate_Stilde}.
\end{proof}

\setcounter{equation}{0}
\renewcommand{\theequation}{C.\arabic{equation}}

\section{Proof of Theorem \ref{thm:theorem_simlutaneous}}
\label{app:statistical_property_AIPW}

This appendix provides the proof of Theorem \ref{thm:theorem_simlutaneous}.

\begin{proof}[Proof of Theorem \ref{thm:theorem_simlutaneous}]
We begin by noting that the objective function $\widehat{W}^{AIPW}(\pi)$ can be expressed as $\widehat{W}^{AIPW}(\pi) = (1/n)\sum_{i=1}^{n}\widehat{V}_{i,1}(\pi)$, where $\widehat{V}_{i,1}(\cdot)$ is defined in Appendix \ref{app:main_proof}. A standard argument from statistical learning theory \cite[e.g.,][]{Lugosi_2002} leads to
\begin{align}
    R(\hat{\pi}^{AIPW})  &=\Delta_{1}\left(\pi^{\ast,opt};\hat{\pi}^{AIPW}\right) 
    \nonumber\\
    &\leq \Delta_{1}\left(\pi^{\ast,opt};\hat{\pi}^{AIPW}\right) - \widehat{\Delta}_{1}\left(\pi^{\ast,opt};\hat{\pi}^{AIPW}\right) \nonumber \\
    &\leq 
    \sup_{\pi^{a},\pi^{b} \in \Pi} |\Delta_{1}(\pi^{a};\pi^{b}) - \widehat{\Delta}_{1}(\pi^{a};\pi^{b})| \nonumber \\
    & \leq   \sup_{\pi^{a},\pi^{b} \in \Pi}|\Delta_{1}(\pi^{a};\pi^{b}) - \widetilde{\Delta}_{1}(\pi^{a};\pi^{b})| 
       + \sup_{\pi^{a},\pi^{b} \in \Pi} |\widehat{\Delta}_{1}(\pi^{a};\pi^{b}) - \widetilde{\Delta}_{1}(\pi^{a};\pi^{b})|, \label{eq:basic_bound_AIPW}
\end{align}
where the first inequality follows because $\hat{\pi}^{AIPW}$ maximizes $(1/n)\sum_{i=1}^{n}\widehat{V}_{i,1}(\pi)$ over $\Pi$; hence, $\widehat{\Delta}_{1}\left(\pi^{\ast,opt};\hat{\pi}^{AIPW}\right) \leq 0$.

Regarding $\sup_{\pi^{a},\pi^{b} \in \Pi}|\Delta_{1}(\pi^{a};\pi^{b}) - \widetilde{\Delta}_{1}(\pi^{a};\pi^{b})|$, under Assumptions~\ref{asm:bounded outcome}, \ref{asm:overlap}, and the result in Lemma~\ref{lem:entropy_integral_bound}, we can apply Lemma~\ref{lem:concentration inequality_influence difference function} to obtain the following result: 
For any $\delta \in (0,1)$, with probability at least $1-2\delta$,
\begin{align}
    \sup_{\pi^{a},\pi^{b} \in \Pi} 
  |\widetilde{\Delta}_{1}(\pi^{a};\pi^{b})  - \Delta_{1}(\pi^{a};\pi^{b})| 
    &\leq \left(54.4 \sqrt{2}\kappa(\Pi) + 435.2 + \sqrt{2 \log \frac{1}{\delta}}\right)\sqrt{\frac{V^{\ast}}{n}} \nonumber \\
    &+ o\left(\frac{1}{\sqrt{n}}\right), \label{eq:bound_influence_difference_function_AIPW}
\end{align}
where $V^{\ast}  \equiv  \sup_{\pi^{a},\pi^{b} \in \Pi}\E
    \left[\left(\widetilde{\Gamma}_{i,1}^{\pi_{2:T}^{a}}(\pi^{a}) - \widetilde{\Gamma}_{i,1}^{\pi_{2:T}^{b}}(\pi^{b}) \right)^2\right]$.
By Assumptions~\ref{asm:bounded outcome} and
\ref{asm:overlap} and the recursive definition of
$\widetilde{\Gamma}_{i,t}$,
\begin{align}
V^{\ast}
\le
 M^2
\mathcal V_{1:T}(\boldsymbol{\eta}).
\end{align}

Regarding $\sup_{\pi^{a},\pi^{b} \in \Pi} |\widehat{\Delta}_{1}(\pi^{a};\pi^{b}) - \widetilde{\Delta}_{1}(\pi^{a};\pi^{b})|$, we have $\widehat{\Delta}_{1}(\pi^{a};\pi^{b}) - \widetilde{\Delta}_{1}(\pi^{a};\pi^{b}) = \sum_{\underline{a}_{T} \in \underline{\MA}_{T}} \widetilde{S}_{1:T}^{\underline{a}_{T}}(\pi^{a},\pi^{b})$.
Applying the result (\ref{eq:sequential_bound}) in Lemma~\ref{lem:convergence_rate_Stilde} to $\widetilde{S}_{1:T}^{\underline{a}_{T}}(\pi^{a},\pi^{b})$ for each $\underline{a}_{T}$ gives
\begin{align}
  \sup_{\pi^{a},\pi^{b} \in \Pi} |\widehat{\Delta}_{1}(\pi^{a};\pi^{b}) - \widetilde{\Delta}_{1}(\pi^{a};\pi^{b})|  =  O_{p}\left(n^{-\min\{1/2,\tau/2\}}\right). \label{eq:convergence_latter_AIPW}
\end{align}
Combining (\ref{eq:basic_bound_AIPW})--(\ref{eq:convergence_latter_AIPW}) leads to the result.
\end{proof}

\setcounter{equation}{0}
\renewcommand{\theequation}{D.\arabic{equation}}

\section{Example: Optimality and Suboptimality of Backward Induction}
\label{app:suboptimallity_backward_induction}

This appendix illustrates that Assumption \ref{asm:first-best} (i.e., the correct specification of $\Pi$) is a sufficient but not necessary condition for the optimality of the backward induction approach. Specifically, we provide two simple examples: one where backward induction results in suboptimality and another where it achieves optimality, both when Assumption \ref{asm:first-best} is not satisfied. These examples, which are adapted from \cite{Sakaguchi_2021}, consider a two-period setting ($T=2$) with binary treatments ($\mathcal{A}_{1} = \mathcal{A}_{2} = \{0,1\}$). 

For the first example, consider a DGP $P$ satisfying the following conditions:
\begin{align}
    & \E[Y_{2}(1,1)]=1.0,\ \E[Y_{2}(1,0)]=0.5,\ \E[Y_{2}(0,1)]=0.0,\ \E[Y_{2}(0,0)]=0.6; \nonumber \\
    & Y_1(0) = Y_1(1) = 0 \mbox{\ a.s.}; \quad \mbox{$A_{1}$ and $A_{2}$ are independently distributed as $\text{Ber}(1/2)$}. \label{eq:example_DGP}
\end{align}
The welfare $W(\pi)$ depends only on the second-stage outcomes. Suppose that $H_1 = \emptyset$ and $H_2 = (A_1)$.

As an example of a misspecified class of DTRs, we consider a class of uniform DTRs; that is, $\Pi_t = \{c_{t}^{0},c_{t}^{1}\}$ for $t=1,2$, where $c_{t}^{0}$ and $c_{t}^{1}$ denote constant functions such that $c_{t}^{0}(h_t)=0$ and $c_{t}^{1}(h_t)=1$ for any $h_t$.
Under the assumed DGP, $\Pi_2 = \{c_{2}^{0},c_{2}^{1}\}$ does not satisfy Assumption \ref{asm:first-best}, because $Q_2(1,c_{2}^{1}) > Q_2(1,c_{2}^{0})$ but $Q_2(0,c_{2}^{0}) > Q_2(0,c_{2}^{1})$.

The optimal DTR over the class of constant DTRs is
\begin{align*}
    (\pi_{1}^{\ast,opt},\pi_{2}^{\ast,opt}) = \argmax_{(\pi_1,\pi_2) \in \{c_{1}^{0},c_{1}^{1}\} \times \{c_{2}^{0},c_{2}^{1}\}} \E\left[Y_2(\pi_1(H_1),\pi_2(\pi_1(H_1)))\right] = (c_{1}^{1},c_{2}^{1}),
\end{align*}
and its welfare is $W(\pi_{1}^{\ast,opt},\pi_{2}^{\ast,opt})=\E[Y_2(1,1)]=1.0$.
However, the solution $(\pi_{1}^{\ast,B},\pi_{2}^{\ast,B})$ of the backward-induction approach is $(c_{1}^{0},c_{2}^{0})$ because
\begin{align*}
    &\mbox{(1st step)\ \ \ \ }\pi_{2}^{\ast,B} = \argmax_{\pi_2 \in \{c_{2}^{0},c_{2}^{1}\}}\E[Q_2(H_2,\pi_2)] = \argmax_{\pi_2 \in \{c_{2}^{0},c_{2}^{1}\}}\E\left[Y_2(A_1,\pi_2)\right] = c_{2}^{0};  \nonumber \\
    &\mbox{(2nd step)\ \ \ \ }\pi_{1}^{\ast,B} = \argmax_{\pi_1 \in \{c_{1}^{0},c_{1}^{1}\}}\E\left[Q_1^{\pi_{2}^{\ast,B}}(H_1,\pi_{1})\right] = \argmax_{\pi_1 \in \{c_{1}^{0},c_{1}^{1}\}}\E\left[Y_2(\pi_1,\pi_{2}^{\ast,B})\right]  = c_{1}^{0}.  
\end{align*}
Hence, the backward-induction solution $\pi^{\ast,B}=(c_{1}^{0},c_{2}^{0})$ differs from the optimal one $\pi^{\ast,opt}=(c_{1}^{1},c_{2}^{1})$ over $\Pi$, and yields suboptimal welfare $W(\pi^{\ast,B})=\E[Y_2(0,0)]=0.6$.
This simple example illustrates that when the DTR class $\Pi$ is not correctly specified, the backward-induction approach does not necessarily yield the optimal DTR.
\par 
Next, we illustrate that misspecification of $\Pi$ does not necessarily make backward induction
suboptimal of the backward-induction approach. Suppose that the DGP $P$ satisfies condition (\ref{eq:example_DGP}) with $\E[Y_2(0,1)] = 0.0$ replaced by $\E[Y_2(0,1)] = 0.4$, where $\Pi_2 = \{c_{2}^{0},c_{2}^{1}\}$ still does not satisfy Assumption \ref{asm:first-best}. In this case, the backward-induction solution becomes $\pi^{\ast,B} = (c_{1}^{1}, c_{2}^{1})$, whereas the optimal DTR $\pi^{\ast,\text{opt}}$ remains unchanged. Therefore, the backward-induction solution coincides with the optimal one, showing that the correct specification of $\Pi$ (Assumption \ref{asm:first-best}) is not a necessary condition for the optimality of the backward-induction approach.

\setcounter{equation}{0}
\renewcommand{\theequation}{E.\arabic{equation}}

\section{Convergence Rate for Sieve-Based Fitted Q-Evaluation}
\label{app:convergence_FQE}

\subsection{B-Spline Sieve Estimation and Primitive Conditions}

This appendix provides a fully worked example of FQE
based on B-spline series least-squares regression. In particular, we
provide primitive sufficient conditions under which
\eqref{eq:Qest_convergence_rate} holds.

Because the conditional mean of future outcomes often does not depend
on every component of $H_t$, we consider a setting in which the relevant Q-functions depend on the
history only through a low-dimensional summary. For each stage $t$, let
$g_t:\mathcal H_t\to\widetilde{\mathcal H}_t$ be a measurable map and, with some abuse of notation, define
$\widetilde{H}_t=g_t(H_t)$.
We refer to $\widetilde{H}_t$ as a sub-history. This notation includes, as a special case,
a subvector of $H_t$, obtained by taking $g_t$ to be the corresponding coordinate
projection.
Let $\tilde{d}_t$ denote the dimension of $\widetilde{H}_t$ and suppose that $\widetilde{H}_t$ has bounded support. Without loss of generality, we assume that $\widetilde{\mathcal H}_t\subseteq [0,1]^{\widetilde d_t}$.

 We assume that, for every $t=1,\ldots,T$, every
$a_t\in\mathcal A_t$, and every future policy sequence
$\pi_{(t+1):T}\in\Pi_{(t+1):T}$, there exists a measurable function
$q_t^{\pi_{(t+1):T}}$ such that
\begin{equation}
    Q_t^{\pi_{(t+1):T}}(H_t,a_t)
    =
    q_t^{\pi_{(t+1):T}}(\widetilde{H}_t,a_t)
    \quad\text{a.s.}
    \label{eq:E_subhistory_representation}
\end{equation}
This representation is imposed only as a dimension-reduction condition for the nuisance
Q-functions. In dynamic treatment problems, the full history $H_t$ grows with $t$, but
the predictive information relevant for future rewards may often be summarized by a
smaller state vector. The condition is not required for the regret theorem in the main
text; it is used here to obtain a transparent FQE convergence rate depending on the
effective dimension $\widetilde d_t$ rather than on the dimension of the full history.

Given an estimator $\widehat q_t^{\pi_{(t+1):T}}$, we write the corresponding estimator
of the original Q-function as
\begin{align*}
    \widehat Q_t^{\pi_{(t+1):T},(n)}(h_t,a_t)
    =
    \widehat q_t^{\pi_{(t+1):T}}(\tilde{h}_t,a_t),
\end{align*}
where $\tilde{h}_{t} = g_t(h_t)$.
To isolate the nonparametric rate calculation, we further restrict
attention to finite policy classes.\footnote{
The finite-class restriction is used only in this appendix to avoid empirical process
complications that are not central to the sieve FQE rate calculation. The main regret
theorem does not require finite policy classes. Extending
this primitive FQE rate verification to entropy-controlled infinite policy classes would
require additional uniform empirical process arguments for the policy-dependent
pseudo-outcomes and is beyond the scope of this appendix.
}
This finite-class setting covers many common implementations of interpretable policy
learning, where the analyst optimizes over a finite collection of candidate rules, such as
shallow trees with pre-specified splitting variables and candidate cutoffs.

We use an action-specific tensor-product B-spline sieve. Let $J_{n,t}$
denote the number of interior knots used in each coordinate direction at stage $t$.
Let $p_{t,n}(\widetilde h_t)\in\mathbb R^{\bar k_{n,t}}$ be the corresponding
normalized tensor-product B-spline basis on $[0,1]^{\widetilde d_t}$. In each coordinate
direction, the basis consists of univariate B-spline basis functions of order $m_t$ based
on quasi-uniform knots. Then the one-action tensor-product sieve dimension satisfies
\[
    \bar k_{n,t}
    :=
    (J_{n,t}+m_t)^{\widetilde d_t}
    \asymp
    J_{n,t}^{\widetilde d_t}.
\]
Define
\[
    \mathcal Q_{t,n}
    =
    \left\{
        q:
        q(\widetilde h_t,a_t)
        =
        p_{t,n}(\widetilde h_t)'\theta_{a_t},
        \quad
        \theta_{a_t}\in\mathbb R^{\bar k_{n,t}},
        \ a_t\in\mathcal A_t
    \right\},
\]
the class of action-specific tensor-product B-spline sieve functions.
Thus the sieve uses the same basis functions of the sub-history for all actions, while
allowing action-specific coefficients. Let $P_{t,n}(\widetilde h_t,a_t)$ be the
corresponding full action-specific block basis vector, so that every
$q\in\mathcal Q_{t,n}$ can be written as
\[
    q(\widetilde h_t,a_t)
    =
    P_{t,n}(\widetilde h_t,a_t)'\theta .
\]
The full action-specific sieve dimension is $k_{n,t}:=|\mathcal A_t|\bar k_{n,t}$.

The FQE estimator is defined recursively. Set $\widehat q_{T+1}^{\pi_{(T+2):T}}\equiv 0$.
For $t=T,\ldots,1$ and each $\pi_{(t+1):T}\in\Pi_{(t+1):T}$, define
\[
    G_{i,t}^{\pi_{(t+1):T}}
    =
    \begin{cases}
    Y_{i,t}
    +
    \widehat q_{t+1}^{\pi_{(t+2):T}}
    \left(
        \widetilde H_{i,t+1},
        \pi_{t+1}(H_{i,t+1})
    \right),
    & t<T, \\[0.7em]
    Y_{i,T},
    & t=T .
    \end{cases}
\]
Then
\begin{align*}
    \widehat q_t^{\pi_{(t+1):T}}
    \in
    \argmin_{q\in\mathcal Q_{t,n}}
    \frac1n
    \sum_{i=1}^n
    \left(
        G_{i,t}^{\pi_{(t+1):T}}
        -
        q(\widetilde H_{i,t},A_{i,t})
    \right)^2 .
\end{align*}
If the least-squares minimizer is not unique, take the minimum Euclidean-norm
coefficient solution.\footnote{Under the empirical Gram well-conditioning established below, the
minimizer is unique with probability approaching one.}

The following proposition establishes a uniform convergence rate for the resulting Q-function estimators. We impose an $L_2$-Sobolev smoothness condition rather than a classical H\"older condition. This formulation is useful because policy-dependent Q-functions may have limited classical differentiability, for example through kinks induced by discrete treatment rules. Sobolev smoothness, formulated in terms of weak derivatives, is well suited to such settings and to the $L_2(P)$ risk considered here.

\medskip

\begin{proposition}\label{prop:uniform_Q_rate_minimal}
Let $T$ be fixed. Suppose that Assumptions~\ref{asm:sequential independence} and \ref{asm:bounded outcome} hold. Suppose also that $\Pi_t$ is finite for all $t=1,\ldots,T$, with cardinalities not depending on $n$. Suppose that the sub-history
representation \eqref{eq:E_subhistory_representation}, the sieve-space definition, and
the FQE definition stated above hold. Suppose further that the following hold:

\begin{itemize}
    \item[(i)] For each $t=1,\ldots,T$, there exists $\beta_t\ge1$ such that
    $q_t^{\pi_{(t+1):T}}(\cdot,a_t)
    \in
    W_2^{\beta_t}\big([0,1]^{\widetilde d_t}\big)$
for every $a_t\in\mathcal A_t$ and every
$\pi_{(t+1):T}\in\Pi_{(t+1):T}$, where
$W_2^{\beta_t}\big([0,1]^{\widetilde d_t}\big)$ denotes the
$L_2$-Sobolev space of order $\beta_t$ on $[0,1]^{\widetilde d_t}$.
\item[(ii)]  The marginal distribution of $\widetilde{H}_t$ is absolutely continuous
with respect to Lebesgue measure on $[0,1]^{\widetilde d_t}$ and has a uniformly bounded density.
\item[(iii)] The
univariate splines have order $m_t$ satisfying $
    m_t\ge \lceil\beta_t\rceil+1$
and are constructed using quasi-uniform knots on $[0,1]$. Suppose also that $ (k_{n,t}\log n)/n\to0$.
\item[(iv)] There exists a constant $\eta_Q>0$ such that, for every
$t=1,\ldots,T$ and every $a_t\in\mathcal A_t$, $ e_t(H_t,a_t)\geq \eta_Q$ a.s.
\item[(v)] Define $ \Gamma_{t,n}^0
    =
    \mathbb E\left[
        p_{t,n}(\widetilde{H}_t)p_{t,n}(\widetilde{H}_t)'
    \right]$.
There exist constants $0<c_\Gamma<C_\Gamma<\infty$ such that
$c_\Gamma
    \le
    \lambda_{\min}(\Gamma_{t,n}^0)
    \le
    \lambda_{\max}(\Gamma_{t,n}^0)
    \le
    C_\Gamma$
uniformly over $t$ and $n$. 
\end{itemize}

Let $\mathbb E_0[\cdot]$ denote the conditional expectation $\E[\cdot | \MS_n]$ given the sample $\MS_n$. Then, for any $t=1,\ldots,T$,
\begin{align*}
&\sup_{a_t\in\mathcal A_t}
\mathbb E_0\left[
    \sup_{\pi_{(t+1):T}\in\Pi_{(t+1):T}}
    \left(
        \widehat Q_t^{\pi_{(t+1):T},(n)}(H_t,a_t)
        -
        Q_t^{\pi_{(t+1):T}}(H_t,a_t)
    \right)^2
\right]
=
O_p\left(
    \sum_{s=t}^{T}
    \left\{
        J_{n,s}^{-2\beta_s}
        +
        \frac{J_{n,s}^{\widetilde d_s}}{n}
    \right\}
\right).
\end{align*}

In particular, if
    $J_{n,s}\asymp n^{1/(2\beta_s+\widetilde d_s)}$
for every $s=t,\ldots,T$, then
\begin{align*}
&\sup_{a_t\in\mathcal A_t}
\mathbb E_0\left[
    \sup_{\pi_{(t+1):T}\in\Pi_{(t+1):T}}
    \left(
        \widehat Q_t^{\pi_{(t+1):T},(n)}(H_t,a_t)
        -
        Q_t^{\pi_{(t+1):T}}(H_t,a_t)
    \right)^2
\right]
 =
O_p\left(
    \sum_{s=t}^{T}
    n^{-2\beta_s/(2\beta_s+\widetilde d_s)}
\right).
\end{align*}
\end{proposition}

\begin{proof}
See Appendix~\ref{app:proof_uniform_Q_rate}.
\end{proof}

\medskip

Proposition~\ref{prop:uniform_Q_rate_minimal} shows that the uniform MSE bound for FQE has the usual sieve least-squares bias--variance form. The term $J_{n,s}^{-2\beta_s}$ is the squared approximation bias of the tensor-product spline sieve for a $\beta_s$-smooth Q-function, while $J_{n,s}^{\widetilde d_s}/n$ is the variance term. The rate depends on the effective dimension $\widetilde d_s$ of the sub-history rather than on the dimension of the full history $H_s$.

The bound also makes clear when the uniform Q-function MSE condition in
\eqref{eq:Qest_convergence_rate}, and hence the Q-function component of
Assumption~\ref{asm:rate_of_convergence_backward}(i), holds at the
$O_p(n^{-1/2})$ rate. For a generic choice of $J_{n,s}$, it is sufficient
that, for every $s=t,\ldots,T$,
\[
    J_{n,s}^{-2\beta_s}=O(n^{-1/2})
    \qquad\text{and}\qquad
    \frac{J_{n,s}^{\widetilde d_s}}{n}=O(n^{-1/2}).
\]
These two requirements are equivalent, up to constant factors, to choosing
$J_{n,s}$ such that
\[
    n^{1/(4\beta_s)}
    \lesssim
    J_{n,s}
    \lesssim
    n^{1/(2\widetilde d_s)}.
\]
Such a choice exists whenever $\beta_s\ge \widetilde d_s/2$.

In particular, under the MSE-optimal choice
$J_{n,s}\asymp n^{1/(2\beta_s+\widetilde d_s)}$, the MSE rate becomes $n^{-2\beta_s/(2\beta_s+\widetilde d_s)}$.
This rate is at least as fast as $n^{-1/2}$ if and only if $\beta_s\ge \widetilde d_s/2$, that is, if the Q-functions are sufficiently smooth in the Sobolev sense relative to the dimension of the sub-history.
Since $T$ is fixed, if this condition holds for all stages $s=1,\ldots,T$, Proposition~\ref{prop:uniform_Q_rate_minimal} implies the uniform Q-function MSE rate $O_p(n^{-1/2})$ at every stage.

\subsection{Proof of Proposition~\ref{prop:uniform_Q_rate_minimal}}
\label{app:proof_uniform_Q_rate}

\begin{proof}[Proof of Proposition~\ref{prop:uniform_Q_rate_minimal}]
For notational simplicity, throughout the proof we write
$q_t^\pi$ for $q_t^{\pi_{(t+1):T}}$ when the future policy sequence is clear.
Let $X_{i,t}:=(\widetilde H_{i,t},A_{i,t})$ and $P_{i,t}:=P_{t,n}(\widetilde H_{i,t},A_{i,t})$.
Throughout the proof, $\|\cdot\|$ denotes the Euclidean norm for vectors, and
$\|\cdot\|_{\mathrm{op}}$ denotes the corresponding operator norm for matrices.


For any measurable function $f$ of $(\widetilde{H}_t,A_t)$, possibly depending on $\MS_n$, define
\[
    \|f\|_{A,t,0}^2
    :=
    \mathbb E_0\left[
        f(\widetilde{H}_t,A_t)^2
    \right],
    \qquad
    \|f\|_{A,t,n}^2
    :=
    \frac1n\sum_{i=1}^n
        f(\widetilde H_{i,t},A_{i,t})^2 .
\]
For a fixed action $a_t\in\mathcal A_t$, define
\[
    \|f(\cdot,a_t)\|_{H,t,0}^2
    :=
    \mathbb E_0\left[
        f(\widetilde{H}_t,a_t)^2
    \right],
    \qquad
    \|f(\cdot,a_t)\|_{H,t,n}^2
    :=
    \frac1n\sum_{i=1}^n
        f(\widetilde H_{i,t},a_t)^2 .
\]
Let $C<\infty$ denote a generic constant that may change from line to line but does not
depend on $n$, the policy sequence, or the treatment arm. Since $T$, $|\mathcal A_t|$,
and the cardinalities of the policy classes are fixed, finite maxima over stages, actions,
and policy sequences do not affect stochastic rates.

Define
\[
    \mathcal R_t
    :=
    \sum_{s=t}^T
    \left\{
        J_{n,s}^{-2\beta_s}
        +
        \frac{k_{n,s}}{n}
    \right\},
    \qquad
    \mathcal R_{T+1}:=0 .
\]
Because $ k_{n,s}=|\mathcal A_s|\bar k_{n,s} \asymp J_{n,s}^{\widetilde d_s}$, this rate is equivalently
\[
    \mathcal R_t
    \asymp
    \sum_{s=t}^T
    \left\{
        J_{n,s}^{-2\beta_s}
        +
        \frac{J_{n,s}^{\widetilde d_s}}{n}
    \right\}.
\]

We prove the slightly stronger statement
\[
\begin{split}
    \mathcal E_t
    :=
    \max_{a_t\in\mathcal A_t}
    \max_{\pi_{(t+1):T}\in\Pi_{(t+1):T}}
    \Big[
        &
        \|
            \widehat q_t^{\pi_{(t+1):T}}(\cdot,a_t)
            -
            q_t^{\pi_{(t+1):T}}(\cdot,a_t)
        \|_{H,t,0}^2
        \\
        &+
        \|
            \widehat q_t^{\pi_{(t+1):T}}(\cdot,a_t)
            -
            q_t^{\pi_{(t+1):T}}(\cdot,a_t)
        \|_{H,t,n}^2
    \Big]
    =
    O_p(\mathcal R_t)
\end{split}
\]
for every $t=1,\ldots,T$. The proposition follows from this bound and the finiteness of
$\Pi_{(t+1):T}$.

First, consider the Gram matrices
\begin{align*}
    \Sigma_{t,n}
    :=
    \mathbb E[P_{i,t}P_{i,t}'],
    \qquad
    \widehat\Sigma_{t,n}
    :=
    \frac1n\sum_{i=1}^n P_{i,t}P_{i,t}' .
\end{align*}
Because the action-specific basis is block diagonal, the $a_t$-block of
$\Sigma_{t,n}$ is
\[
    \mathbb E\left[
        \mathbf 1\{A_t=a_t\}
        p_{t,n}(\widetilde{H}_t)p_{t,n}(\widetilde{H}_t)'
    \right]
    =
    \mathbb E\left[
        e_t(H_t,a_t)
        p_{t,n}(\widetilde{H}_t)p_{t,n}(\widetilde{H}_t)'
    \right].
\]
The overlap condition assumed in this proposition implies
$e_t(H_t,a_t)\ge \eta_Q$ for all $t$ and $a_t$. Hence
\[
    \lambda_{\min}(\Sigma_{t,n})
    \ge
    \eta_Q\lambda_{\min}(\Gamma_{t,n}^0)
    \ge
    \eta_Q c_\Gamma,
    \qquad
    \lambda_{\max}(\Sigma_{t,n})
    \le
    C_\Gamma .
\]
Moreover, by the boundedness of the normalized tensor-product B-spline basis,
\[
    \sup_i \|P_{i,t}\|^2
    =
    \sup_i \|p_{t,n}(\widetilde H_{i,t})\|^2
    \le
    C\bar k_{n,t}
    \le
    C k_{n,t}.
\]

Let $ \widetilde{X}_{i,t}
    :=
    P_{i,t}P_{i,t}'-\Sigma_{t,n}$.
Then $\|\widetilde{X}_{i,t}\|_{\mathrm{op}}\le Ck_{n,t}$. Moreover,
\[
\begin{aligned}
\left\|
    \sum_{i=1}^n \E[\widetilde{X}_{i,t}^2]
\right\|_{\mathrm{op}}
&\le
n\left\|
    \E\!\left[
        \|P_{i,t}\|^2P_{i,t}P_{i,t}'
    \right]
\right\|_{\mathrm{op}}  \\
&\le
Cnk_{n,t}\|\Sigma_{t,n}\|_{\mathrm{op}}
\le
Cnk_{n,t}.
\end{aligned}
\]
Hence, a standard matrix Bernstein inequality gives
\[
    \|\widehat\Sigma_{t,n}-\Sigma_{t,n}\|_{\mathrm{op}}
    =
    O_p\left(
        \sqrt{\frac{k_{n,t}\log n}{n}}
        +
        \frac{k_{n,t}\log n}{n}
    \right)
    =
    o_p(1),
\]
where the last equality follows from
$k_{n,t}\log n/n\to0$.

By Weyl's inequality,
\[
\lambda_{\min}(\widehat\Sigma_{t,n})
\ge
\lambda_{\min}(\Sigma_{t,n})
-
\|\widehat\Sigma_{t,n}-\Sigma_{t,n}\|_{\mathrm{op}},
\]
and
\[
\lambda_{\max}(\widehat\Sigma_{t,n})
\le
\lambda_{\max}(\Sigma_{t,n})
+
\|\widehat\Sigma_{t,n}-\Sigma_{t,n}\|_{\mathrm{op}}.
\]
Since $T$ is fixed and
$\|\widehat\Sigma_{t,n}-\Sigma_{t,n}\|_{\mathrm{op}}=o_p(1)$
for every $t$, it follows that, for every $t$, with probability approaching one,
\[
    0<c
    \le
    \lambda_{\min}(\widehat\Sigma_{t,n})
    \le
    \lambda_{\max}(\widehat\Sigma_{t,n})
    \le
    C.
\]

Applying the same argument to
\[
    \widehat\Gamma_{t,n}^0
    :=
    \frac1n\sum_{i=1}^n
    p_{t,n}(\widetilde H_{i,t})p_{t,n}(\widetilde H_{i,t})'
\]
shows that the empirical sub-history Gram matrix is also uniformly well-conditioned.
Consequently, on an event whose probability tends to one, for every
$f\in\mathcal Q_{t,n}$,
\begin{equation}
    \sum_{a_t\in\mathcal A_t}
    \|f(\cdot,a_t)\|_{H,t,0}^2
    +
    \sum_{a_t\in\mathcal A_t}
    \|f(\cdot,a_t)\|_{H,t,n}^2
    \le
    C\|f\|_{A,t,n}^2 .
    \label{eq:E_norm_comparison}
\end{equation}

Second, by the tensor-product B-spline approximation theorem (see, e.g., \cite{schumaker_1981}), the spline order condition
$m_t\ge\lceil\beta_t\rceil+1$ implies that, for every $a_t\in\mathcal A_t$ and every
$\pi_{(t+1):T}\in\Pi_{(t+1):T}$, there exists a spline approximation
$q_{t,n}^{\pi,o}\in\mathcal Q_{t,n}$ such that
\begin{equation}
    \max_{a_t\in\mathcal A_t}
    \max_{\pi_{(t+1):T}\in\Pi_{(t+1):T}}
    \|
        q_{t,n}^{\pi,o}(\cdot,a_t)
        -
        q_t^\pi(\cdot,a_t)
    \|_{H,t,0}^2
    \le
    C J_{n,t}^{-2\beta_t}.
    \label{eq:E_spline_approx_population}
\end{equation}
The bounded-density assumption converts the usual Lebesgue $L_2$ spline approximation
bound into the $L_2(P_{\widetilde{H}_t})$ bound above. Because the action sets and policy classes are finite and fixed, Markov's inequality also gives
\begin{equation}
    \max_{a_t\in\mathcal A_t}
    \max_{\pi_{(t+1):T}\in\Pi_{(t+1):T}}
    \|
        q_{t,n}^{\pi,o}(\cdot,a_t)
        -
        q_t^\pi(\cdot,a_t)
    \|_{H,t,n}^2
    =
    O_p(J_{n,t}^{-2\beta_t}).
    \label{eq:E_spline_approx_empirical}
\end{equation}

Third, fix $t$ and a future policy sequence $\pi_{(t+1):T}$. For $t<T$, define the
oracle pseudo-outcome
\[
    U_{i,t}^{\pi}
    :=
    Y_{i,t}
    +
    q_{t+1}^{\pi_{(t+2):T}}
    \left(
        \widetilde H_{i,t+1},
        \pi_{t+1}(H_{i,t+1})
    \right),
\]
and for $t=T$, define $U_{i,T}^{\pi}:=Y_{i,T}$. Let $U_t^\pi$ denote a generic copy of $U_{i,t}^\pi$. By the recursive definition of the
Q-function and the sub-history representation,
\[
    q_t^\pi(\widetilde{H}_t,a_t)
    =
    \mathbb E[
        U_t^\pi
        \mid
        \widetilde{H}_t,\ A_t=a_t
    ].
\]
Indeed,
\[
    \mathbb E[U_t^\pi\mid H_t,A_t=a_t]
    =
    Q_t^{\pi_{(t+1):T}}(H_t,a_t),
\]
and the latter is equal to $q_t^\pi(\widetilde{H}_t,a_t)$ by
\eqref{eq:E_subhistory_representation}. 

Let
$   \varepsilon_{i,t}^{\pi}
    :=
    U_{i,t}^{\pi}
    -
    q_t^\pi(\widetilde H_{i,t},A_{i,t})$.
Then
\begin{equation}
    \mathbb E[
        \varepsilon_{i,t}^{\pi}
        \mid
        \widetilde H_{i,t},A_{i,t}
    ]
    =
    0 .
    \label{eq:E_conditional_mean_zero}
\end{equation}
Assumption~\ref{asm:bounded outcome} also implies that all true Q-functions, and
therefore $\varepsilon_{i,t}^{\pi}$, are uniformly bounded.

The actual FQE pseudo-outcome can be written as
\[
    \widehat U_{i,t}^{\pi}
    =
    U_{i,t}^{\pi}
    +
    \Delta_{i,t+1}^{\pi},
\]
where
$
    \Delta_{i,t+1}^{\pi}
    :=
    \widehat q_{t+1}^{\pi_{(t+2):T}}
    \left(
        \widetilde H_{i,t+1},
        \pi_{t+1}(H_{i,t+1})
    \right)
    -
    q_{t+1}^{\pi_{(t+2):T}}
    \left(
        \widetilde H_{i,t+1},
        \pi_{t+1}(H_{i,t+1})
    \right),
$
with the convention $\Delta_{i,T+1}^{\pi}=0$.

Let
$
    b_{i,t}^{\pi}
    :=
    q_t^\pi(\widetilde H_{i,t},A_{i,t})
    -
    q_{t,n}^{\pi,o}(\widetilde H_{i,t},A_{i,t}).
$
On the Gram well-conditioning event, the least-squares minimizer is unique. 
For any scalar sequence $\{v_{i,t}\}_{i=1}^n$, write
$v_t=(v_{1,t},\ldots,v_{n,t})'$, and let $\Pi_{t,n}v_t$
denote its empirical least-squares projection onto
$\mathcal Q_{t,n}$.
Since
$\widehat q_t^\pi$ is the empirical least-squares projection of
$\widehat U_{i,t}^{\pi}$ onto $\mathcal Q_{t,n}$, the linearity of projection gives
\[
    \widehat q_t^\pi - q_{t,n}^{\pi,o}
    =
    \Pi_{t,n}
    \left(
        \varepsilon_t^\pi
        +
        b_t^\pi
        +
        \Delta_{t+1}^\pi
    \right).
\]
Since the empirical least-squares projection is non-expansive,
\begin{equation}
    \|
        \widehat q_t^\pi-q_{t,n}^{\pi,o}
    \|_{A,t,n}^2
    \le
    3
    \|\Pi_{t,n}\varepsilon_t^\pi\|_{A,t,n}^2
    +
    3
    \|b_t^\pi\|_{A,t,n}^2
    +
    3
    \|\Delta_{t+1}^\pi\|_n^2 ,
    \label{eq:E_basic_projection_bound}
\end{equation}
where
$
    \|\Delta_{t+1}^\pi\|_n^2
    :=
    (1/n)\sum_{i=1}^n
        (\Delta_{i,t+1}^\pi)^2 .
$

Let
$
    f_t^\pi
    :=
    \widehat q_t^\pi-q_{t,n}^{\pi,o}$
and
$\rho_t^\pi
    :=
    q_{t,n}^{\pi,o}-q_t^\pi.
$
Then $f_t^\pi\in\mathcal Q_{t,n}$ and
$\widehat q_t^\pi-q_t^\pi=f_t^\pi+\rho_t^\pi$. Combining
\eqref{eq:E_norm_comparison},
\eqref{eq:E_spline_approx_population},
\eqref{eq:E_spline_approx_empirical}, and
\eqref{eq:E_basic_projection_bound}, we obtain, uniformly over
$\pi_{(t+1):T}$,
\begin{align}
&
    \max_{a_t\in\mathcal A_t}
    \|
        \widehat q_t^\pi(\cdot,a_t)
        -
        q_t^\pi(\cdot,a_t)
    \|_{H,t,0}^2
    +
    \max_{a_t\in\mathcal A_t}
    \|
        \widehat q_t^\pi(\cdot,a_t)
        -
        q_t^\pi(\cdot,a_t)
    \|_{H,t,n}^2
    \nonumber\\
&\le
    C
    \left\{
        \sum_{a_t\in\mathcal A_t}
        \|f_t^\pi(\cdot,a_t)\|_{H,t,0}^2
        +
        \sum_{a_t\in\mathcal A_t}
        \|f_t^\pi(\cdot,a_t)\|_{H,t,n}^2
    \right\}
    +
    O_p(J_{n,t}^{-2\beta_t})
    \nonumber\\
&\le
    C\|f_t^\pi\|_{A,t,n}^2
    +
    O_p(J_{n,t}^{-2\beta_t})
    \nonumber\\
&=
    C
    \left\|
        \Pi_{t,n}
        \left(
            \varepsilon_t^\pi
            +
            b_t^\pi
            +
            \Delta_{t+1}^\pi
        \right)
    \right\|_{A,t,n}^2
    +
    O_p(J_{n,t}^{-2\beta_t})
    \nonumber\\
&\le
    C\|\Pi_{t,n}\varepsilon_t^\pi\|_{A,t,n}^2
    +
    C\|b_t^\pi\|_{A,t,n}^2
    +
    C\|\Delta_{t+1}^{\pi}\|_n^2
    +
    O_p(J_{n,t}^{-2\beta_t})
    \nonumber\\
&\le
    C\|\Pi_{t,n}\varepsilon_t^\pi\|_{A,t,n}^2
    +
    C\|\Delta_{t+1}^{\pi}\|_n^2
    +
    O_p(J_{n,t}^{-2\beta_t}).
    \label{eq:E_one_step_bound}
\end{align}
In the last inequality, we used
\[
    \|b_t^\pi\|_{A,t,n}^2
    \le
    \sum_{a_t\in\mathcal A_t}
    \|\rho_t^\pi(\cdot,a_t)\|_{H,t,n}^2
    =
    O_p(J_{n,t}^{-2\beta_t}).
\]

Fourth, we bound the projection of the regression noise. Let
$
    Z_{t,n}^{\pi}
    :=
    (1/n)
    \sum_{i=1}^n
    P_{i,t}\varepsilon_{i,t}^{\pi}.
$
Then
\[
    \|\Pi_{t,n}\varepsilon_t^\pi\|_{A,t,n}^2
    =
    (Z_{t,n}^{\pi})'
    \widehat\Sigma_{t,n}^{-1}
    Z_{t,n}^{\pi}.
\]
On the Gram well-conditioning event,
\[
    \|\Pi_{t,n}\varepsilon_t^\pi\|_{A,t,n}^2
    \le
    C\|Z_{t,n}^{\pi}\|^2 .
\]
By \eqref{eq:E_conditional_mean_zero},
$\mathbb E[Z_{t,n}^{\pi}]=0$. Since $\varepsilon_{i,t}^{\pi}$ is uniformly bounded and
$\|P_{i,t}\|^2\le C k_{n,t}$,
\[
    \mathbb E\|Z_{t,n}^{\pi}\|^2
    =
    \frac1n
    \mathbb E\left[
        (\varepsilon_{i,t}^{\pi})^2
        \|P_{i,t}\|^2
    \right]
    \le
    C\frac{k_{n,t}}{n}.
\]
Therefore,
\begin{equation}
    \|\Pi_{t,n}\varepsilon_t^\pi\|_{A,t,n}^2
    =
    O_p\left(\frac{k_{n,t}}{n}\right),
    \label{eq:E_noise_projection_rate}
\end{equation}
uniformly over $\pi_{(t+1):T}$, because $\Pi_{(t+1):T}$ is finite with cardinality not
depending on $n$.

Fifth, we control the propagation of future-stage estimation error. For $t=T$,
$\Delta_{T+1}^\pi=0$. For $t<T$,
\[
\begin{split}
    \|\Delta_{t+1}^\pi\|_n^2
    =
    \frac1n\sum_{i=1}^n
    \Big[
        &
        \widehat q_{t+1}^{\pi_{(t+2):T}}
        \left(
            \widetilde H_{i,t+1},
            \pi_{t+1}(H_{i,t+1})
        \right)
        \\
        &-
        q_{t+1}^{\pi_{(t+2):T}}
        \left(
            \widetilde H_{i,t+1},
            \pi_{t+1}(H_{i,t+1})
        \right)
    \Big]^2 .
\end{split}
\]
Since $\pi_{t+1}(H_{i,t+1})\in\mathcal A_{t+1}$,
\begin{equation}
    \|\Delta_{t+1}^\pi\|_n^2
    \le
    \sum_{a_{t+1}\in\mathcal A_{t+1}}
    \|
        \widehat q_{t+1}^{\pi_{(t+2):T}}(\cdot,a_{t+1})
        -
        q_{t+1}^{\pi_{(t+2):T}}(\cdot,a_{t+1})
    \|_{H,t+1,n}^2 .
    \label{eq:E_future_error_empirical_bound}
\end{equation}
Hence,
\begin{equation}
    \max_{\pi_{(t+1):T}\in\Pi_{(t+1):T}}
    \|\Delta_{t+1}^\pi\|_n^2
    \le
    C\mathcal E_{t+1}.
    \label{eq:E_future_error_bound}
\end{equation}

Combining \eqref{eq:E_one_step_bound},
\eqref{eq:E_noise_projection_rate}, and
\eqref{eq:E_future_error_bound} yields
\begin{equation}
    \mathcal E_t
    =
    O_p\left(
        J_{n,t}^{-2\beta_t}
        +
        \frac{k_{n,t}}{n}
        +
        \mathcal E_{t+1}
    \right).
    \label{eq:E_backward_recursion}
\end{equation}
At $t=T$, $\mathcal E_{T+1}=0$, so
\[
    \mathcal E_T
    =
    O_p\left(
        J_{n,T}^{-2\beta_T}
        +
        \frac{k_{n,T}}{n}
    \right).
\]
Applying \eqref{eq:E_backward_recursion} recursively backward from $T$ to $t$ gives
\[
    \mathcal E_t
    =
    O_p\left(
        \sum_{s=t}^T
        \left\{
            J_{n,s}^{-2\beta_s}
            +
            \frac{k_{n,s}}{n}
        \right\}
    \right).
\]
Since
\[
    k_{n,s}
    =
    |\mathcal A_s|\bar k_{n,s}
    \asymp
    J_{n,s}^{\widetilde d_s},
\]
and $|\mathcal A_s|$ is fixed, this is
\[
    \mathcal E_t
    =
    O_p\left(
        \sum_{s=t}^T
        \left\{
            J_{n,s}^{-2\beta_s}
            +
            \frac{J_{n,s}^{\widetilde d_s}}{n}
        \right\}
    \right).
\]

Finally, for a fixed $a_t\in\mathcal A_t$, the sub-history representation implies
\[
\begin{split}
&
\mathbb E_0\left[
    \sup_{\pi_{(t+1):T}\in\Pi_{(t+1):T}}
    \left\{
        \widehat Q_t^{\pi_{(t+1):T},(n)}(H_t,a_t)
        -
        Q_t^{\pi_{(t+1):T}}(H_t,a_t)
    \right\}^2
\right]
\\
&\qquad
=
\mathbb E_0\left[
    \sup_{\pi_{(t+1):T}\in\Pi_{(t+1):T}}
    \left\{
        \widehat q_t^{\pi_{(t+1):T}}(\widetilde{H}_t,a_t)
        -
        q_t^{\pi_{(t+1):T}}(\widetilde{H}_t,a_t)
    \right\}^2
\right].
\end{split}
\]
Because $\Pi_{(t+1):T}$ is finite with fixed cardinality,
\[
\begin{split}
&
\mathbb E_0\left[
    \sup_{\pi_{(t+1):T}\in\Pi_{(t+1):T}}
    \left\{
        \widehat q_t^{\pi_{(t+1):T}}(\widetilde{H}_t,a_t)
        -
        q_t^{\pi_{(t+1):T}}(\widetilde{H}_t,a_t)
    \right\}^2
\right]
\\
&\qquad
\le
|\Pi_{(t+1):T}|
\max_{\pi_{(t+1):T}\in\Pi_{(t+1):T}}
\|
    \widehat q_t^{\pi_{(t+1):T}}(\cdot,a_t)
    -
    q_t^{\pi_{(t+1):T}}(\cdot,a_t)
\|_{H,t,0}^2
=
O_p(\mathcal R_t).
\end{split}
\]
Taking the supremum over $a_t\in\mathcal A_t$ gives
\[
\begin{split}
&\sup_{a_t\in\mathcal A_t}
\mathbb E_0\left[
    \sup_{\pi_{(t+1):T}\in\Pi_{(t+1):T}}
    \left\{
        \widehat Q_t^{\pi_{(t+1):T},(n)}(H_t,a_t)
        -
        Q_t^{\pi_{(t+1):T}}(H_t,a_t)
    \right\}^2
\right]
=
O_p\left(
    \sum_{s=t}^T
    \left\{
        J_{n,s}^{-2\beta_s}
        +
        \frac{J_{n,s}^{\widetilde d_s}}{n}
    \right\}
\right).
\end{split}
\]
This proves the first claim.

For the final statement, if $J_{n,s}\asymp n^{1/(2\beta_s+\widetilde d_s)}$, then
\[
    J_{n,s}^{-2\beta_s}
    \asymp
    n^{-2\beta_s/(2\beta_s+\widetilde d_s)},
    \qquad
    \frac{J_{n,s}^{\widetilde d_s}}{n}
    \asymp
    n^{-2\beta_s/(2\beta_s+\widetilde d_s)}.
\]
Substituting these relations into the preceding bound gives
\[
\begin{split}
&\sup_{a_t\in\mathcal A_t}
\mathbb E_0\left[
    \sup_{\pi_{(t+1):T}\in\Pi_{(t+1):T}}
    \left\{
        \widehat Q_t^{\pi_{(t+1):T},(n)}(H_t,a_t)
        -
        Q_t^{\pi_{(t+1):T}}(H_t,a_t)
    \right\}^2
\right]
=
O_p\left(
    \sum_{s=t}^T
    n^{-2\beta_s/(2\beta_s+\widetilde d_s)}
\right).
\end{split}
\]
The proof is complete.
\end{proof}

\setcounter{equation}{0}
\setcounter{table}{0}
\renewcommand{\theequation}{\Alph{section}.\arabic{equation}}
\renewcommand{\thetable}{\Alph{section}.\arabic{table}}

\section{Additional Simulation Results}
\label{app:additional_simulation_results}

This appendix presents three additional simulation studies:
(i) misspecification of either the Q-functions or the propensity scores;
(ii) comparison with the simultaneous optimization approach under
linear policies; and
(iii) long treatment horizons and weak policy-relevant overlap.

\subsection{Misspecification}
\label{app:misspecification}

This appendix presents additional simulation results that examine the effects of misspecification of either the Q-functions or propensity scores. We use the same stylized DGPs 1 and 2 and the same DTR class as in Section \ref{sec:simulation}. For both the Q-functions and the propensity scores, we consider two types of estimators. The first type, as in Section \ref{sec:simulation}, is based on generalized random forests \citep{Athey_et_al_2019} and represents correctly specified models. The second type uses misspecified models. For Q-learning with and
without policy search, the Q-functions are estimated by linear
regression using the predictor sets
$(A_2,A_2\cdot H_2,H_2)$ and
$(A_1,A_1\cdot H_1,H_1)$ at stages 2 and 1, respectively. The
propensity scores $e_2$ and $e_1$ are estimated by probit regression
using $H_1$ as the predictor set in both cases.

Tables~\ref{table:simulatin_results_dgp1_miss} and~\ref{table:simulatin_results_dgp2_miss} report the results from 500 Monte Carlo replications for DGPs 1 and 2, respectively, using sample sizes $n = 250$, $500$, $1000$, $2000$, and $4000$. In each simulation, welfare and regret are evaluated on a test sample of 50,000 observations independently drawn from the same DGP.
The results indicate that Q-learning, Q-search, and IPW are adversely affected by misspecification of the nuisance functions -- often at larger sample sizes -- in both DGPs. In contrast, DR consistently outperforms the other methods, even when either the Q-functions or the propensity scores are misspecified. Its performance remains largely robust to such misspecification in each DGP. These findings underscore the doubly robust property of the proposed method.
\medskip

\begin{table}[htbp]
\centering
\caption{Monte Carlo simulation results for DGP1 under misspecification}
\label{table:simulatin_results_dgp1_miss}
\bigskip
\scalebox{0.82}{
\begin{tabular}{@{}llllccccc@{}}
\toprule
& \multicolumn{2}{c}{Specification}
& & \multicolumn{5}{c}{Sample Size} \\
\cmidrule(lr){2-3}
\cmidrule(lr){5-9}
Method
& Q-function
& PS
& Statistic
& 250
& 500
& 1000
& 2000
& 4000 \\
\midrule

\multirow{4}{*}{Q-learn}
& \multirow{2}{*}{correct}
& \multirow{2}{*}{--}
& Welfare (SD)
& 0.20 (0.12)
& 0.31 (0.11)
& 0.43 (0.10)
& 0.57 (0.07)
& 0.68 (0.03) \\
&&& Regret
& 0.54
& 0.43
& 0.31
& 0.17
& 0.06 \\

\cdashline{2-9}

& \multirow{2}{*}{miss}
& \multirow{2}{*}{--}
& Welfare (SD)
& 0.25 (0.09)
& 0.28 (0.07)
& 0.31 (0.05)
& 0.33 (0.04)
& 0.33 (0.03) \\
&&& Regret
& 0.49
& 0.46
& 0.43
& 0.41
& 0.41 \\

\midrule

\multirow{4}{*}{Q-search}
& \multirow{2}{*}{correct}
& \multirow{2}{*}{--}
& Welfare (SD)
& 0.17 (0.19)
& 0.29 (0.21)
& 0.40 (0.21)
& 0.53 (0.16)
& 0.64 (0.10) \\
&&& Regret
& 0.57
& 0.45
& 0.34
& 0.21
& 0.10 \\

\cdashline{2-9}

& \multirow{2}{*}{miss}
& \multirow{2}{*}{--}
& Welfare (SD)
& 0.23 (0.16)
& 0.23 (0.16)
& 0.23 (0.16)
& 0.22 (0.16)
& 0.18 (0.15) \\
&&& Regret
& 0.51
& 0.51
& 0.51
& 0.52
& 0.56 \\

\midrule

\multirow{4}{*}{IPW}
& \multirow{2}{*}{--}
& \multirow{2}{*}{correct}
& Welfare (SD)
& 0.24 (0.13)
& 0.29 (0.14)
& 0.37 (0.16)
& 0.45 (0.14)
& 0.55 (0.13) \\
&&& Regret
& 0.50
& 0.45
& 0.37
& 0.29
& 0.19 \\

\cdashline{2-9}

& \multirow{2}{*}{--}
& \multirow{2}{*}{miss}
& Welfare (SD)
& 0.24 (0.11)
& 0.29 (0.12)
& 0.32 (0.12)
& 0.39 (0.13)
& 0.44 (0.14) \\
&&& Regret
& 0.50
& 0.45
& 0.42
& 0.35
& 0.30 \\

\midrule

\multirow{6}{*}{DR}
& \multirow{2}{*}{correct}
& \multirow{2}{*}{miss}
& Welfare (SD)
& 0.24 (0.14)
& 0.37 (0.16)
& 0.54 (0.15)
& 0.67 (0.08)
& 0.72 (0.03) \\
&&& Regret
& 0.50
& 0.37
& 0.20
& 0.07
& 0.02 \\

\cdashline{2-9}

& \multirow{2}{*}{miss}
& \multirow{2}{*}{correct}
& Welfare (SD)
& 0.34 (0.17)
& 0.46 (0.15)
& 0.59 (0.11)
& 0.67 (0.04)
& 0.69 (0.03) \\
&&& Regret
& 0.40
& 0.28
& 0.15
& 0.07
& 0.05 \\

\cdashline{2-9}

& \multirow{2}{*}{correct}
& \multirow{2}{*}{correct}
& Welfare (SD)
& 0.29 (0.17)
& 0.44 (0.17)
& 0.61 (0.11)
& 0.70 (0.04)
& 0.72 (0.02) \\
&&& Regret
& 0.45
& 0.30
& 0.13
& 0.04
& 0.02 \\

\bottomrule
\end{tabular}
}
\smallskip

\begin{tablenotes}
\footnotesize
\item Notes: The columns under ``Specification'' indicate the
specifications of the Q-functions and propensity scores (PS),
where ``correct'' and ``miss'' denote correct specification and
misspecification, respectively. The rows labeled ``Welfare (SD)''
report the mean welfare, with the Monte Carlo standard deviation
in parentheses. Regret is calculated as the difference between
the optimal welfare, 0.74484, and the mean welfare of the estimated
policy. All regret values are rounded to two decimal places.
The results are based on 500 Monte Carlo replications using a test
sample of 50,000 observations randomly drawn from DGP1.
\end{tablenotes}
\end{table}

\begin{table}[htbp]
\centering
\caption{Monte Carlo simulation results for DGP2 under misspecification}
\label{table:simulatin_results_dgp2_miss}
\bigskip
\scalebox{0.82}{
\begin{tabular}{@{}llllccccc@{}}
\toprule
& \multicolumn{2}{c}{Specification}
& & \multicolumn{5}{c}{Sample Size} \\
\cmidrule(lr){2-3}
\cmidrule(lr){5-9}
Method
& Q-function
& PS
& Statistic
& 250
& 500
& 1000
& 2000
& 4000 \\
\midrule

\multirow{4}{*}{Q-learn}
& \multirow{2}{*}{correct}
& \multirow{2}{*}{--}
& Welfare (SD)
& 1.31 (0.36)
& 1.69 (0.11)
& 1.78 (0.08)
& 1.84 (0.05)
& 1.89 (0.05) \\
&&& Regret
& 0.75
& 0.37
& 0.28
& 0.22
& 0.17 \\

\cdashline{2-9}

& \multirow{2}{*}{miss}
& \multirow{2}{*}{--}
& Welfare (SD)
& 1.46 (0.07)
& 1.50 (0.05)
& 1.52 (0.04)
& 1.53 (0.03)
& 1.54 (0.03) \\
&&& Regret
& 0.60
& 0.56
& 0.54
& 0.53
& 0.52 \\

\midrule

\multirow{4}{*}{Q-search}
& \multirow{2}{*}{correct}
& \multirow{2}{*}{--}
& Welfare (SD)
& 1.26 (0.48)
& 1.58 (0.18)
& 1.54 (0.09)
& 1.52 (0.04)
& 1.54 (0.04) \\
&&& Regret
& 0.80
& 0.48
& 0.52
& 0.54
& 0.52 \\

\cdashline{2-9}

& \multirow{2}{*}{miss}
& \multirow{2}{*}{--}
& Welfare (SD)
& 1.44 (0.15)
& 1.44 (0.12)
& 1.44 (0.10)
& 1.43 (0.08)
& 1.42 (0.08) \\
&&& Regret
& 0.62
& 0.62
& 0.62
& 0.63
& 0.64 \\

\midrule

\multirow{4}{*}{IPW}
& \multirow{2}{*}{--}
& \multirow{2}{*}{correct}
& Welfare (SD)
& 1.30 (0.21)
& 1.42 (0.17)
& 1.56 (0.19)
& 1.68 (0.21)
& 1.81 (0.21) \\
&&& Regret
& 0.76
& 0.64
& 0.50
& 0.38
& 0.25 \\

\cdashline{2-9}

& \multirow{2}{*}{--}
& \multirow{2}{*}{miss}
& Welfare (SD)
& 1.32 (0.18)
& 1.41 (0.13)
& 1.49 (0.10)
& 1.51 (0.07)
& 1.52 (0.07) \\
&&& Regret
& 0.74
& 0.65
& 0.57
& 0.55
& 0.54 \\

\midrule

\multirow{6}{*}{DR}
& \multirow{2}{*}{correct}
& \multirow{2}{*}{miss}
& Welfare (SD)
& 1.44 (0.21)
& 1.63 (0.23)
& 1.78 (0.22)
& 1.87 (0.19)
& 1.99 (0.19) \\
&&& Regret
& 0.62
& 0.43
& 0.28
& 0.19
& 0.07 \\

\cdashline{2-9}

& \multirow{2}{*}{miss}
& \multirow{2}{*}{correct}
& Welfare (SD)
& 1.56 (0.23)
& 1.69 (0.25)
& 1.85 (0.23)
& 1.95 (0.16)
& 2.02 (0.16) \\
&&& Regret
& 0.50
& 0.37
& 0.21
& 0.11
& 0.04 \\

\cdashline{2-9}

& \multirow{2}{*}{correct}
& \multirow{2}{*}{correct}
& Welfare (SD)
& 1.51 (0.26)
& 1.77 (0.24)
& 1.95 (0.14)
& 2.01 (0.07)
& 2.04 (0.07) \\
&&& Regret
& 0.55
& 0.29
& 0.11
& 0.05
& 0.02 \\

\bottomrule
\end{tabular}
}
\smallskip

\begin{tablenotes}
\footnotesize
\item Notes: The columns under ``Specification'' indicate the
specifications of the Q-functions and propensity scores (PS),
where ``correct'' and ``miss'' denote correct specification and
misspecification, respectively. The rows labeled ``Welfare (SD)''
report the mean welfare, with the Monte Carlo standard deviation
in parentheses. Regret is calculated as the difference between
the optimal welfare, 2.062189, and the mean welfare of the estimated
policy. All regret values are rounded to two decimal places.
The results are based on 500 Monte Carlo replications using a test
sample of 50,000 observations randomly drawn from DGP2.
\end{tablenotes}
\end{table}

\newpage

\subsection{Computational Comparison with Simultaneous Optimization under Linear Policies}
\label{app:linear_computational_comparison}

This subsection provides a simulation comparison between the proposed
backward-induction approach and the simultaneous AIPW optimization
approach discussed in Section~\ref{sec:existing_approach}. The purpose of
this exercise is to isolate the computational consequences of stagewise
versus joint policy optimization. The simultaneous approach directly
maximizes an estimated welfare criterion over the DTR class and thus
jointly searches over the product of the stage-specific policy classes.
By contrast, the proposed approach solves a sequence of stage-specific
optimization problems.

To focus on differences in policy-optimization costs, we use the
heuristic simultaneous implementation described in
Footnote~\ref{fn:heuristic_simultaneous}, which avoids estimating
policy-specific Q-functions for every candidate DTR. The comparison
therefore abstracts from the additional computational burden associated
with the general simultaneous AIPW implementation and is, in this sense,
favorable to the simultaneous approach.

We consider a two-stage binary-treatment setting ($\MA_1=\MA_2=\{0,1\}$). The state variables and potential outcomes are generated according to
\begin{align*}
&(S_1^{(1)},\ldots,S_1^{(20)})^\prime \sim N(0,I_{20}),\
S_2(a_1)
=
\phi_{1,1}
+\phi_{1,2}S_{1}^{(1)}
+\left(\phi_{1,3}+\phi_{1,4}S_{1}^{(2)}\right)a_1
+U_1,\\
&Y_2(a_1,a_2)
=
\phi_{2,1}
+\phi_{2,2}S_2(a_1)
+
\left(
\phi_{2,3}
+\phi_{2,4}a_1
+\phi_{2,5}S_2(a_1)
\right)a_2
+S_{1}^{(4)}-\bigl(S_{1}^{(5)}\bigr)^2+S_{1}^{(6)}+U_2,
\end{align*}
where $(\phi_{1,1},\ldots,\phi_{1,4})=(0.5,-0.7,0.8,2.0)$,
$(\phi_{2,1},\ldots,\phi_{2,5})=(0.5,0.3,-0.2,1.0,0.8)$,
and $U_1\sim N(0,1)$ and $U_2\sim N(0,1)$.

Treatments are assigned according to the logistic propensity scores:
\begin{align*}
A_1
&\sim
\operatorname{Bernoulli}
\left(
\frac{1}{
1+\exp\left(0.5S_{1}^{(2)}-0.5S_{1}^{(3)}-0.5S_{1}^{(5)}\right)
}
\right),\\
A_2
&\sim
\operatorname{Bernoulli}
\left(
\frac{1}{
1+\exp\left(0.5S_{1}^{(5)}+0.5S_2-0.3A_1\right)
}
\right).
\end{align*}

We consider the class of DTRs $\Pi=\Pi_1\times\Pi_2$, where each stage-specific policy class consists of linear decision rules:
\begin{align*}
\Pi_1
&=
\left\{
\I\left\{\alpha_1+\beta_1 s_{1}^{(1)}\geq 0\right\}
:
(\alpha_1,\beta_1)\in\Real^2
\right\},\\
\Pi_2
&=
\left\{
\I\left\{\alpha_2+\beta_{2,1}a_1 + \beta_{2,2}s_2 \geq 0\right\}
:
(\alpha_2,\beta_{2,1},\beta_{2,2})\in\Real^3
\right\}.
\end{align*}

This design facilitates a clean computational comparison because both the stage-specific optimization problems of the proposed approach and the joint optimization problem of the simultaneous approach can be formulated
as mixed-integer linear programs (see the supplement to \citet{Sakaguchi_2021} for details). The resulting comparison therefore
highlights the difference in computational cost between optimizing
sequentially over the stage-specific policy classes and jointly over
their product.

For each Monte Carlo replication, we estimate the DTR using five
methods: the proposed backward-induction doubly robust estimator
(DR-BI), the simultaneous AIPW estimator (DR-Sim), Q-learn, Q-search,
and IPW. We use 5-fold cross-fitting for both DR-BI and DR-Sim. We consider sample sizes $n\in\{250,500,1000,2000\}$.

Table~\ref{table:mc_dgp1_linear_integrated} presents the simulation results. The results are based on 500 Monte Carlo replications, and welfare and regret are evaluated using an independent test sample generated from the same DGP. DR-BI and DR-Sim attain very similar welfare and regret across all sample sizes. For example, when $n=2000$, both estimators have regret of approximately $0.01$.

Their computational burdens, however, differ substantially. Across the sample sizes considered, DR-Sim requires roughly four to five times as much computation time as DR-BI. At $n=2000$, for example, the average computation time is $132.63$ seconds for DR-BI and $681.29$ seconds for DR-Sim. Thus, simultaneous optimization is considerably more
computationally demanding even in this relatively favorable setting with only two stages and low-dimensional linear policy classes.

This numerical pattern reflects the same computational distinction as the formal comparison in Remark~\ref{rem:computational_burden}. Under direct enumeration of finite policy classes, if the stage-$t$ policy class contains $L_t$ candidate policies, the backward approach evaluates $\sum_{t=1}^{T}L_t$ stage-specific candidates, whereas simultaneous optimization evaluates $\prod_{t=1}^{T}L_t$ candidate DTRs. Although the optimization problems in the present simulation are solved by mixed-integer linear programming rather than direct enumeration, the same distinction between stagewise and joint optimization contributes to their computational burden. This distinction suggests that the computational
gap may become larger as the treatment horizon or the complexity of the stage-specific policy classes increases. Accordingly, the present two-stage linear design provides a relatively conservative illustration of the computational advantage of backward optimization.

\begin{table}[!htbp]
\centering
\caption{Monte Carlo results under linear specifications: welfare, regret, and computation time}
\label{table:mc_dgp1_linear_integrated}
\begin{threeparttable}
\setlength{\tabcolsep}{4.5pt}
\begin{tabular}{llcccc}
\toprule
Method & Metric
& \multicolumn{4}{c}{Sample size $(n)$} \\
\cmidrule(lr){3-6}
& & 250 & 500 & 1000 & 2000 \\
\midrule
\multirow{3}{*}{Q-learn}
& Welfare (SD) & 1.82 (0.45) & 2.12 (0.24) & 2.30 (0.07) & 2.33 (0.02) \\
& Regret       & -- & -- & -- & -- \\
& Time (sec.)  & 0.28 & 0.60 & 1.25 & 2.78 \\
\addlinespace[0.3em]

\multirow{3}{*}{Q-search}
& Welfare (SD) & 1.74 (0.69) & 1.94 (0.56) & 2.12 (0.36) & 2.22 (0.12) \\
& Regret       & 0.50 & 0.30 & 0.13 & 0.03 \\
& Time (sec.)  & 0.78 & 1.73 & 4.99 & 33.50 \\
\addlinespace[0.3em]

\multirow{3}{*}{IPW}
& Welfare (SD) & 1.91 (0.35) & 2.08 (0.20) & 2.17 (0.10) & 2.22 (0.02) \\
& Regret       & 0.34 & 0.17 & 0.07 & 0.02 \\
& Time (sec.)  & 1.39 & 5.91 & 24.15 & 177.43 \\
\addlinespace[0.3em]

\multirow{3}{*}{DR-BI}
& Welfare (SD) & 1.95 (0.33) & 2.12 (0.16) & 2.20 (0.06) & 2.23 (0.02) \\
& Regret       & 0.29 & 0.12 & 0.04 & 0.01 \\
& Time (sec.)  & 2.40 & 7.32 & 26.79 & 132.63 \\
\addlinespace[0.3em]

\multirow{3}{*}{DR-Sim}
& Welfare (SD) & 1.89 (0.36) & 2.10 (0.16) & 2.19 (0.07) & 2.23 (0.02) \\
& Regret       & 0.35 & 0.14 & 0.05 & 0.01 \\
& Time (sec.)  & 9.83 & 33.52 & 118.21 & 681.29 \\
\bottomrule
\end{tabular}
\begin{tablenotes}\footnotesize
\item Notes: Welfare, regret, and computation time are averaged over 500 Monte Carlo replications.
Standard deviations of welfare are reported in parentheses. Regret is defined as
$R_\Pi(\hat\pi)=W(\pi_\Pi^\ast)-W(\hat\pi)$, where $\pi_\Pi^\ast$ is the oracle welfare-maximizing DTR within the policy class $\Pi$.
DR-BI denotes the proposed backward-induction doubly robust estimator, while DR-Sim denotes the simultaneous AIPW estimator optimized jointly over the full DTR class.
Time reports the average elapsed computation
time per replication in seconds. Simulations were conducted on a machine with a 13th Gen Intel(R) Core(TM) i9-13900 (2.00 GHz) processor and 64 GB of RAM. Monte
Carlo replications were not parallelized. The MILP optimization problems
were solved using Gurobi (version 12.0.1) with its default threading settings.
\end{tablenotes}
\end{threeparttable}
\end{table}

\newpage
\subsection{Long Horizons and Weak Policy-Relevant Overlap}
\label{app:large_T_simulation}
\label{sec:simulation_long_overlap}

This appendix presents an additional simulation study that examines the finite-sample performance of the proposed method when the treatment horizon is long and policy-relevant overlap is weak. The purpose of the simulation is to complement the theoretical discussion on the dependence of the regret bound on the number of stages and the overlap constant. We vary both the horizon length and the severity of weak overlap while keeping the training sample
size fixed at \(n=500\).

\subsubsection{DGP}

We consider binary treatment at each stage,
\(A_t\in\{0,1\}\), and let $T\in\{2,4,6,8,10\}$.
The state vector is
\(S_t=(S_t^{(1)},\ldots,S_t^{(p)})\), with \(p=10\). The initial state
is generated as $S_1\sim N(0,I_p)$.
For \(t<T\), the state evolves according to
\[
S_{t+1}
=
0.5S_t
+
0.25(2A_t-1)e_3
+
\varepsilon_{t+1},
\]
where \(e_3\) denotes the third standard basis vector in $\Real^p$ and
\(\varepsilon_{t+1}\sim N(0,I_p)\) independently over time and across
individuals. Thus, treatment affects the third component of the subsequent state.

To isolate the effects of long horizons and weak overlap from policy-class
misspecification, we specify the DGP so that the oracle stage-specific
policy is representable by the depth-2 policy-tree class used by the
proposed method, IPW, and Q-search.
Define $\pi_t^\ast(S_t)
=
\I\{S_t^{(1)}>0,\ S_t^{(2)}>0\}$
and
$\tau(S_t)
=
2\pi_t^\ast(S_t)-1$.
All intermediate outcomes are set equal to zero, and the final outcome
is generated according to
\[
Y_T
=
\sum_{t=1}^T
\gamma^{T-t}(2A_t-1)\tau(S_t)
+
0.5S_T^{(3)}
+
3S_1^{(4)}
+
\varepsilon_Y,
\]
where \(\gamma=0.8\) and
\(\varepsilon_Y\sim N(0,1)\).

The treatment-induced shift in \(S_{t+1}^{(3)}\) generates an additional
dynamic effect through the terminal component \(0.5S_T^{(3)}\), but its
magnitude is chosen so that it does not alter the sign of the
stage-specific treatment effect. Hence, the oracle policy remains
$
\pi_t^\ast(S_t)
=
1\{S_t^{(1)}>0,\ S_t^{(2)}>0\},
$
and therefore belongs to the depth-2 tree class used by the tree-based
policy-learning methods.

To control policy-relevant overlap directly, treatment is assigned
according to
\[
P(A_t=1\mid H_t)
=
\begin{cases}
\eta,
&
\pi_t^\ast(S_t)=1,\\[1mm]
1/2,
&
\pi_t^\ast(S_t)=0,
\end{cases}
\]
where $\eta\in\{0.50,0.20,0.05\}$.
Thus, in the region where treatment 1 is optimal, the optimal action is
observed with probability $\eta$, so decreasing $\eta$ weakens
policy-relevant overlap by making the oracle-prescribed action
increasingly unlikely in this region. Because the probability of
observing a treatment trajectory compatible with a dynamic policy
involves products of stage-specific propensity scores, weak overlap at
individual stages can result in very small probabilities of observing
policy-compatible trajectories as the treatment horizon increases.

\subsubsection{Methods and implementation}

We compare Q-learning, Q-search, IPW, and the proposed doubly robust
method (DR). For numerical stability, the baseline DR estimator clips
the estimated stage-specific propensity scores using a threshold of
$c=0.01$. We additionally consider DR estimators with clipping
thresholds $c=0.025$ and $c=0.05$, where
\[
\widehat e_{t,c}(h_t,1)
=
\min\bigl\{
    \max\{\widehat e_t(h_t,1),c\},
    1-c
\bigr\}.
\]
We set
$\widehat e_{t,c}(h_t,0)
=
1-\widehat e_{t,c}(h_t,1)$.
The IPW estimator also uses the baseline clipping threshold $c=0.01$.

For DR, IPW, and Q-search, the stage-specific policy class consists of
depth-2 decision trees based on the two current policy-relevant state
variables $(S_t^{(1)},S_t^{(2)})$. All nuisance functions are estimated
using generalized random forests. For DR, both the Q-functions and the
propensity scores are cross-fitted using five folds.

For each pair $(T,\eta)$, we conduct 500 Monte Carlo replications.
In each replication, the learned DTRs are evaluated on an independently
generated test sample of 50,000 observations.

\subsubsection{Results}

To make regret comparable across different horizon lengths, we report
normalized regret. Let $\pi_t^{\mathrm{opp}}(S_t)
=
1-\pi_t^\ast(S_t)$
denote the policy that takes the action opposite to the oracle action
at every stage. We define the normalized regret as
\[
R^{\mathrm{norm}}(\widehat\pi)
=
\frac{
W(\pi^\ast)-W(\widehat\pi)
}{
W(\pi^\ast)-W(\pi^{\mathrm{opp}})
}.
\]
This normalization scales regret by the horizon-specific welfare gap
between the oracle policy and its opposite-action benchmark, facilitating
comparisons across different values of $T$.

Table~\ref{table:mc_long_horizon_overlap} reports mean normalized
regret, with Monte Carlo standard deviations in parentheses, and the
average computation time per replication.
The results reveal several clear patterns. First, normalized regret generally increases with the treatment horizon for all methods. This illustrates the increasing difficulty of learning optimal DTRs when the sample size is held fixed while the number of stages grows. Second, weaker policy-relevant overlap further deteriorates performance, with the deterioration becoming more pronounced at longer horizons.

The proposed DR method generally achieves lower normalized regret than
Q-learning and Q-search and remains competitive with IPW, although IPW
performs better in some of the weakest-overlap, longer-horizon settings.
Additional clipping of the estimated propensity scores has little effect
under strong overlap and produces modest improvements in some of the
weakest-overlap settings. This suggests that clipping can help mitigate
the finite-sample consequences of extreme inverse-probability weights.
Q-learning and Q-search avoid explicit inverse-probability weighting but
nevertheless exhibit increasing regret as the horizon grows, consistent
with the increasing difficulty of Q-function estimation over longer
horizons.


\begin{table}[!htbp]
\centering
\caption{Monte Carlo results for long horizons and weak overlap}
\label{table:mc_long_horizon_overlap}
\begin{threeparttable}
\scriptsize
\setlength{\tabcolsep}{2.5pt}
\begin{tabular}{llccccc}
\toprule
Method & Metric
& \multicolumn{5}{c}{Number of stages $(T)$} \\
\cmidrule(lr){3-7}
& & 2 & 4 & 6 & 8 & 10 \\
\midrule

\multicolumn{7}{l}{\textit{Panel A: $\eta=0.50$}} \\[0.2em]

\multirow{2}{*}{Q-learn}
& Normalized regret (SD)
& 0.342 (0.020) & 0.417 (0.017) & 0.443 (0.016) & 0.459 (0.019) & 0.470 (0.018) \\
& Time (sec.)
& 1.00 & 2.57 & 4.35 & 6.25 & 8.12 \\
\addlinespace[0.2em]

\multirow{2}{*}{Q-search}
& Normalized regret (SD)
& 0.310 (0.052) & 0.381 (0.057) & 0.409 (0.049) & 0.420 (0.051) & 0.426 (0.049) \\
& Time (sec.)
& 9.90 & 21.76 & 34.21 & 47.18 & 60.24 \\
\addlinespace[0.2em]

\multirow{2}{*}{IPW}
& Normalized regret (SD)
& 0.120 (0.050) & 0.209 (0.049) & 0.253 (0.046) & 0.280 (0.039) & 0.292 (0.041) \\
& Time (sec.)
& 1.48 & 3.63 & 5.94 & 8.28 & 10.57 \\
\addlinespace[0.2em]

\multirow{2}{*}{DR}
& Normalized regret (SD)
& 0.039 (0.017) & 0.132 (0.033) & 0.192 (0.037) & 0.233 (0.034) & 0.258 (0.034) \\
& Time (sec.)
& 9.73 & 25.52 & 42.36 & 59.78 & 77.57 \\
\addlinespace[0.2em]

\multirow{2}{*}{DR (clip=0.05)}
& Normalized regret (SD)
& 0.038 (0.017) & 0.132 (0.033) & 0.193 (0.035) & 0.232 (0.034) & 0.259 (0.037) \\
& Time (sec.)
& 10.35 & 25.63 & 42.38 & 60.17 & 77.88 \\
\addlinespace[0.2em]

\multirow{2}{*}{DR (clip=0.025)}
& Normalized regret (SD)
& 0.037 (0.016) & 0.131 (0.032) & 0.193 (0.037) & 0.232 (0.035) & 0.256 (0.035) \\
& Time (sec.)
& 10.28 & 25.75 & 42.39 & 59.90 & 76.82 \\

\addlinespace[0.4em]
\midrule

\multicolumn{7}{l}{\textit{Panel B: $\eta=0.20$}} \\[0.2em]

\multirow{2}{*}{Q-learn}
& Normalized regret (SD)
& 0.343 (0.022) & 0.417 (0.018) & 0.445 (0.017) & 0.458 (0.019) & 0.471 (0.018) \\
& Time (sec.)
& 0.88 & 2.21 & 3.73 & 5.51 & 7.39 \\
\addlinespace[0.2em]

\multirow{2}{*}{Q-search}
& Normalized regret (SD)
& 0.316 (0.055) & 0.382 (0.057) & 0.413 (0.051) & 0.429 (0.049) & 0.442 (0.045) \\
& Time (sec.)
& 6.36 & 18.21 & 33.22 & 47.25 & 60.92 \\
\addlinespace[0.2em]

\multirow{2}{*}{IPW}
& Normalized regret (SD)
& 0.156 (0.060) & 0.261 (0.051) & 0.307 (0.044) & 0.332 (0.041) & 0.343 (0.040) \\
& Time (sec.)
& 1.45 & 3.91 & 5.79 & 7.67 & 9.65 \\
\addlinespace[0.2em]

\multirow{2}{*}{DR}
& Normalized regret (SD)
& 0.086 (0.041) & 0.221 (0.043) & 0.279 (0.039) & 0.317 (0.041) & 0.339 (0.040) \\
& Time (sec.)
& 8.00 & 22.68 & 38.76 & 55.77 & 72.87 \\
\addlinespace[0.2em]

\multirow{2}{*}{DR (clip=0.05)}
& Normalized regret (SD)
& 0.086 (0.042) & 0.218 (0.044) & 0.281 (0.041) & 0.317 (0.041) & 0.340 (0.040) \\
& Time (sec.)
& 10.91 & 25.84 & 39.98 & 57.57 & 74.23 \\
\addlinespace[0.2em]

\multirow{2}{*}{DR (clip=0.025)}
& Normalized regret (SD)
& 0.086 (0.043) & 0.219 (0.043) & 0.280 (0.038) & 0.318 (0.039) & 0.339 (0.040) \\
& Time (sec.)
& 9.83 & 24.18 & 40.99 & 57.15 & 72.91 \\

\addlinespace[0.4em]
\midrule

\multicolumn{7}{l}{\textit{Panel C: $\eta=0.05$}} \\[0.2em]

\multirow{2}{*}{Q-learn}
& Normalized regret (SD)
& 0.343 (0.022) & 0.419 (0.017) & 0.445 (0.016) & 0.458 (0.016) & 0.470 (0.017) \\
& Time (sec.)
& 0.91 & 2.19 & 3.67 & 5.32 & 7.03 \\
\addlinespace[0.2em]

\multirow{2}{*}{Q-search}
& Normalized regret (SD)
& 0.320 (0.062) & 0.408 (0.060) & 0.425 (0.053) & 0.435 (0.047) & 0.436 (0.047) \\
& Time (sec.)
& 5.06 & 15.62 & 30.61 & 46.11 & 62.35 \\
\addlinespace[0.2em]

\multirow{2}{*}{IPW}
& Normalized regret (SD)
& 0.170 (0.055) & 0.275 (0.052) & 0.320 (0.044) & 0.349 (0.041) & 0.358 (0.038) \\
& Time (sec.)
& 1.16 & 3.51 & 6.09 & 8.28 & 9.64 \\
\addlinespace[0.2em]

\multirow{2}{*}{DR}
& Normalized regret (SD)
& 0.140 (0.043) & 0.265 (0.043) & 0.327 (0.042) & 0.364 (0.040) & 0.384 (0.038) \\
& Time (sec.)
& 7.79 & 20.58 & 36.68 & 52.91 & 69.47 \\
\addlinespace[0.2em]

\multirow{2}{*}{DR (clip=0.05)}
& Normalized regret (SD)
& 0.142 (0.044) & 0.264 (0.044) & 0.330 (0.039) & 0.363 (0.041) & 0.383 (0.039) \\
& Time (sec.)
& 8.97 & 26.75 & 40.95 & 56.49 & 72.73 \\
\addlinespace[0.2em]

\multirow{2}{*}{DR (clip=0.025)}
& Normalized regret (SD)
& 0.142 (0.044) & 0.260 (0.044) & 0.327 (0.039) & 0.363 (0.040) & 0.383 (0.038) \\
& Time (sec.)
& 8.64 & 23.40 & 39.83 & 55.76 & 71.66 \\

\bottomrule
\end{tabular}

\begin{tablenotes}\footnotesize
\item Notes:
The results are based on 500 Monte Carlo replications.
The rows labeled ``Normalized regret (SD)'' report mean normalized
regret, with Monte Carlo standard deviations in parentheses.
Time reports the average elapsed computation time per replication in
seconds. This simulation was run in the same computing environment as
the main simulations, with Monte Carlo replications parallelized across
10 workers.
\end{tablenotes}
\end{threeparttable}
\end{table}

\newpage
\section{Details of the WGAN-based Empirical Monte Carlo Design}
\label{app:wgan_dgp}

This appendix describes the construction of DGP3 in the simulation study. The goal of DGP3 is to evaluate the proposed method in a finite-sample environment calibrated to the Project STAR data used in the empirical application. Our construction follows the empirical Monte Carlo procedure of \cite{Athey_et_al_2024}, who use WGANs to generate artificial data that closely mimic a real dataset and then evaluate competing estimators using repeated samples from a large synthetic population. We adapt their procedure to the dynamic treatment setting.

The main difference from the static treatment setting considered by \cite{Athey_et_al_2024} is that, in our setting, the data-generating process must respect the timing of a dynamic treatment problem. In a static setting, it is sufficient to generate potential outcomes under alternative treatment values conditional on pre-treatment covariates. In our two-stage setting, however, the first-stage treatment affects the intermediate state, and the second-stage treatment is assigned after that intermediate state is realized. Therefore, we generate not only potential final outcomes but also counterfactual intermediate states. Specifically, for each synthetic individual, we generate $S_2(a_1)$ for each first-stage treatment $a_1\in\{0,1\}$ and $Y_2(a_1,a_2)$ for each treatment history $(a_1,a_2)\in\{0,1\}^2$. We then evaluate each DTR using these generated counterfactual objects.

In the Project STAR application, the two treatment variables are the kindergarten and grade-1 class types, denoted by $A_1$ and $A_2$, respectively. We set $A_t=0$ for a regular-size class with a full-time teacher aide and $A_t=1$ for a small-size class without a teacher aide. The first-stage state vector $S_1$ consists of the following seven variables: student gender, student ethnicity, eligibility for free or reduced-price school lunch, school location type, teacher's degree, years of teaching experience, and teacher ethnicity. The intermediate state vector $S_2$ contains the kindergarten mathematics test score. The final outcome $Y_2$ is the cost-adjusted grade-1 outcome.

We construct the WGAN-based empirical Monte Carlo design in four steps.

First, we create a balanced training sample from the Project STAR data. This step extends the training-sample balancing used in the WGAN implementation of \cite{Athey_et_al_2024} to the dynamic treatment setting. In their binary-treatment setting, a balanced dataset is constructed by sampling treated and control observations with similar probabilities, and the WGAN is trained on this balanced dataset so that the quality of the generated outcomes is similar for both treatment groups. In our two-stage setting, the analogous objects are the four treatment histories rather than a single binary treatment. Let $p(a_1,a_2)$ be the empirical frequency of the treatment path $(A_1,A_2)=(a_1,a_2)$. We resample observations with replacement using weights proportional to $1/p(A_1,A_2)$. The size of the balanced training sample is four times the size of the original empirical sample. This balancing step is used only for training the conditional WGANs so that all four treatment histories are sufficiently represented.

Second, we train three conditional WGANs. Specifically, we approximate the empirical distribution of the observed dynamic trajectory by decomposing it into three conditional components: the distribution of first-stage state variables $S_1$ given the kindergarten class type $A_1$, the distribution of intermediate state variables $S_2$ and the grade-1 class type $A_2$ given first-stage information $(S_1,A_1)$, and the distribution of the final outcome $Y_2$ given the second-stage history. This conditional decomposition adapts the conditional WGAN approach of \cite{Athey_et_al_2024} to our two-stage dynamic treatment setting. Rather than training a single WGAN for the entire observed trajectory, we train separate conditional WGANs for these components of the dynamic data-generating process.

The first WGAN generates the first-stage state variables conditional on the kindergarten class type, denoted by $S_1 \sim \widehat P_0(\cdot\mid A_1)$. We use this factorization as a statistical device for approximating the empirical joint distribution of the Project STAR data. It should not be interpreted as a causal ordering in which the kindergarten class type determines the first-stage state variables.

The second WGAN generates the intermediate state variables and the grade-1 class type conditional on first-stage information, denoted by $(S_2,A_2) \sim \widehat P_1(\cdot\mid S_1,A_1)$. The third WGAN generates the final outcome conditional on the second-stage history and the grade-1 class type, denoted by $Y_2 \sim \widehat P_2(\cdot\mid S_1,A_1,S_2,A_2)$. This final step is the dynamic analogue of generating potential outcomes conditional on common pre-treatment covariates in the static setting. For the continuous variables generated by the WGANs, we impose nonnegativity restrictions when appropriate.

For the WGAN implementation, we use the same feed-forward neural network architecture as in \cite{Athey_et_al_2024} for all three conditional WGANs, while allowing the input and output dimensions to vary across the three components of the dynamic data-generating process.\footnote{Specifically, as in \cite{Athey_et_al_2024}, both the generator and the critic have three hidden layers with 128 units in each hidden layer, and the hidden layers use rectified linear unit activations. The generator output layer uses transformations appropriate for the variable type: sigmoid transformations for binary or categorical variables, rectified linear transformations for nonnegative continuous variables, and identity transformations for unrestricted continuous variables. The critic has the same hidden-layer architecture and a scalar linear output. In our implementation, the three WGANs are trained using the same tuning parameters: batch size 470, a maximum of 1000 epochs, and learning rates $10^{-3}$ for both the critic and the generator.}

Third, following \cite{Athey_et_al_2024}, we use the fitted WGANs to generate a large synthetic population calibrated to the Project STAR data. This synthetic population operationally defines DGP3 for the Monte Carlo experiment, and the Monte Carlo training and evaluation samples are drawn from it. We first draw observations with replacement from the Project STAR data to obtain empirical draws of the auxiliary conditioning variables and the first-stage treatment $A_1$. We then apply the first WGAN to generate $S_1$ conditional on $A_1$. Next, we apply the second WGAN to generate the observed pair $(S_2,A_2)$ conditional on $(S_1,A_1)$, so that the second-stage treatment assignment is also simulated. Finally, we apply the third WGAN to generate the observed outcome $Y_2$ conditional on $(S_1,A_1,S_2,A_2)$. This gives a synthetic observational dataset whose joint distribution is calibrated to the Project STAR application.

Fourth, we augment the synthetic population with counterfactual intermediate states and counterfactual final outcomes, which are used only for welfare evaluation. For each $a_1\in\{0,1\}$, we set $A_1=a_1$ and use the second WGAN to generate $S_2(a_1) \sim \widehat P_1(\cdot\mid S_1,A_1=a_1)$. In this counterfactual step, we retain only the generated intermediate state; the generated second-stage treatment from the second WGAN is not used. We generate $S_2(0)$ and $S_2(1)$ only once for each synthetic individual. Then, for each treatment history $(a_1,a_2)\in\{0,1\}^2$, we set $A_1=a_1$ and $A_2=a_2$, set the intermediate state equal to $S_2(a_1)$, and use the third WGAN to generate $Y_2(a_1,a_2) \sim \widehat P_2(\cdot\mid S_1,A_1=a_1,S_2(a_1),A_2=a_2)$. Thus, $Y_2(a_1,0)$ and $Y_2(a_1,1)$ are generated using the same counterfactual intermediate state $S_2(a_1)$. This construction respects the timing of the dynamic treatment problem, because $S_2(a_1)$ is realized after the first-stage treatment and before the second-stage treatment.

The resulting synthetic population contains both the simulated observed
trajectory $(S_1,A_1,S_2,A_2,Y_2)$ and the counterfactual objects needed
for policy evaluation. For each synthetic individual, we generate
$S_2(0)$, $S_2(1)$, and the four potential final outcomes
$\{Y_2(0,0),Y_2(0,1),Y_2(1,0),Y_2(1,1)\}$. 

For each Monte Carlo replication, we draw a training sample of size $n$
from the simulated observational component of DGP3 and estimate a DTR
using each policy-learning method. We numerically approximate the oracle
welfare-maximizing DTR within $\Pi$ using an independent sample of
50,000 observations from DGP3. The welfare of both the resulting oracle
policy and each estimated policy is evaluated on a separate independent
test sample of 50,000 observations from DGP3. The same test sample is
used to compute regret.


\section{Costs and Monetary Benefits of Teacher Aides and Small-Size
Classes \label{appendix:cost_benefit}}

This appendix outlines the calculation of the costs associated with a
full-time teacher aide and small-size classes in our empirical application
(Section~\ref{sec:empirical_illustration}) and evaluates the monetary
benefits of the estimated DTRs. Since the outcome measure in our study is
test scores, we convert the monetary costs per student into equivalent
test-score units, following the discussion in
\citeauthor{krueger_1999} (\citeyear{krueger_1999}, Section IV). In line
with \cite{krueger_1999}, we apply a 3\% annual discount rate,
discounting all monetary benefit and cost streams back to age 6, the
point of intervention.

We begin by translating first-grade mathematics test scores into the
present value of lifetime income gains, following the discussion in
\citeauthor{krueger_1999} (\citeyear{krueger_1999}, Section IV).
Estimates from the High School and Beyond sample in
\cite{Murnane_et_al_1995} suggest that male high school seniors who
scored one standard deviation higher on the basic math achievement test
in 1980 earned 7.93\% more six years later, while the corresponding
figure for females was 10.98\%. Average annual earnings in the United
States in 1996 were \$34,705 for men and \$20,570 for women among
workers aged 18 and older \citep{US_Census}. Assuming that real earnings
remain constant and that the estimates in \cite{Murnane_et_al_1995}
apply to first-grade test scores in the STAR experiment, the present
value of the average earnings gain associated with a one-point increase
in test scores (0.025 standard deviations) is \$1,125.26 for men and
\$923.53 for women. This calculation assumes that students enter the
workforce at age 20 and retire at age 65 and applies the 3\% annual
discount rate.\footnote{\cite{Chetty_et_al_2011} also estimate the
average lifetime earnings gain associated with test-score improvements.
However, since their analysis focuses on the sum of reading and
mathematics test scores rather than mathematics scores alone, we use the
results from \cite{Murnane_et_al_1995} to estimate the average earnings
gain from increasing mathematics test scores.}

Next, we assess the monetary costs of the two class types, drawing on \cite{Word_et_al_1990} and \cite{krueger_1999}. \cite{krueger_1999} estimated that reducing class sizes by one-third would increase per-student costs by roughly \$2,151 per year, including capital costs. We use this figure as the annual per-student cost of assigning a student to a small-size class rather than to a regular-size class without a teacher aide. Since grade K is the point of intervention and grade 1 occurs one year later, applying a 3\% real discount rate gives present values of approximately \$2,151 for grade K and \$2,088.35 ($=2,151/1.03$) for grade 1.

We then convert the cost of small-size classes into test score points. For kindergarten, the cost of a small-size class corresponds to 1.912 points ($\approx 2,151/1,125.26$) for each male student and 2.329 points ($\approx 2,151/923.53$) for each female student on the mathematics test. In grade 1, the cost corresponds to 1.856 points ($\approx 2,088.35/1,125.26$) for males and 2.261 points ($\approx 2,088.35/923.53$) for females. Assuming equal proportions of male and female students, we estimate the average cost of a small-size class per student to be equivalent to 2.120 points in kindergarten and 2.059 points in grade 1.

We also assess the cost of employing a full-time teacher aide. According to \cite{Word_et_al_1990}, adding a full-time aide in Grades K--3 across Tennessee cost approximately 75 million dollars annually, while reducing class sizes by one-third cost around 196--205 million dollars per year. Thus, the cost of a full-time teacher aide was at most 38\% of the cost of class-size reduction. Based on this upper-bound calculation and the per-student cost estimate
in \citet{krueger_1999}, we estimate the cost of a teacher aide to be \$817.4 ($\approx 2,151 \times 0.38$) per student per grade, assuming that kindergarten and grade 1 have equal undiscounted costs. Applying a 3\% real discount rate, the present values of the costs of a full-time teacher aide for grades K and 1 are approximately \$817.40 and \$793.60, respectively.

Using these estimates, we convert the cost of a full-time teacher aide into test score points. For kindergarten, the cost of a full-time aide corresponds to 0.726 points ($\approx 817.40/1,125.26$) for each male student and 0.885 points ($\approx 817.40/923.53$) for each female student on the mathematics test. In grade 1, the cost corresponds to 0.705 points ($\approx 793.60/1,125.26$) for males and 0.859 points ($\approx 793.60/923.53$) for females. Assuming an equal ratio of male and female students, we estimate the average cost of a full-time aide per student to be equivalent to 0.806 points in kindergarten and 0.782 points in grade~1.

Because the empirical choice set consists of a regular-size class with a
full-time teacher aide and a small class without one, the relevant cost
adjustment in the welfare function is the incremental cost of assigning
a student to a small class rather than to a regular-size class with a
full-time teacher aide. Therefore, in our empirical analysis, we set the
incremental costs to
$c_1 = 2.120 - 0.806 \approx 1.31$
for kindergarten and
$c_2 = 2.059 - 0.782 \approx 1.28$
for grade 1. We use these cost values in the empirical analysis in
Section~\ref{sec:empirical_illustration}.

Based on these calculations, we estimate that the DTR $\hat{\pi}$
presented in Table~\ref{table:empirical_results} would increase the
present value of each student's lifetime earnings, net of the associated
class-type costs, by an average of \$12,026 relative to assigning a
regular-size class with a full-time teacher aide at both stages and by
\$3,268 relative to assigning a small class at both stages.

We are aware that this cost-benefit calculation relies on several assumptions. In particular, real earnings may increase or decrease over time; the relationship between test scores and future earnings may differ from that assumed above; and general-equilibrium effects may arise. For example, an increase in the overall education level of the population could expand the supply of skilled labor and thereby lower wage rates for more educated individuals.


\section{Estimated DTR under the Depth-3 Grade-1 Policy Class}
\label{app:depth3_tree}

Figure~\ref{fig:estimated_decision_trees_depth3} displays the estimated
DTR when the grade-1 policy class $\Pi_2$ is expanded from depth-2 to
depth-3 trees, while the grade-K policy class $\Pi_1$ remains at the
baseline depth-1 specification. The estimated grade-K policy is unchanged,
whereas the grade-1 allocation rule changes under the richer policy class.
As shown in Table~\ref{table:empirical_results}, however, the depth-3
specification does not substantially change estimated welfare.

\begin{figure}[!htbp]
\centering
\caption{Estimated DTR for class-type allocation in grades K and 1 under the depth-3 grade-1 policy-class specification}
\label{fig:estimated_decision_trees_depth3}
\vspace{0.3cm}

\medskip

\centering
{\small (a) Policy for grade K}\\
\vspace{0.4cm}

\begin{tikzpicture}
[   scale = 1.1,
every node/.style={font=\small, align=center},
    level 1/.style = {sibling distance = 3cm},
    level 2/.style = {sibling distance = 3cm},
    edge from parent/.style={draw, -latex},
]
 
\node {Free/reduced-price lunch eligibility}
    child {node {Small} 
    edge from parent node [left] {True}} 
    child {node {Aide}
    edge from parent node [right] {False}
    };
 
\end{tikzpicture}

\bigskip

\centering
{\small (b) Policy for grade 1}\\
\vspace{0.4cm}
\begin{tikzpicture}[
    scale=0.95,
    transform shape,
    every node/.style={font=\small, align=center},
    edge from parent/.style={draw, -latex},
    level distance=18mm,
    level 1/.style={sibling distance=90mm},
    level 2/.style={sibling distance=48mm},
    level 3/.style={sibling distance=24mm}
]

\node {Math test score $\leq 463$}
    child {
        node {Math test score $\leq 429$}
        child {
            node {Math test score $\leq 423$}
            child {
                node {Aide}
                edge from parent node[left] {True}
            }
            child {
                node {Small}
                edge from parent node[right] {False}
            }
            edge from parent node[left] {True}
        }
        child {
            node {Grade-K type = Aide}
            child {
                node {Small}
                edge from parent node[left] {True}
            }
            child {
                node {Aide}
                edge from parent node[right] {False}
            }
            edge from parent node[right] {False}
        }
        edge from parent node[left] {True}
    }
    child {
        node {Math test score $\leq 494$}
        child {
            node {Math test score $\leq 478$}
            child {
                node {Small}
                edge from parent node[left] {True}
            }
            child {
                node {Aide}
                edge from parent node[right] {False}
            }
            edge from parent node[left] {True}
        }
        child {
            node {Math test score $\leq 528$}
            child {
                node {Small}
                edge from parent node[left] {True}
            }
            child {
                node {Aide}
                edge from parent node[right] {False}
            }
            edge from parent node[right] {False}
        }
        edge from parent node[right] {False}
    };

\end{tikzpicture}

\vspace{0.3cm}
\begin{minipage}{0.92\textwidth}
\footnotesize
\textit{Notes:}
This figure displays the estimated DTR from the sensitivity analysis in 
Section~\ref{sec:empirical_illustration}. The grade-K policy class $\Pi_1$ is the 
same as in the baseline specification and is restricted to depth-1 trees. The grade-1 
policy class $\Pi_2$ is expanded from depth-2 to depth-3 trees. Panels (a) and (b) 
show the estimated class-type allocation rules for grade K and grade 1, respectively. 
``Aide'' denotes assignment to a regular-size class with a full-time teacher aide, and 
``Small'' denotes assignment to a small-size class without a teacher aide.
\end{minipage}
\end{figure}


\end{document}